\documentclass[aps,prb,twocolumn,notitlepage,citeautoscript,superscriptaddress,10pt,footinbib]{revtex4-1}

\usepackage{times}
\usepackage{amsmath,amssymb,mathrsfs,bm,setspace}
\usepackage{dcolumn}% Align table columns on decimal point
\usepackage{graphicx}
\usepackage{latexsym,hyperref}
\usepackage{xspace}

\usepackage{color}

\usepackage{bm}
\usepackage{epsfig}
\usepackage{amsmath}
\usepackage{color}

\newcommand{\jeff}{j_{\textrm{eff}}}
\newcommand{\threet}{(\frac{3}{2},\frac{3}{2})}
\newcommand{\threeh}{(\frac{3}{2},\frac{1}{2})}
\newcommand{\halfh}{(\frac{1}{2},\frac{1}{2})}

\newcommand{\threehpm}{(\frac{3}{2},\pm\frac{1}{2})}

\newcommand{\SRO}{Sr$_2$RuO$_4$\xspace}

\newcommand{\ttgf}{$t_{2g}^4$\xspace}

\newcommand{\tfl}{$T_\mathrm{FL}$\xspace}

\newcommand{\lsoc}{\lambda^{}_\mathrm{SOC}\xspace}

\begin{document}

\title{Interplay between spin-orbit coupling and van Hove singularity in the Hund's metallicity of \SRO}

\author{Hyeong~Jun \surname{Lee}}

\affiliation{Center for Theoretical Physics of Complex Systems, Institute for Basic Science (IBS), Daejeon 34126, Republic of Korea}
\affiliation{Center for Correlated Electron Systems, Institute for Basic Science (IBS), Seoul 08826, Republic of Korea}
\affiliation{Department of Physics and Astronomy, Seoul National University, Seoul 08826, Republic of Korea}

\author{Choong~H. \surname{Kim}}
\email[]{chkim82@snu.ac.kr}

\affiliation{Center for Correlated Electron Systems, Institute for Basic Science (IBS), Seoul 08826, Republic of Korea}
\affiliation{Department of Physics and Astronomy, Seoul National University, Seoul 08826, Republic of Korea}

\author{Ara \surname{Go}}
\email[]{arago@ibs.re.kr}

\affiliation{Center for Theoretical Physics of Complex Systems, Institute for Basic Science (IBS), Daejeon 34126, Republic of Korea}
\affiliation{Department of Physics, Chonnam National University, Gwangju 61186, Republic of Korea}

\begin{abstract}
We investigate the roles of spin-orbit coupling and the van Hove singularity in the dynamical properties of \SRO, which become prominent at zero and very low temperature,
by means of density functional theory plus dynamical mean-field theory with an exact diagonalization solver.
We examine the crossover between a Fermi liquid and a Hund's metal for a wide range of temperatures and Hund's coupling.
In the absence of doping, we confirm that the Fermi liquid persists at zero temperature even with nonzero Hund's coupling.
The freezing-moment mechanism suggests that thermal fluctuations lead to a suppression of the Fermi liquid phase and promote Hund's metallicity with incoherence.
We show that the van Hove singularity is an additional key ingredient to drive the suppression at very low temperature by observing a doping dependence of the freezing or long-lived paramagnetic moments.
The role of spin-orbit coupling is marked by an amplified Van Vleck contribution of spin susceptibility, significantly promoting Hund's metallicity.
Together with the known doping dependence of Hund's metallicity, the additional VHS doping dependence found here may allow for the control of the Hund's metallicity of \SRO (\ttgf) by a fine tuning of the doping or possibly even strain.
\end{abstract}

\date{\today}
\maketitle

\section{Introduction}
The recent discovery of Hund's metals promotes the importance of Hund's coupling in correlated multi-orbital systems~\cite{Haule2009,Yin2011,Georges2013}.
While Hund's coupling can appear to reduce correlations by suppressing the Mott phase, it also enhances another type of correlation that exhibits various characteristic behaviors---for example, large renormalization masses depending on filling~\cite{DeMedici2011PRL,DeMedici2011PRB}, spin (or $J$) freezing~\cite{Werner2008,AJKim2017}, and non-Fermi liquid~\cite{Werner2008}.
Furthermore, a strong Hund's coupling in multi-orbital systems induces significant orbital-selective behaviors~\cite{Koga2004,*Anisimov2002, Werner2007,Werner2009,DeMedici2009,Werner2016}.
As the name implies, the aforementioned characteristics of Hund's metals originate from local correlation effects.

Strontium ruthenate, \SRO, is a well-known correlated metal~\cite{Mackenzie1996,Bergemann2000,Bergemann2003,Damascelli2000,Shen2001,Iwasawa2005,Ingle2005,Kidd2005,Shen2007,Iwasawa2010,Iwasawa2012,Stricker2014,Tamai2019}
that undergoes a crossover from Hund's metal to Fermi liquid as the temperature drops~\cite{Maeno1997,Hussey1998,Tyler1998}.
It exhibits unconventional superconductivity below 1.5 K~\cite{Maeno1994,Rice1995,Mackenzie1998,Ishida1998,Mackenzie2003}, the precise mechanism of which is still under debate~\cite{Mackenzie2017,Pustogow2019}.
Recent experiments observing that $T_c$ is enhanced by strain~\cite{Hicks2014,Steppke2017,Barber2018,Luo2019,Barber2019,Pustogow2019} have puzzled the relation between the van Hove singularity (VHS) and the superconductivity of \SRO.
Since the VHS in ruthenates lies in close vicinity to their Fermi levels, even a tiny tuning of its electronic structure may result in a drastic change by amplifying correlation-induced instability.
One example of this phenomenon among the perovskite ruthenates is SrRuO$_3$, with ferromagnetism arising from a high density of states (DOS) at its Fermi level due to a nearby VHS~\cite{Han2016}.
To clarify the effects of the VHS in \SRO, while there have been intensive studies on how the normal state reacts when the VHS approaches the Fermi level~\cite{Kikugawa2004PRBR,Kikugawa2004,Shen2007,Burganov2016,Herman2019}, the correlation effects across the VHS have not received much attention yet.

Pioneering studies have adequately explained the normal state properties of \SRO via three-band models 
~\cite{Liebsch2000,Pchelkina2007,Mravlje2011,Deng2016,Mravlje2016,Zhang2016,Sarvestani2018,MKim2018,Facio2018,Deng2019,Zingl2019,Strand2019,Gingras2019,Kugler2020,Linden2020,Karp2020,Ryee2016,Cobo2016}
mostly employing dynamical mean-field theory (DMFT) 
in combination with density functional theory (DFT).
Research has shown that Hund's coupling causes orbital-dependent correlations~\cite{Mackenzie1996,Bergemann2000,Bergemann2003} and bad metallic behaviors at nonzero temperatures, which are prototypical features of Hund's metals.
Another important local interaction in \SRO is the spin-orbit coupling (SOC) that lifts the degeneracy between $j_\mathrm{eff}=1/2$ and $3/2$~\cite{Veenstra2014} (where $\jeff$ is the effective total angular momentum~\cite{BJKim2008})
and also modifies the Fermi surface~\cite{Pavarini2006,Haverkort2008,Zhang2016,MKim2018,Facio2018,Sarvestani2018,Tamai2019}.
Since the continuous-time Monte Carlo solver~\cite{Gull2011}, which has been the most widely used in this system, can hardly reach $T$$<$300 K with SOC,
understanding of the nonperturbative nature of electronic correlations at very low temperature is not yet satisfactory.
As low temperatures can manifest a coherence of electrons and amplify the effects of the VHS lying near the Fermi level, studying the low temperature behavior of \SRO with various electron configurations is expected to fill this gap and may even provide insight into its superconductivity.

In this paper, we systematically investigate how shifting the VHS affects the spin-freezing phenomena by observing single-particle responses and long-lived paramagnetic moments represented as long-time correlators via the DFT+DMFT method~\cite{Georges1996,Kotliar2006}.
We construct a comprehensive phase diagram to clarify the effect of Hund's coupling at various temperatures in the presence of SOC, and further investigate how the system evolves across the VHS by applying electron doping.
We report a strong suppression of the Fermi liquid behavior by Hund's coupling, where the suppression is reinforced by the VHS upon doping.

\section{Model and Method}
We obtain a tight-binding Hamiltonian \cite{supp} of DFT-based maximally localized Wannier orbitals~\cite{Mostofi2014} from the Vienna \textit{Ab-initio} Simulation Package~\cite{Kresse1999}.
We then include the rotationally invariant Slater--Kanamori interaction~\cite{Kanamori1963} with a SOC of $\lsoc=0.1$ eV on top of the tight-binding Hamiltonian.
We use an exact diagonalization (ED) solver~\cite{Caffarel1994}
with three (six, considering spin) correlated and nine (eighteen) bath orbitals to satisfy the minimum requirement to describe multi-orbital physics~\cite{Koch2008,Senechal2010,Liebsch2012,Go2015,Fertitta2018}.
An advantage of the ED is that it allows us to approach zero and very low temperature in the presence of SOC without the fermionic sign problem~\cite{supp}.
While there exist finite-size effects coming from discrete bath energy levels~\cite{Go2015,Go2017},
the ED solver can provide a valuable complementary point of view inaccessible to any other impurity solver.

\begin{figure}[tb]
        \includegraphics[width=\columnwidth]{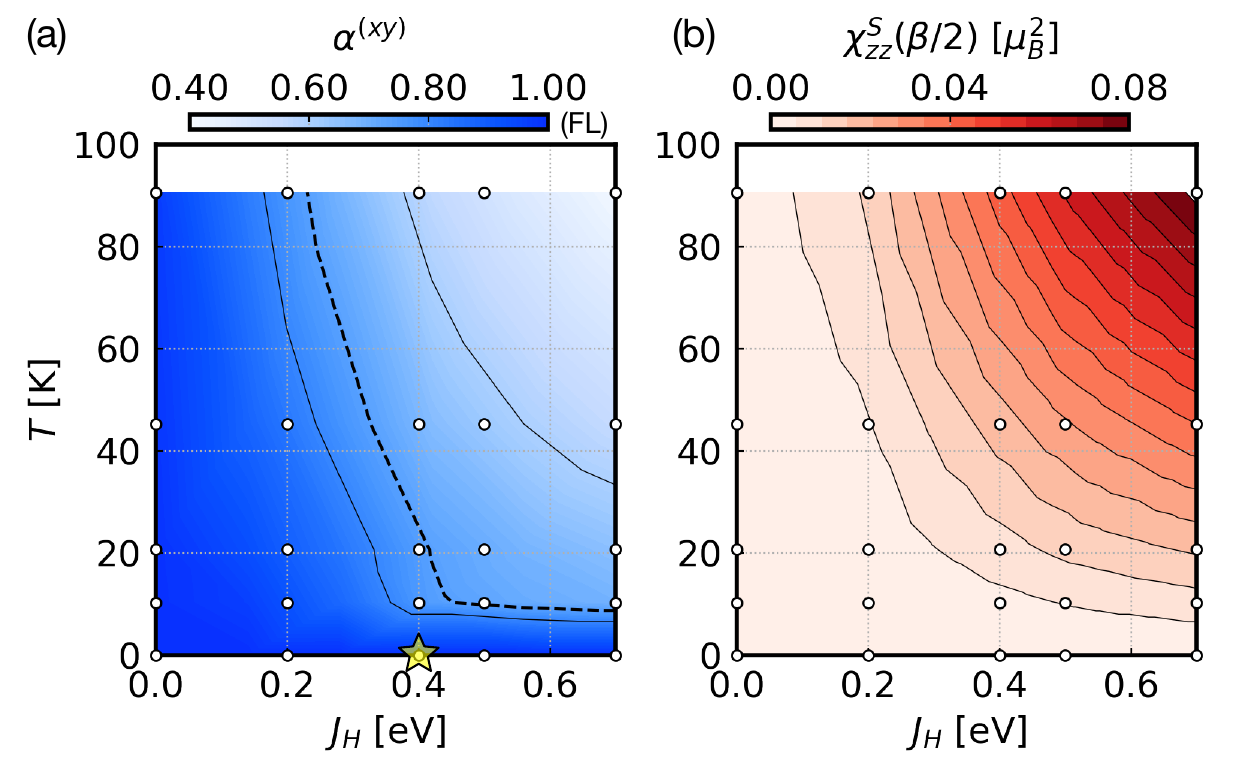}
        \caption{
        (a) Power exponent $\alpha$ of the imaginary part of the Matsubara self-energy and
        (b) long-time correlator $\chi^{S}_{zz}(\tau=\beta/2)$ in the $T$-$J_H$ plane.
	The dashed line in (a) denotes the crossover between Fermi liquid and Hund's metal regimes, and
	the yellow star marks a realistic value of $J_H$ for \SRO.
        %The data are collected around $n_{\mathrm{el}}=4$ (circles) and interpolated.
        }
\label{fig:pd}
\end{figure}

\section{Results}
%\subsection{Hund's metal crossover in \SRO}
We first perform calculations on bare \SRO to test whether the ED solver can reproduce the well-known consensus in this system.
The layered structure of \SRO induces a strong orbital selectivity with the aid of Hund's coupling and the VHS~\cite{Mravlje2011,Karp2020}.
Based on the self-energy, the $d_{xy}$ orbital is more strongly correlated in comparison to the other two $t_{2g}$ orbitals (see Supplementary Information for more details~\cite{supp}).
Since the $d_{xy}$ orbital experiences a drastic change in correlations over temperature~\cite{Mravlje2011,Linden2020,Karp2020} as well as Hund's coupling, its self-energy $\Sigma_{xy}(i\omega)$ is a good measure for identifying
whether it follows Fermi liquid behavior or not~\cite{supp}.
We present the power exponent of the imaginary part of the self-energies in Fig.~\ref{fig:pd}(a) as Hund's coupling and temperature vary.
We obtain power exponent $\alpha$ by fitting the self-energies to $\mathrm{Im}\Sigma(i\omega_n) \sim \omega_n^{\alpha}$ with the two lowest values of $\omega_n$.
The linear power of the self-energy [marked in blue in Fig.~\ref{fig:pd}(a)] with respect to the Matsubara frequency, i.e. $\mathrm{Im}\Sigma_{xy}(i\omega)\sim\omega$,
indicates that the system is in the Fermi liquid phase.
This phase is stable over the entire range of temperature that we can access when Hund's coupling is zero.
We define the Fermi liquid temperature $T_\mathrm{FL}$ for a given $J_H$ as the temperature whose corresponding value of $\alpha$ equals the $\alpha$ at $T$ = 25 K for $J_H$ = 0.4 eV,
where $T$ = 25 K is based on earlier experimental works~\cite{Maeno1997,Hussey1998,Tyler1998}.
To present how the Fermi liquid regime evolves as a function of $J_H$, we mark $T_\mathrm{FL}$ with a dashed line in Fig.~\ref{fig:pd}(a).
$T_\mathrm{FL}$ decreases once Hund's coupling is turned on, but the Fermi liquid ground state persists up to $J_H=0.7$ eV.
At $J_H$ = 0.4 eV, which is believed to be a typical Hund's coupling strength in \SRO~\cite{Mravlje2011,MKim2018}, the linear power seems to hold below 10 K in our result.
This scale is comparable with previous experimental results from resistivity curves~\cite{Maeno1997,Hussey1998,Tyler1998}.

Deviation from Fermi liquid behavior has been extensively studied with three-band models, where the long-lived local magnetic moment may promote the scattering rate of electrons~\cite{Werner2008,Liebsch2012,Stadler2015,Kowalski2019}.
Development of the long-lived paramagnetic moment (or so-called moment freezing) can be observed in the related magnetic susceptibility
as a non-vanishing long-time correlator.
The long-time correlator is defined as
\begin{align}
	\chi^{S}_{zz}(\tau=\beta/2)
	&=  \langle \hat{S}_z (\beta/2) \hat{S}_z \rangle  ,
\end{align}
where $\beta$ is the inverse temperature, and $\hat{S}_z$ is the $z$-component of the total angular momentum operator.
To examine the origin of the power reduction in \SRO, we compute the magnetic susceptibility as displayed in Fig.~\ref{fig:pd}(b).
We observe a clear similarity of the $T$- and $J_H$-dependencies between the two quantities in Fig.~\ref{fig:pd} (a) and (b).
When the exponent indicates Fermi liquid, the long-time correlator is vanishingly small. 
We note here that the mere existence of the long-lived moment does not necessarily lead to time-reversal symmetry breaking; rather, the long-lived moment coexists with its time-reversal partner and the system remains paramagnetic.

\begin{figure}[tb]
        \includegraphics[width=\columnwidth]{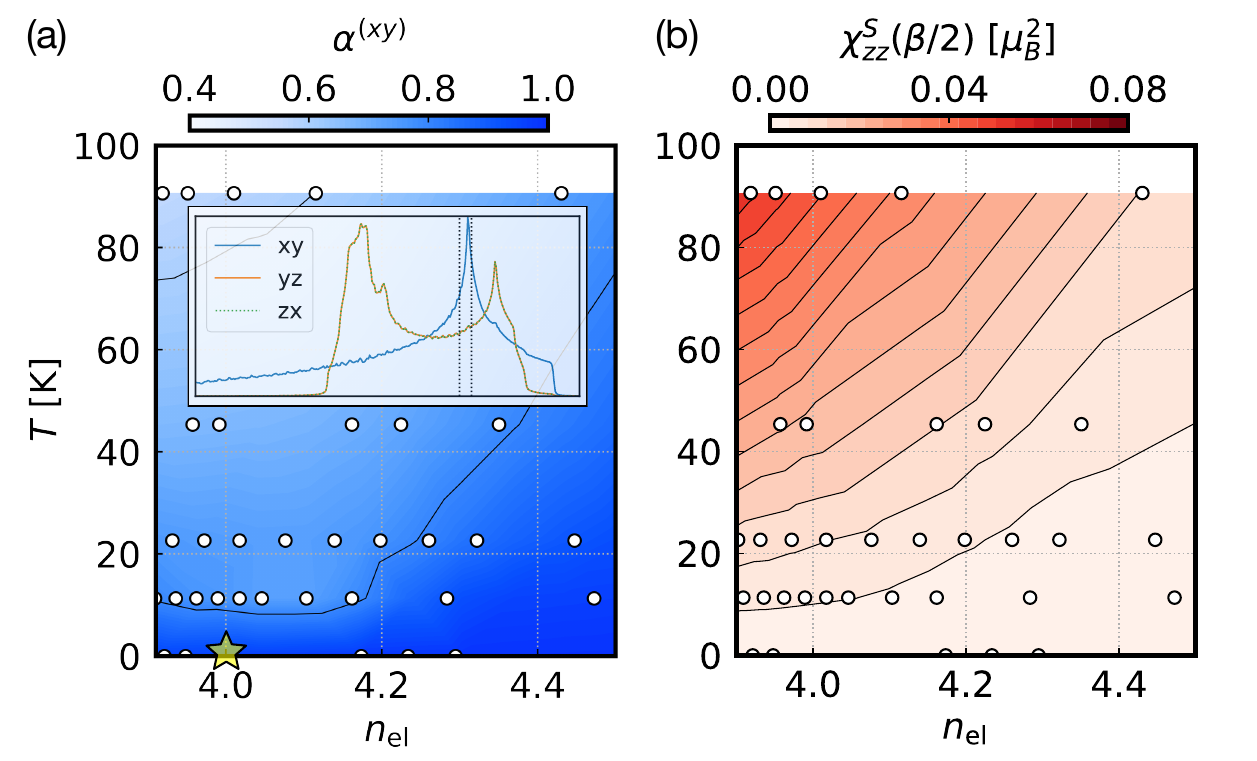}%
        \caption{
        Doping dependence around the van Hove singularity.
        (a) Power exponent of the imaginary part of the Matsubara self-energy and
        (b) long-time correlator $\chi^{S}_{zz}(\tau=\beta/2)$ in the $T$-$n_{\mathrm{el}}$ plane.
        Inset: Projected density of states on $t_{2g}$ for the non-interacting case.
        Vertical dotted lines denote the Fermi level when $n_{\mathrm{el}}=4$ and 4.35 around the VHS of the $d_{xy}$ band.
        }
\label{fig:doping}
\end{figure}

%\subsection{Effects of van Hove singularity with doping}
We have observed that Hund's coupling suppresses the Fermi liquid regime at \ttgf, showing consistent behavior with the long-time correlator.
This power exponent deviation and \tfl reduction in \SRO are reminiscent of the $J_H$-induced spin-freezing crossover in the three-band model at \ttgf~\cite{Werner2008,AJKim2017}.
However, there is a prominent difference upon electron doping due to the VHS of \SRO.
Previous model calculations for the Bethe lattice showed that Fermi liquid behavior is enhanced by electron doping on top of \ttgf~\cite{Werner2008,AJKim2017},
while earlier DFT+DMFT works focused on the Hund's metallicity of bare \SRO
~\cite{Mravlje2011,Deng2016,Mravlje2016,Zhang2016,Sarvestani2018,MKim2018,Facio2018,Deng2019,Zingl2019,Strand2019,Gingras2019,Kugler2020}.
In \SRO, though, there is a nontrivial doping dependence of the power exponent.
In Fig.~\ref{fig:doping}, we compare the power exponent from $\Sigma_{xy}$ and the long-time correlator for electron-doped cases.
Unlike the \ttgf case shown in Fig.~\ref{fig:pd}, the two quantities reveal a quantitatively distinct behavior for $n_{\mathrm{el}}>4$ (where $n_{\mathrm{el}}$ is electron occupancy).
Here, $T_\mathrm{FL}$ becomes minimal around $n_{\mathrm{el}} \sim 4.1$, while the long-time correlator monotonically decreases in the regime shown in Fig.~\ref{fig:doping}.
This can be attributed to the electronic structure of \SRO that has a VHS slightly above the Fermi level in the absence of doping~\cite{Oguchi1995,Singh1995}.
This supports that a mechanism separate from spin-freezing leads the power deviation.
The VHS strongly suppresses hybridizations and amplifies the correlation effect delivered to the self-energy~\cite{Mravlje2011,Kugler2020}.
The lowest $T_\mathrm{FL}$ is achieved when a Lifshitz transition appears and the VHS is expected to cross the Fermi level.
Its doping concentration is, however, a bit less than one previously reported in experiments and in a non-interacting case~\cite{Kikugawa2004PRBR,Shen2007}.
The lowest $T_\mathrm{FL}$ moves to a higher $n_{\mathrm{el}}$ if SOC is turned off~\cite{supp};
this originates from the VHS and the enhanced magnetic fluctuations by SOC, which will be discussed in more detail later.

\begin{figure}[tb]
	\centering
	\includegraphics[width=\columnwidth]{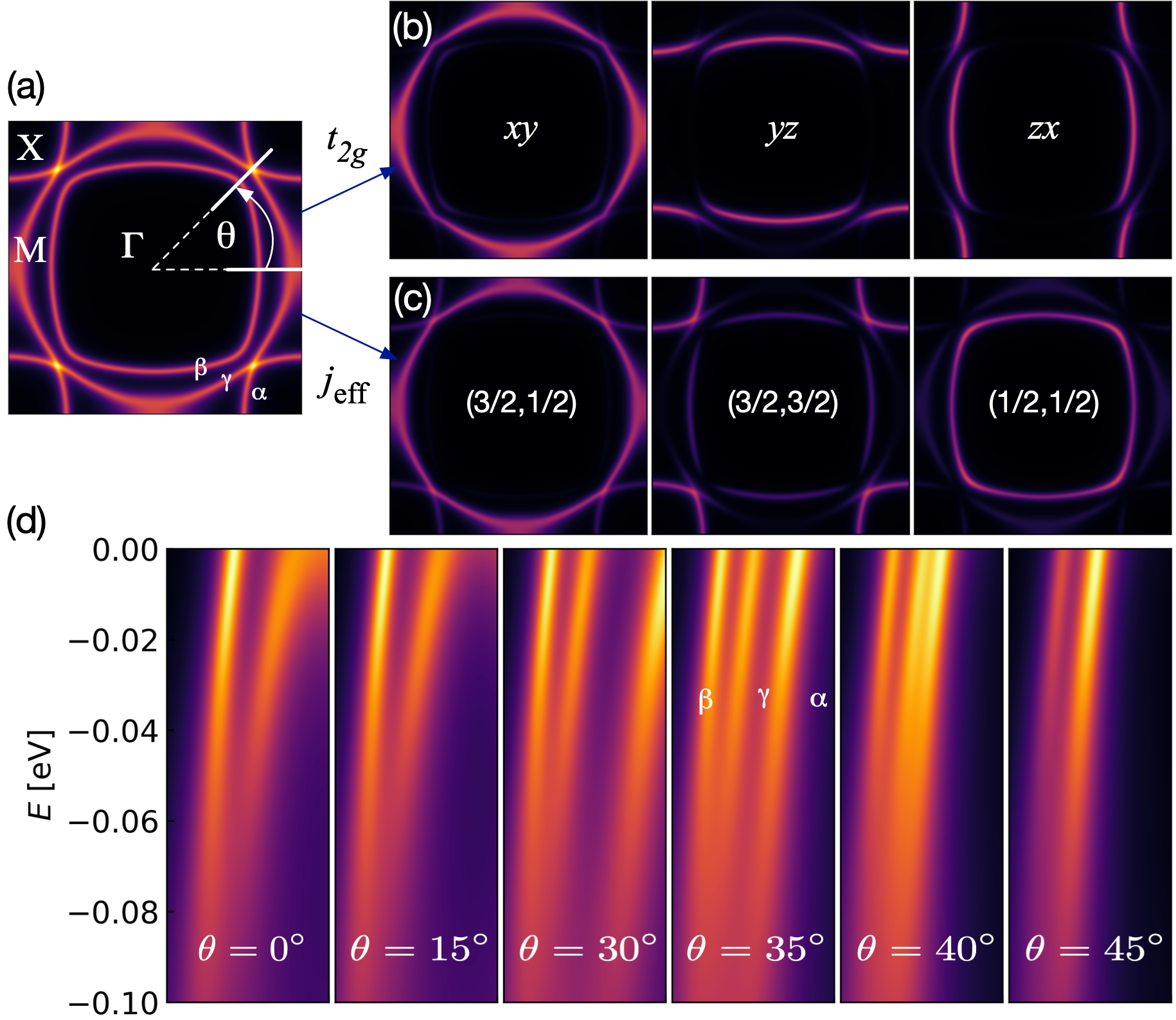}%
	\caption{
	Spectral density of (a) the Fermi surface and its projection onto the (b) $t_{2g}$- and (c) $j_{\textrm{eff}}$-basis states.
	(d) Dispersion along different angles $\theta$ between the $\Gamma$M and $\Gamma$X lines depicted in (a). 
	All data are obtained with $U$ = 2.5 eV, $J_H$ = 0.4 eV, and $\lsoc$ = 0.1 eV at $T$ = 23 K.
	}
\label{fig:fs}
\end{figure}

%\subsection{Spectral decomposition in \SRO}
With realistic parameters of \SRO~\cite{Mravlje2011,Tamai2019}, we compute the spectral functions on the Fermi level and compare them to experiments.
We present the Fermi surface and low energy hole excitation spectra of \SRO in Fig.~\ref{fig:fs}, with parameters $U$ = 2.5 eV and $J_H=0.4$ eV.
This reproduces the similar Fermi surface geometry previously obtained from ARPES spectra~\cite{Damascelli2000,Shen2001,Ingle2005,Iwasawa2005,Kidd2005,Shen2007,Iwasawa2010,Iwasawa2012,Tamai2019}, which is consistent with earlier theoretical works~\cite{Zhang2016,MKim2018,Tamai2019}.
The strong correlation effects have an intimate connection with the Fermi surface spectra.
The effective masses from our results also satisfy $m^*_\gamma > m^*_\beta > m^*_\alpha$ 
\footnote{We obtain $(m^*_{\threeh} , m^*_{\halfh} , m^*_{\threet})\simeq$ (5.60, 5.09, 3.94) at $T=0$ K and (4.56, 4.27, 3.61) at $T=11$ K.
We use a fourth order polynomial with lowest six Matsubara frequencies and fit imaginary part of self-energy. See Supplementary Information for more details.}
in good agreement with experimental results~\cite{Bergemann2003,Shen2007,Kondo2016}
and theoretical works~\cite{Mravlje2011,Deng2016,MKim2018,Sarvestani2018}.
Orbital selectivity near the Fermi level brings us the pocket-dependent scattering of three sheets, $\alpha, \beta$, and $\gamma$, as marked in Fig.~\ref{fig:fs}(a).
But inside the pockets, especially $\beta$ and $\gamma$, their effective masses are known to vary strongly near the $\Gamma$X line~\cite{Veenstra2014,Tamai2019}.
This is also consistent with our spectra, as shown in Fig.~\ref{fig:fs}.

Another indispensable ingredient to explain the spectral properties here is the SOC~\cite{Veenstra2014,MKim2018}.
While SOC is known to be essential to produce the three pockets on the Fermi surface as discussed in previous studies~\cite{Pavarini2006,Haverkort2008},
the role of SOC in describing low-lying excitation has yet to receive much focus.
As SOC introduces a mixing between $t_{2g}$ orbitals,
the strong orbital dependence of the self-energy appears controversial to underpin the importance of SOC.
In spite of this, here we find that the enhanced SOC affects the character of the Fermi surface pocket from the orbital projection of the spectra.

We project the spectral weights of the Fermi level in Fig.~\ref{fig:fs}(a) onto the $t_{2g}$ and spin-orbit coupled ($j_\mathrm{eff}$) basis states in Fig.~\ref{fig:fs}(b) and (c), respectively.
The two 1D-like spectra associated with the $yz$ and $zx$ are discontinuous near the $\Gamma$X lines due to the orbital mixing by SOC~\cite{Veenstra2014}.
Interestingly, the Fermi pockets are well-defined on the spin-orbit eigenstates in the realistic parameter regime.
The clear decomposition by the $\jeff$ basis is not only limited to the Fermi level but also extends to an energy window of a few hundred meV.
In Fig.~\ref{fig:fs}(d),
we present spectral weights as a function of energy along a few different paths between $\Gamma$M and $\Gamma$X [marked in Fig.~\ref{fig:fs}(a)], 
which are directly related to a recent ARPES experiment~\cite{Tamai2019}.
The scattering of the $\gamma$ pocket is the largest because of the strong $xy$ components in $(j_\mathrm{eff}, j_z)= \threehpm$.
The strong $j_\mathrm{eff}$ character remains by the enhanced SOC from correlation effects.
While the bare SOC is approximately 0.1 eV in Ru, the local interaction boosts the splitting induced by SOC~\cite{Liu2008,Bunemann2016,Zhang2016,MKim2018,Linden2020}.
The effective SOC is thus twice as large as the bare value~\cite{Zhang2016}, which is comparable to the Hund's coupling of realistic parameters in our calculation.
This indicates that the low energy excitation near the Fermi level should be described based on $\jeff$ rather than on $t_{2g}$ degrees of freedom.

\begin{figure}[tb]
        \includegraphics[width=\columnwidth]{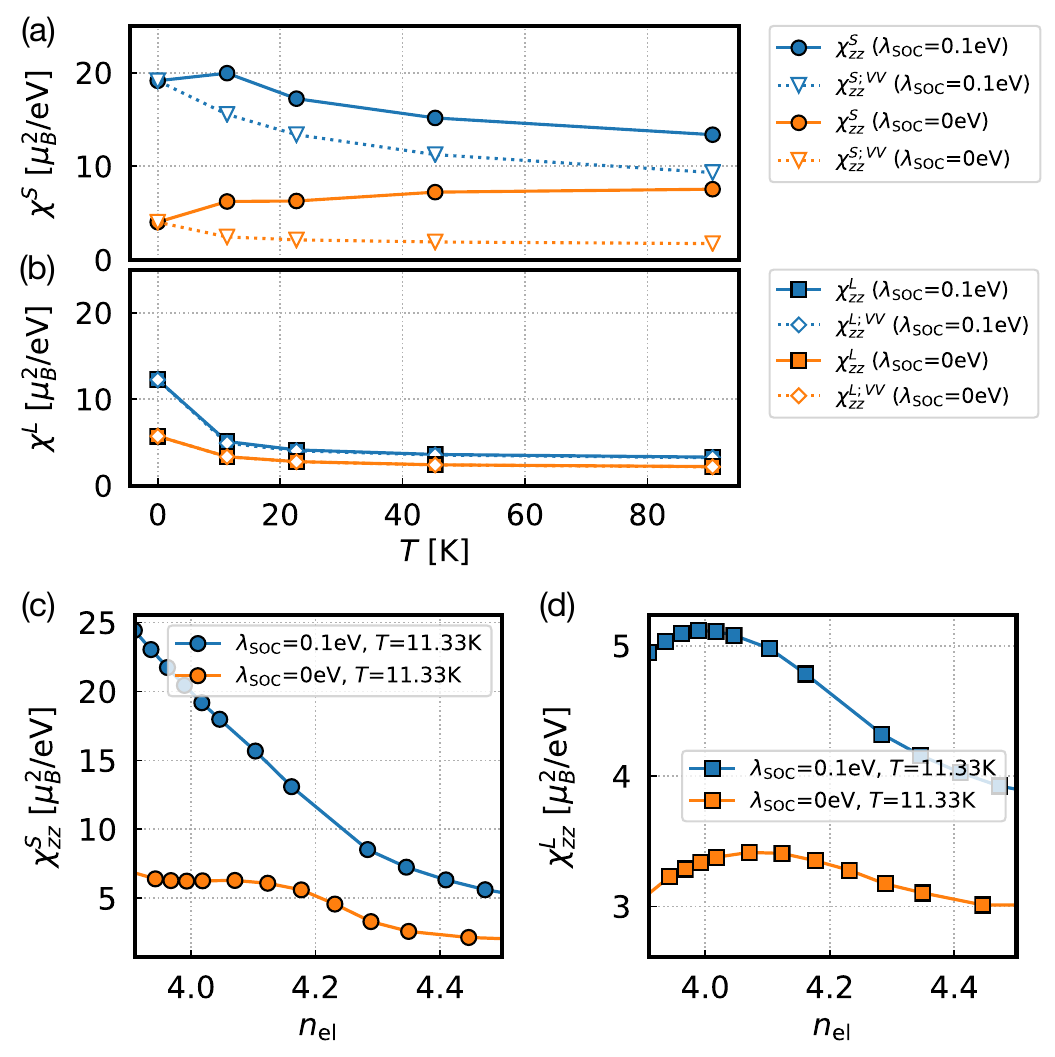}%
        \caption{
	Static spin susceptibilities $\chi^{S}_{zz}$ in the presence or absence of SOC 
	(a) as a function of temperature for $n_{\mathrm{el}}=4$
	and (c) as a function of $n_{\mathrm{el}}$ at $T=11.33$ K.
	Static orbital susceptibilities $\chi^{L}_{zz}$ are plotted in (b) and (d).
	We also plot their Van Vleck contributions $\chi^{S/L;\mathrm{VV}}_{zz}$ in (a) and (b).
	The data are from $(U,J_H)=(2.5,0.4)$ eV.
        }
\label{fig:sus}
\end{figure}

Spin-orbit coupling is also vital to understand the magnetic response of this system.
We compute the static susceptibility $\chi^{M}_{zz}(\omega=0)$ as
\begin{align}
	\chi^{M}_{zz}(\omega) 
	=&   \beta \delta(\omega) \sum_{a,b} e^{-\beta E_a} |\langle a | \hat{M}_z | b \rangle |^2 \delta(E_a-E_b) \nonumber \\
	+& \sum_{a,b} e^{-\beta E_a}  |\langle a | \hat{M}_z | b \rangle |^2\bigg(\frac{1}{\omega+\omega_{ba}}-\frac{1}{\omega-\omega_{ba}}\bigg), \label{eq:sus}
\end{align}
where $M$ is angular momentum such as $S$ or $L$, summations run over the energy eigenstates, $E_a$ is the eigenvalue of the $a$th energy eigenstate, and $\omega_{ba} = E_b - E_a$.
The first term in Eq.~(\ref{eq:sus}) reflects the contribution from the long-lived moment, which is stated as the Curie term,
whereas the second term expresses the Van Vleck (VV) component.
Figure~\ref{fig:sus}(a) displays the long-time correlator that governs the Curie term in the susceptibility.
The Curie term is generated by energetically degenerate eigenstates and directly affected by the long-lived magnetic moment.
SOC lifts the degeneracy of the eigenstates and is supposed to reduce the Curie term and the long-time correlator in \ttgf.
This results in a reduction of the long-time correlator in the Hund's metal regime ($T$$>$$T_\mathrm{FL}$), as shown in Fig.~\ref{fig:sus}(a).
On the other hand, the degeneracy-lifting introduced by the SOC has an opposite effect on the VV term.
The VV contribution comes from the angular-momentum-mediated virtual excitations between the eigenstates.
The newly contributing excitations between the SOC-lifted multiplets increases the VV term, as presented in Fig.~\ref{fig:sus}(b).
In the range of $T$ we consider, susceptibility is boosted by SOC due to the prominent VV contribution.
This is in contrast to its effect on the single-particle response.
The SOC slightly changes the self-energy, to which its renormalized masses are not very sensitive~\cite{Linden2020}.
We note that the VV contribution dominates $\Delta \chi = \chi^{}(\omega=0) - \beta \chi(\tau=\beta/2)$, which measures the strength of fluctuations~\cite{Hoshino2015}.
Hence, the enhanced VV at low temperature shows the significance of SOC in the correlated nature of \SRO.
Combined with the VHS, spin-orbit coupling makes the system reach its lowest $T_\mathrm{FL}$ at smaller $n_{\mathrm{el}}$ than expected when based solely on the VHS.

The temperature dependence of the two contributions of Eq.~(\ref{eq:sus}) shows a sharp contrast.
As $T$ increases, the long-time correlator rises while the VV counterpart decreases.
Overall, the enhancement of the total susceptibility by SOC decreases as a function of $T$.
Based on the envelope, SOC appears to suppress the susceptibility at higher temperatures.
This would be consistent with a previous study that reported a reduced spin susceptibility by SOC at $T\gtrsim300$ K~\cite{Strand2019}.

The VHS of \SRO modifies its Hund's metallic behavior, as noted in the comparison to the three-band model of the Bethe lattice.
Upon electron doping of \ttgf with the elliptical DOS of the Bethe lattice, the susceptibility monotonically decreases~\cite{Hoshino2015}.
However, in \SRO, the VHS manifests its singular occupancy dependence in the absence of SOC, as shown in Fig.~\ref{fig:sus}(c,d).
Both $\chi^S$ and $\chi^L$ show a peak near $n_\mathrm{el}$ where the Fermi level locates at the VHS, reflecting a promoted moment response due to the singularity.
(Practically, the DMFT self-consistency equation recognizes the existence of the VHS by a dip in the noninteracting hybridization function.)
Nonzero SOC significantly enhances the susceptibility, especially the spin responses around \ttgf [see Fig.~\ref{fig:sus}(c)].
This implies the importance of SOC in detecting spin-freezing driven phenomena in theoretical calculations.
Further, engineering the VHS closer to \ttgf would amplify $\chi^{S,L}$, thereby reinforcing the Hund's metallic behavior.

\section{Conclusions}
In this work, we performed DFT+DMFT calculations with an exact diagonalization solver on \SRO. 
Considering the rotationally invariant Slater--Kanamori interaction as well as spin-orbit coupling,
we constructed a comprehensive phase diagram on the $T-J_H$ plane by extracting power exponents of the self-energy.
The ground state is a Fermi liquid, and the system enters the Hund's metal regime as the temperature increases, with Hund's coupling suppressing the Fermi liquid temperature.
We identified the particular parameter set that reproduces experimentally observed spectral functions.
The ground state of \SRO turned out to be very close to the Hund's metal regime in the phase diagram, reflecting its correlated nature.
We then applied electron doping to study how the VHS affects the correlated metal, and found that the VHS extends the Hund's metal regime, enhancing the scattering rate of the electrons, whereas the long-time correlator weakens as the system is doped.

We also investigated the role of SOC by comparing two independent calculations, namely with and without SOC included in the DMFT loop.
When SOC is included, the Fermi surface and low lying excitations are clearly decomposed in the $\jeff$ basis, indicating the fundamental importance of SOC in the low energy excitation of this system.
Furthermore, SOC revives the orbital degrees of freedom to affect the magnetic responses.
More specifically, it reduces the long-time correlator and amplifies the VV contribution in the susceptibility.
Combining these two effects, SOC increases susceptibility at sufficiently low temperature and reinforces Hund's metallicity by enhancing the magnetic fluctuations near $n_{\mathrm{el}}$ = 4.
This points to the inevitability of including SOC in clarifying the roles of Hund's coupling and the VHS in \SRO.

Our results call for further studies in various directions.
We observed that VV susceptibility with SOC outperforms the Curie part as the temperature decreases;
this reveals genuine effects of SOC in correlated multi-orbital systems.
By analyzing the VHS and static susceptibility, we also showed the importance that the electronic structure has on the correlation effects in \SRO.
Investigation into the interplay between the VHS and superconductivity would be another promising direction for future works.
From a technical perspective, it would be interesting to test the role of additional bath orbitals by employing adaptively truncated Hilbert spaces~\cite{Zgid2011,Lu2014,Go2017} in highly correlated regimes where Hund's coupling is large.

\begin{acknowledgments}
We thank Andrew J. Millis and Philipp Werner for insightful comments and discussion.
This work was supported by the Institute for Basic Science under Grants No. IBS-R024-D1 (HJL and AG) and IBS-R009-D1 (CHK).
\end{acknowledgments}

\bibliography{SRO214}

%merlin.mbs apsrev4-1.bst 2010-07-25 4.21a (PWD, AO, DPC) hacked
%Control: key (0)
%Control: author (0) dotless jnrlst
%Control: editor formatted (1) identically to author
%Control: production of article title (0) allowed
%Control: page (1) range
%Control: year (0) verbatim
%Control: production of eprint (0) enabled
\begin{thebibliography}{93}%
\makeatletter
\providecommand \@ifxundefined [1]{%
 \@ifx{#1\undefined}
}%
\providecommand \@ifnum [1]{%
 \ifnum #1\expandafter \@firstoftwo
 \else \expandafter \@secondoftwo
 \fi
}%
\providecommand \@ifx [1]{%
 \ifx #1\expandafter \@firstoftwo
 \else \expandafter \@secondoftwo
 \fi
}%
\providecommand \natexlab [1]{#1}%
\providecommand \enquote  [1]{``#1''}%
\providecommand \bibnamefont  [1]{#1}%
\providecommand \bibfnamefont [1]{#1}%
\providecommand \citenamefont [1]{#1}%
\providecommand \href@noop [0]{\@secondoftwo}%
\providecommand \href [0]{\begingroup \@sanitize@url \@href}%
\providecommand \@href[1]{\@@startlink{#1}\@@href}%
\providecommand \@@href[1]{\endgroup#1\@@endlink}%
\providecommand \@sanitize@url [0]{\catcode `\\12\catcode `\$12\catcode
  `\&12\catcode `\#12\catcode `\^12\catcode `\_12\catcode `\%12\relax}%
\providecommand \@@startlink[1]{}%
\providecommand \@@endlink[0]{}%
\providecommand \url  [0]{\begingroup\@sanitize@url \@url }%
\providecommand \@url [1]{\endgroup\@href {#1}{\urlprefix }}%
\providecommand \urlprefix  [0]{URL }%
\providecommand \Eprint [0]{\href }%
\providecommand \doibase [0]{http://dx.doi.org/}%
\providecommand \selectlanguage [0]{\@gobble}%
\providecommand \bibinfo  [0]{\@secondoftwo}%
\providecommand \bibfield  [0]{\@secondoftwo}%
\providecommand \translation [1]{[#1]}%
\providecommand \BibitemOpen [0]{}%
\providecommand \bibitemStop [0]{}%
\providecommand \bibitemNoStop [0]{.\EOS\space}%
\providecommand \EOS [0]{\spacefactor3000\relax}%
\providecommand \BibitemShut  [1]{\csname bibitem#1\endcsname}%
\let\auto@bib@innerbib\@empty
%</preamble>
\bibitem [{\citenamefont {Haule}\ and\ \citenamefont
  {Kotliar}(2009)}]{Haule2009}%
  \BibitemOpen
  \bibfield  {author} {\bibinfo {author} {\bibfnamefont {Kristjan}\
  \bibnamefont {Haule}}\ and\ \bibinfo {author} {\bibfnamefont {Gabriel}\
  \bibnamefont {Kotliar}},\ }\bibfield  {title} {\enquote {\bibinfo {title}
  {{Coherence–incoherence crossover in the normal state of iron oxypnictides
  and importance of Hund's rule coupling}},}\ }\href
  {http://stacks.iop.org/1367-2630/11/i=2/a=025021} {\bibfield  {journal}
  {\bibinfo  {journal} {New J. Phys.}\ }\textbf {\bibinfo {volume} {11}},\
  \bibinfo {pages} {025021} (\bibinfo {year} {2009})}\BibitemShut {NoStop}%
\bibitem [{\citenamefont {Yin}\ \emph {et~al.}(2011)\citenamefont {Yin},
  \citenamefont {Haule},\ and\ \citenamefont {Kotliar}}]{Yin2011}%
  \BibitemOpen
  \bibfield  {author} {\bibinfo {author} {\bibfnamefont {Z~P}\ \bibnamefont
  {Yin}}, \bibinfo {author} {\bibfnamefont {Kristjan}\ \bibnamefont {Haule}}, \
  and\ \bibinfo {author} {\bibfnamefont {Gabriel}\ \bibnamefont {Kotliar}},\
  }\bibfield  {title} {\enquote {\bibinfo {title} {{Kinetic frustration and the
  nature of the magnetic and paramagnetic states in iron pnictides and iron
  chalcogenides}},}\ }\href {http://dx.doi.org/10.1038/nmat3120
  http://10.0.4.14/nmat3120
  https://www.nature.com/articles/nmat3120{\#}supplementary-information}
  {\bibfield  {journal} {\bibinfo  {journal} {Nat. Mater.}\ }\textbf {\bibinfo
  {volume} {10}},\ \bibinfo {pages} {932} (\bibinfo {year} {2011})}\BibitemShut
  {NoStop}%
\bibitem [{\citenamefont {Georges}\ \emph {et~al.}(2013)\citenamefont
  {Georges}, \citenamefont {de' Medici},\ and\ \citenamefont
  {Mravlje}}]{Georges2013}%
  \BibitemOpen
  \bibfield  {author} {\bibinfo {author} {\bibfnamefont {Antoine}\ \bibnamefont
  {Georges}}, \bibinfo {author} {\bibfnamefont {Luca}\ \bibnamefont {de'
  Medici}}, \ and\ \bibinfo {author} {\bibfnamefont {Jernej}\ \bibnamefont
  {Mravlje}},\ }\bibfield  {title} {\enquote {\bibinfo {title} {{Strong
  Correlations from Hund's Coupling}},}\ }\href {\doibase
  10.1146/annurev-conmatphys-020911-125045} {\bibfield  {journal} {\bibinfo
  {journal} {Annu. Rev. Condens. Matter Phys.}\ }\textbf {\bibinfo {volume}
  {4}},\ \bibinfo {pages} {137} (\bibinfo {year} {2013})}\BibitemShut {NoStop}%
\bibitem [{\citenamefont {de' Medici}\ \emph {et~al.}(2011)\citenamefont {de'
  Medici}, \citenamefont {Mravlje},\ and\ \citenamefont
  {Georges}}]{DeMedici2011PRL}%
  \BibitemOpen
  \bibfield  {author} {\bibinfo {author} {\bibfnamefont {Luca}\ \bibnamefont
  {de' Medici}}, \bibinfo {author} {\bibfnamefont {Jernej}\ \bibnamefont
  {Mravlje}}, \ and\ \bibinfo {author} {\bibfnamefont {Antoine}\ \bibnamefont
  {Georges}},\ }\bibfield  {title} {\enquote {\bibinfo {title} {{Janus-Faced
  Influence of Hund's Rule Coupling in Strongly Correlated Materials}},}\
  }\href {https://link.aps.org/doi/10.1103/PhysRevLett.107.256401} {\bibfield
  {journal} {\bibinfo  {journal} {Phys. Rev. Lett.}\ }\textbf {\bibinfo
  {volume} {107}},\ \bibinfo {pages} {256401} (\bibinfo {year}
  {2011})}\BibitemShut {NoStop}%
\bibitem [{\citenamefont {de' Medici}(2011)}]{DeMedici2011PRB}%
  \BibitemOpen
  \bibfield  {author} {\bibinfo {author} {\bibfnamefont {Luca}\ \bibnamefont
  {de' Medici}},\ }\bibfield  {title} {\enquote {\bibinfo {title} {{Hund's
  coupling and its key role in tuning multiorbital correlations}},}\ }\href
  {https://link.aps.org/doi/10.1103/PhysRevB.83.205112} {\bibfield  {journal}
  {\bibinfo  {journal} {Phys. Rev. B}\ }\textbf {\bibinfo {volume} {83}},\
  \bibinfo {pages} {205112} (\bibinfo {year} {2011})}\BibitemShut {NoStop}%
\bibitem [{\citenamefont {Werner}\ \emph {et~al.}(2008)\citenamefont {Werner},
  \citenamefont {Gull}, \citenamefont {Troyer},\ and\ \citenamefont
  {Millis}}]{Werner2008}%
  \BibitemOpen
  \bibfield  {author} {\bibinfo {author} {\bibfnamefont {Philipp}\ \bibnamefont
  {Werner}}, \bibinfo {author} {\bibfnamefont {Emanuel}\ \bibnamefont {Gull}},
  \bibinfo {author} {\bibfnamefont {Matthias}\ \bibnamefont {Troyer}}, \ and\
  \bibinfo {author} {\bibfnamefont {Andrew~J}\ \bibnamefont {Millis}},\
  }\bibfield  {title} {\enquote {\bibinfo {title} {{Spin Freezing Transition
  and Non-Fermi-Liquid Self-Energy in a Three-Orbital Model}},}\ }\href
  {https://link.aps.org/doi/10.1103/PhysRevLett.101.166405} {\bibfield
  {journal} {\bibinfo  {journal} {Phys. Rev. Lett.}\ }\textbf {\bibinfo
  {volume} {101}},\ \bibinfo {pages} {166405} (\bibinfo {year}
  {2008})}\BibitemShut {NoStop}%
\bibitem [{\citenamefont {Kim}\ \emph {et~al.}(2017)\citenamefont {Kim},
  \citenamefont {Jeschke}, \citenamefont {Werner},\ and\ \citenamefont
  {Valent{\'{i}}}}]{AJKim2017}%
  \BibitemOpen
  \bibfield  {author} {\bibinfo {author} {\bibfnamefont {Aaram~J}\ \bibnamefont
  {Kim}}, \bibinfo {author} {\bibfnamefont {Harald~O}\ \bibnamefont {Jeschke}},
  \bibinfo {author} {\bibfnamefont {Philipp}\ \bibnamefont {Werner}}, \ and\
  \bibinfo {author} {\bibfnamefont {Roser}\ \bibnamefont {Valent{\'{i}}}},\
  }\bibfield  {title} {\enquote {\bibinfo {title} {{$\mathbf{J}$ Freezing and
  Hund's Rules in Spin-Orbit-Coupled Multiorbital Hubbard Models}},}\ }\href
  {\doibase 10.1103/PhysRevLett.118.086401} {\bibfield  {journal} {\bibinfo
  {journal} {Phys. Rev. Lett.}\ }\textbf {\bibinfo {volume} {118}},\ \bibinfo
  {pages} {086401} (\bibinfo {year} {2017})}\BibitemShut {NoStop}%
\bibitem [{\citenamefont {Koga}\ \emph {et~al.}(2004)\citenamefont {Koga},
  \citenamefont {Kawakami}, \citenamefont {Rice},\ and\ \citenamefont
  {Sigrist}}]{Koga2004}%
  \BibitemOpen
  \bibfield  {author} {\bibinfo {author} {\bibfnamefont {Akihisa}\ \bibnamefont
  {Koga}}, \bibinfo {author} {\bibfnamefont {Norio}\ \bibnamefont {Kawakami}},
  \bibinfo {author} {\bibfnamefont {T~M}\ \bibnamefont {Rice}}, \ and\ \bibinfo
  {author} {\bibfnamefont {Manfred}\ \bibnamefont {Sigrist}},\ }\bibfield
  {title} {\enquote {\bibinfo {title} {{Orbital-Selective Mott Transitions in
  the Degenerate Hubbard Model}},}\ }\href
  {https://link.aps.org/doi/10.1103/PhysRevLett.92.216402} {\bibfield
  {journal} {\bibinfo  {journal} {Phys. Rev. Lett.}\ }\textbf {\bibinfo
  {volume} {92}},\ \bibinfo {pages} {216402} (\bibinfo {year}
  {2004})}\BibitemShut {NoStop}%
\bibitem [{\citenamefont {Anisimov}\ \emph {et~al.}(2002)\citenamefont
  {Anisimov}, \citenamefont {Nekrasov}, \citenamefont {Kondakov}, \citenamefont
  {Rice},\ and\ \citenamefont {Sigrist}}]{Anisimov2002}%
  \BibitemOpen
  \bibfield  {author} {\bibinfo {author} {\bibfnamefont {V~I}\ \bibnamefont
  {Anisimov}}, \bibinfo {author} {\bibfnamefont {I~A}\ \bibnamefont
  {Nekrasov}}, \bibinfo {author} {\bibfnamefont {D~E}\ \bibnamefont
  {Kondakov}}, \bibinfo {author} {\bibfnamefont {T~M}\ \bibnamefont {Rice}}, \
  and\ \bibinfo {author} {\bibfnamefont {M}~\bibnamefont {Sigrist}},\
  }\bibfield  {title} {\enquote {\bibinfo {title} {{Orbital-selective
  Mott-insulator transition in
  ${\mathrm{Ca}}_{2-x}{\mathrm{Sr}}_{x}{\mathrm{RuO}}_{4}$}},}\ }\href
  {\doibase 10.1140/epjb/e20020021} {\bibfield  {journal} {\bibinfo  {journal}
  {Eur. Phys. J. B}\ }\textbf {\bibinfo {volume} {25}},\ \bibinfo {pages}
  {191--201} (\bibinfo {year} {2002})}\BibitemShut {NoStop}%
\bibitem [{\citenamefont {Werner}\ and\ \citenamefont
  {Millis}(2007)}]{Werner2007}%
  \BibitemOpen
  \bibfield  {author} {\bibinfo {author} {\bibfnamefont {Philipp}\ \bibnamefont
  {Werner}}\ and\ \bibinfo {author} {\bibfnamefont {Andrew~J}\ \bibnamefont
  {Millis}},\ }\bibfield  {title} {\enquote {\bibinfo {title} {{High-Spin to
  Low-Spin and Orbital Polarization Transitions in Multiorbital Mott
  Systems}},}\ }\href {\doibase 10.1103/PhysRevLett.99.126405} {\bibfield
  {journal} {\bibinfo  {journal} {Phys. Rev. Lett.}\ }\textbf {\bibinfo
  {volume} {99}},\ \bibinfo {pages} {126405} (\bibinfo {year}
  {2007})}\BibitemShut {NoStop}%
\bibitem [{\citenamefont {Werner}\ \emph {et~al.}(2009)\citenamefont {Werner},
  \citenamefont {Gull},\ and\ \citenamefont {Millis}}]{Werner2009}%
  \BibitemOpen
  \bibfield  {author} {\bibinfo {author} {\bibfnamefont {Philipp}\ \bibnamefont
  {Werner}}, \bibinfo {author} {\bibfnamefont {Emanuel}\ \bibnamefont {Gull}},
  \ and\ \bibinfo {author} {\bibfnamefont {Andrew~J}\ \bibnamefont {Millis}},\
  }\bibfield  {title} {\enquote {\bibinfo {title} {{Metal-insulator phase
  diagram and orbital selectivity in three-orbital models with rotationally
  invariant Hund coupling}},}\ }\href
  {https://link.aps.org/doi/10.1103/PhysRevB.79.115119} {\bibfield  {journal}
  {\bibinfo  {journal} {Phys. Rev. B}\ }\textbf {\bibinfo {volume} {79}},\
  \bibinfo {pages} {115119} (\bibinfo {year} {2009})}\BibitemShut {NoStop}%
\bibitem [{\citenamefont {de' Medici}\ \emph {et~al.}(2009)\citenamefont {de'
  Medici}, \citenamefont {Hassan}, \citenamefont {Capone},\ and\ \citenamefont
  {Dai}}]{DeMedici2009}%
  \BibitemOpen
  \bibfield  {author} {\bibinfo {author} {\bibfnamefont {Luca}\ \bibnamefont
  {de' Medici}}, \bibinfo {author} {\bibfnamefont {S~R}\ \bibnamefont
  {Hassan}}, \bibinfo {author} {\bibfnamefont {Massimo}\ \bibnamefont
  {Capone}}, \ and\ \bibinfo {author} {\bibfnamefont {Xi}~\bibnamefont {Dai}},\
  }\bibfield  {title} {\enquote {\bibinfo {title} {{Orbital-Selective Mott
  Transition out of Band Degeneracy Lifting}},}\ }\href
  {https://link.aps.org/doi/10.1103/PhysRevLett.102.126401} {\bibfield
  {journal} {\bibinfo  {journal} {Phys. Rev. Lett.}\ }\textbf {\bibinfo
  {volume} {102}},\ \bibinfo {pages} {126401} (\bibinfo {year}
  {2009})}\BibitemShut {NoStop}%
\bibitem [{\citenamefont {Werner}\ \emph {et~al.}(2016)\citenamefont {Werner},
  \citenamefont {Hoshino},\ and\ \citenamefont {Shinaoka}}]{Werner2016}%
  \BibitemOpen
  \bibfield  {author} {\bibinfo {author} {\bibfnamefont {Philipp}\ \bibnamefont
  {Werner}}, \bibinfo {author} {\bibfnamefont {Shintaro}\ \bibnamefont
  {Hoshino}}, \ and\ \bibinfo {author} {\bibfnamefont {Hiroshi}\ \bibnamefont
  {Shinaoka}},\ }\bibfield  {title} {\enquote {\bibinfo {title} {Spin-freezing
  perspective on cuprates},}\ }\href {\doibase 10.1103/PhysRevB.94.245134}
  {\bibfield  {journal} {\bibinfo  {journal} {Phys. Rev. B}\ }\textbf {\bibinfo
  {volume} {94}},\ \bibinfo {pages} {245134} (\bibinfo {year}
  {2016})}\BibitemShut {NoStop}%
\bibitem [{\citenamefont {Mackenzie}\ \emph {et~al.}(1996)\citenamefont
  {Mackenzie}, \citenamefont {Julian}, \citenamefont {Diver}, \citenamefont
  {McMullan}, \citenamefont {Ray}, \citenamefont {Lonzarich}, \citenamefont
  {Maeno}, \citenamefont {Nishizaki},\ and\ \citenamefont
  {Fujita}}]{Mackenzie1996}%
  \BibitemOpen
  \bibfield  {author} {\bibinfo {author} {\bibfnamefont {A~P}\ \bibnamefont
  {Mackenzie}}, \bibinfo {author} {\bibfnamefont {S~R}\ \bibnamefont {Julian}},
  \bibinfo {author} {\bibfnamefont {A~J}\ \bibnamefont {Diver}}, \bibinfo
  {author} {\bibfnamefont {G~J}\ \bibnamefont {McMullan}}, \bibinfo {author}
  {\bibfnamefont {M~P}\ \bibnamefont {Ray}}, \bibinfo {author} {\bibfnamefont
  {G~G}\ \bibnamefont {Lonzarich}}, \bibinfo {author} {\bibfnamefont
  {Y}~\bibnamefont {Maeno}}, \bibinfo {author} {\bibfnamefont {S}~\bibnamefont
  {Nishizaki}}, \ and\ \bibinfo {author} {\bibfnamefont {T}~\bibnamefont
  {Fujita}},\ }\bibfield  {title} {\enquote {\bibinfo {title} {{Quantum
  Oscillations in the Layered Perovskite Superconductor
  S${\mathrm{r}}_{2}$Ru${\mathrm{O}}_{4}$}},}\ }\href
  {https://link.aps.org/doi/10.1103/PhysRevLett.76.3786} {\bibfield  {journal}
  {\bibinfo  {journal} {Phys. Rev. Lett.}\ }\textbf {\bibinfo {volume} {76}},\
  \bibinfo {pages} {3786--3789} (\bibinfo {year} {1996})}\BibitemShut {NoStop}%
\bibitem [{\citenamefont {Bergemann}\ \emph {et~al.}(2000)\citenamefont
  {Bergemann}, \citenamefont {Julian}, \citenamefont {Mackenzie}, \citenamefont
  {NishiZaki},\ and\ \citenamefont {Maeno}}]{Bergemann2000}%
  \BibitemOpen
  \bibfield  {author} {\bibinfo {author} {\bibfnamefont {C}~\bibnamefont
  {Bergemann}}, \bibinfo {author} {\bibfnamefont {S~R}\ \bibnamefont {Julian}},
  \bibinfo {author} {\bibfnamefont {A~P}\ \bibnamefont {Mackenzie}}, \bibinfo
  {author} {\bibfnamefont {S}~\bibnamefont {NishiZaki}}, \ and\ \bibinfo
  {author} {\bibfnamefont {Y}~\bibnamefont {Maeno}},\ }\bibfield  {title}
  {\enquote {\bibinfo {title} {{Detailed Topography of the Fermi Surface of
  ${\mathrm{Sr}}_{2}{\mathrm{RuO}}_{4}$}},}\ }\href {\doibase
  10.1103/PhysRevLett.84.2662} {\bibfield  {journal} {\bibinfo  {journal}
  {Phys. Rev. Lett.}\ }\textbf {\bibinfo {volume} {84}},\ \bibinfo {pages}
  {2662--2665} (\bibinfo {year} {2000})}\BibitemShut {NoStop}%
\bibitem [{\citenamefont {Bergemann}\ \emph {et~al.}(2003)\citenamefont
  {Bergemann}, \citenamefont {Mackenzie}, \citenamefont {Julian}, \citenamefont
  {Forsythe},\ and\ \citenamefont {Ohmichi}}]{Bergemann2003}%
  \BibitemOpen
  \bibfield  {author} {\bibinfo {author} {\bibfnamefont {C.}~\bibnamefont
  {Bergemann}}, \bibinfo {author} {\bibfnamefont {A.~P.}\ \bibnamefont
  {Mackenzie}}, \bibinfo {author} {\bibfnamefont {S.~R.}\ \bibnamefont
  {Julian}}, \bibinfo {author} {\bibfnamefont {D.}~\bibnamefont {Forsythe}}, \
  and\ \bibinfo {author} {\bibfnamefont {E.}~\bibnamefont {Ohmichi}},\
  }\bibfield  {title} {\enquote {\bibinfo {title} {{Quasi-two-dimensional Fermi
  liquid properties of the unconventional superconductor Sr$_2$RuO$_4$}},}\
  }\href {\doibase 10.1080/00018730310001621737} {\bibfield  {journal}
  {\bibinfo  {journal} {Adv. Phys.}\ }\textbf {\bibinfo {volume} {52}},\
  \bibinfo {pages} {639--725} (\bibinfo {year} {2003})}\BibitemShut {NoStop}%
\bibitem [{\citenamefont {Damascelli}\ \emph {et~al.}(2000)\citenamefont
  {Damascelli}, \citenamefont {Lu}, \citenamefont {Shen}, \citenamefont
  {Armitage}, \citenamefont {Ronning}, \citenamefont {Feng}, \citenamefont
  {Kim}, \citenamefont {Shen}, \citenamefont {Kimura}, \citenamefont {Tokura},
  \citenamefont {Mao},\ and\ \citenamefont {Maeno}}]{Damascelli2000}%
  \BibitemOpen
  \bibfield  {author} {\bibinfo {author} {\bibfnamefont {A}~\bibnamefont
  {Damascelli}}, \bibinfo {author} {\bibfnamefont {D~H}\ \bibnamefont {Lu}},
  \bibinfo {author} {\bibfnamefont {K~M}\ \bibnamefont {Shen}}, \bibinfo
  {author} {\bibfnamefont {N~P}\ \bibnamefont {Armitage}}, \bibinfo {author}
  {\bibfnamefont {F}~\bibnamefont {Ronning}}, \bibinfo {author} {\bibfnamefont
  {D~L}\ \bibnamefont {Feng}}, \bibinfo {author} {\bibfnamefont
  {C}~\bibnamefont {Kim}}, \bibinfo {author} {\bibfnamefont {Z.-X.}\
  \bibnamefont {Shen}}, \bibinfo {author} {\bibfnamefont {T}~\bibnamefont
  {Kimura}}, \bibinfo {author} {\bibfnamefont {Y}~\bibnamefont {Tokura}},
  \bibinfo {author} {\bibfnamefont {Z~Q}\ \bibnamefont {Mao}}, \ and\ \bibinfo
  {author} {\bibfnamefont {Y}~\bibnamefont {Maeno}},\ }\bibfield  {title}
  {\enquote {\bibinfo {title} {{Fermi Surface, Surface States, and Surface
  Reconstruction in ${\mathrm{Sr}}_{2}{\mathrm{RuO}}_{4}$}},}\ }\href
  {https://link.aps.org/doi/10.1103/PhysRevLett.85.5194} {\bibfield  {journal}
  {\bibinfo  {journal} {Phys. Rev. Lett.}\ }\textbf {\bibinfo {volume} {85}},\
  \bibinfo {pages} {5194--5197} (\bibinfo {year} {2000})}\BibitemShut {NoStop}%
\bibitem [{\citenamefont {Shen}\ \emph {et~al.}(2001)\citenamefont {Shen},
  \citenamefont {Damascelli}, \citenamefont {Lu}, \citenamefont {Armitage},
  \citenamefont {Ronning}, \citenamefont {Feng}, \citenamefont {Kim},
  \citenamefont {Shen}, \citenamefont {Singh}, \citenamefont {Mazin},
  \citenamefont {Nakatsuji}, \citenamefont {Mao}, \citenamefont {Maeno},
  \citenamefont {Kimura},\ and\ \citenamefont {Tokura}}]{Shen2001}%
  \BibitemOpen
  \bibfield  {author} {\bibinfo {author} {\bibfnamefont {K~M}\ \bibnamefont
  {Shen}}, \bibinfo {author} {\bibfnamefont {A}~\bibnamefont {Damascelli}},
  \bibinfo {author} {\bibfnamefont {D~H}\ \bibnamefont {Lu}}, \bibinfo {author}
  {\bibfnamefont {N~P}\ \bibnamefont {Armitage}}, \bibinfo {author}
  {\bibfnamefont {F}~\bibnamefont {Ronning}}, \bibinfo {author} {\bibfnamefont
  {D~L}\ \bibnamefont {Feng}}, \bibinfo {author} {\bibfnamefont
  {C}~\bibnamefont {Kim}}, \bibinfo {author} {\bibfnamefont {Z.-X.}\
  \bibnamefont {Shen}}, \bibinfo {author} {\bibfnamefont {D~J}\ \bibnamefont
  {Singh}}, \bibinfo {author} {\bibfnamefont {I~I}\ \bibnamefont {Mazin}},
  \bibinfo {author} {\bibfnamefont {S}~\bibnamefont {Nakatsuji}}, \bibinfo
  {author} {\bibfnamefont {Z~Q}\ \bibnamefont {Mao}}, \bibinfo {author}
  {\bibfnamefont {Y}~\bibnamefont {Maeno}}, \bibinfo {author} {\bibfnamefont
  {T}~\bibnamefont {Kimura}}, \ and\ \bibinfo {author} {\bibfnamefont
  {Y}~\bibnamefont {Tokura}},\ }\bibfield  {title} {\enquote {\bibinfo {title}
  {{Surface electronic structure of ${\mathrm{Sr}}_{2}{\mathrm{RuO}}_{4}$}},}\
  }\href {https://link.aps.org/doi/10.1103/PhysRevB.64.180502} {\bibfield
  {journal} {\bibinfo  {journal} {Phys. Rev. B}\ }\textbf {\bibinfo {volume}
  {64}},\ \bibinfo {pages} {180502(R)} (\bibinfo {year} {2001})}\BibitemShut
  {NoStop}%
\bibitem [{\citenamefont {Iwasawa}\ \emph {et~al.}(2005)\citenamefont
  {Iwasawa}, \citenamefont {Aiura}, \citenamefont {Saitoh}, \citenamefont
  {Hase}, \citenamefont {Ikeda}, \citenamefont {Yoshida}, \citenamefont
  {Bando}, \citenamefont {Higashiguchi}, \citenamefont {Miura}, \citenamefont
  {Cui}, \citenamefont {Shimada}, \citenamefont {Namatame},\ and\ \citenamefont
  {Taniguchi}}]{Iwasawa2005}%
  \BibitemOpen
  \bibfield  {author} {\bibinfo {author} {\bibfnamefont {H}~\bibnamefont
  {Iwasawa}}, \bibinfo {author} {\bibfnamefont {Y}~\bibnamefont {Aiura}},
  \bibinfo {author} {\bibfnamefont {T}~\bibnamefont {Saitoh}}, \bibinfo
  {author} {\bibfnamefont {I}~\bibnamefont {Hase}}, \bibinfo {author}
  {\bibfnamefont {S~I}\ \bibnamefont {Ikeda}}, \bibinfo {author} {\bibfnamefont
  {Y}~\bibnamefont {Yoshida}}, \bibinfo {author} {\bibfnamefont
  {H}~\bibnamefont {Bando}}, \bibinfo {author} {\bibfnamefont {M}~\bibnamefont
  {Higashiguchi}}, \bibinfo {author} {\bibfnamefont {Y}~\bibnamefont {Miura}},
  \bibinfo {author} {\bibfnamefont {X~Y}\ \bibnamefont {Cui}}, \bibinfo
  {author} {\bibfnamefont {K}~\bibnamefont {Shimada}}, \bibinfo {author}
  {\bibfnamefont {H}~\bibnamefont {Namatame}}, \ and\ \bibinfo {author}
  {\bibfnamefont {M}~\bibnamefont {Taniguchi}},\ }\bibfield  {title} {\enquote
  {\bibinfo {title} {{Orbital selectivity of the kink in the dispersion of
  ${\mathrm{Sr}}_{2}{\mathrm{RuO}}_{4}$}},}\ }\href {\doibase
  10.1103/PhysRevB.72.104514} {\bibfield  {journal} {\bibinfo  {journal} {Phys.
  Rev. B}\ }\textbf {\bibinfo {volume} {72}},\ \bibinfo {pages} {104514}
  (\bibinfo {year} {2005})}\BibitemShut {NoStop}%
\bibitem [{\citenamefont {Ingle}\ \emph {et~al.}(2005)\citenamefont {Ingle},
  \citenamefont {Shen}, \citenamefont {Baumberger}, \citenamefont {Meevasana},
  \citenamefont {Lu}, \citenamefont {Shen}, \citenamefont {Damascelli},
  \citenamefont {Nakatsuji}, \citenamefont {Mao}, \citenamefont {Maeno},
  \citenamefont {Kimura},\ and\ \citenamefont {Tokura}}]{Ingle2005}%
  \BibitemOpen
  \bibfield  {author} {\bibinfo {author} {\bibfnamefont {N~J~C}\ \bibnamefont
  {Ingle}}, \bibinfo {author} {\bibfnamefont {K~M}\ \bibnamefont {Shen}},
  \bibinfo {author} {\bibfnamefont {F}~\bibnamefont {Baumberger}}, \bibinfo
  {author} {\bibfnamefont {W}~\bibnamefont {Meevasana}}, \bibinfo {author}
  {\bibfnamefont {D~H}\ \bibnamefont {Lu}}, \bibinfo {author} {\bibfnamefont
  {Z.-X.}\ \bibnamefont {Shen}}, \bibinfo {author} {\bibfnamefont
  {A}~\bibnamefont {Damascelli}}, \bibinfo {author} {\bibfnamefont
  {S}~\bibnamefont {Nakatsuji}}, \bibinfo {author} {\bibfnamefont {Z~Q}\
  \bibnamefont {Mao}}, \bibinfo {author} {\bibfnamefont {Y}~\bibnamefont
  {Maeno}}, \bibinfo {author} {\bibfnamefont {T}~\bibnamefont {Kimura}}, \ and\
  \bibinfo {author} {\bibfnamefont {Y}~\bibnamefont {Tokura}},\ }\bibfield
  {title} {\enquote {\bibinfo {title} {{Quantitative analysis of
  ${\mathrm{Sr}}_{2}{\mathrm{RuO}}_{4}$ angle-resolved photoemission spectra:
  Many-body interactions in a model Fermi liquid}},}\ }\href {\doibase
  10.1103/PhysRevB.72.205114} {\bibfield  {journal} {\bibinfo  {journal} {Phys.
  Rev. B}\ }\textbf {\bibinfo {volume} {72}},\ \bibinfo {pages} {205114}
  (\bibinfo {year} {2005})}\BibitemShut {NoStop}%
\bibitem [{\citenamefont {Kidd}\ \emph {et~al.}(2005)\citenamefont {Kidd},
  \citenamefont {Valla}, \citenamefont {Fedorov}, \citenamefont {Johnson},
  \citenamefont {Cava},\ and\ \citenamefont {Haas}}]{Kidd2005}%
  \BibitemOpen
  \bibfield  {author} {\bibinfo {author} {\bibfnamefont {T~E}\ \bibnamefont
  {Kidd}}, \bibinfo {author} {\bibfnamefont {T}~\bibnamefont {Valla}}, \bibinfo
  {author} {\bibfnamefont {A~V}\ \bibnamefont {Fedorov}}, \bibinfo {author}
  {\bibfnamefont {P~D}\ \bibnamefont {Johnson}}, \bibinfo {author}
  {\bibfnamefont {R~J}\ \bibnamefont {Cava}}, \ and\ \bibinfo {author}
  {\bibfnamefont {M~K}\ \bibnamefont {Haas}},\ }\bibfield  {title} {\enquote
  {\bibinfo {title} {{Orbital Dependence of the Fermi Liquid State in
  ${\mathrm{S}\mathrm{r}}_{2}{\mathrm{R}\mathrm{u}\mathrm{O}}_{4}$}},}\ }\href
  {\doibase 10.1103/PhysRevLett.94.107003} {\bibfield  {journal} {\bibinfo
  {journal} {Phys. Rev. Lett.}\ }\textbf {\bibinfo {volume} {94}},\ \bibinfo
  {pages} {107003} (\bibinfo {year} {2005})}\BibitemShut {NoStop}%
\bibitem [{\citenamefont {Shen}\ \emph {et~al.}(2007)\citenamefont {Shen},
  \citenamefont {Kikugawa}, \citenamefont {Bergemann}, \citenamefont {Balicas},
  \citenamefont {Baumberger}, \citenamefont {Meevasana}, \citenamefont {Ingle},
  \citenamefont {Maeno}, \citenamefont {Shen},\ and\ \citenamefont
  {Mackenzie}}]{Shen2007}%
  \BibitemOpen
  \bibfield  {author} {\bibinfo {author} {\bibfnamefont {K~M}\ \bibnamefont
  {Shen}}, \bibinfo {author} {\bibfnamefont {N}~\bibnamefont {Kikugawa}},
  \bibinfo {author} {\bibfnamefont {C}~\bibnamefont {Bergemann}}, \bibinfo
  {author} {\bibfnamefont {L}~\bibnamefont {Balicas}}, \bibinfo {author}
  {\bibfnamefont {F}~\bibnamefont {Baumberger}}, \bibinfo {author}
  {\bibfnamefont {W}~\bibnamefont {Meevasana}}, \bibinfo {author}
  {\bibfnamefont {N~J~C}\ \bibnamefont {Ingle}}, \bibinfo {author}
  {\bibfnamefont {Y}~\bibnamefont {Maeno}}, \bibinfo {author} {\bibfnamefont
  {Z.-X.}\ \bibnamefont {Shen}}, \ and\ \bibinfo {author} {\bibfnamefont {A~P}\
  \bibnamefont {Mackenzie}},\ }\bibfield  {title} {\enquote {\bibinfo {title}
  {{Evolution of the Fermi Surface and Quasiparticle Renormalization through a
  van Hove Singularity in
  ${\mathrm{Sr}}_{2\ensuremath{-}y}{\mathrm{La}}_{y}{\mathrm{RuO}}_{4}$}},}\
  }\href {https://link.aps.org/doi/10.1103/PhysRevLett.99.187001} {\bibfield
  {journal} {\bibinfo  {journal} {Phys. Rev. Lett.}\ }\textbf {\bibinfo
  {volume} {99}},\ \bibinfo {pages} {187001} (\bibinfo {year}
  {2007})}\BibitemShut {NoStop}%
\bibitem [{\citenamefont {Iwasawa}\ \emph {et~al.}(2010)\citenamefont
  {Iwasawa}, \citenamefont {Yoshida}, \citenamefont {Hase}, \citenamefont
  {Koikegami}, \citenamefont {Hayashi}, \citenamefont {Jiang}, \citenamefont
  {Shimada}, \citenamefont {Namatame}, \citenamefont {Taniguchi},\ and\
  \citenamefont {Aiura}}]{Iwasawa2010}%
  \BibitemOpen
  \bibfield  {author} {\bibinfo {author} {\bibfnamefont {H}~\bibnamefont
  {Iwasawa}}, \bibinfo {author} {\bibfnamefont {Y}~\bibnamefont {Yoshida}},
  \bibinfo {author} {\bibfnamefont {I}~\bibnamefont {Hase}}, \bibinfo {author}
  {\bibfnamefont {S}~\bibnamefont {Koikegami}}, \bibinfo {author}
  {\bibfnamefont {H}~\bibnamefont {Hayashi}}, \bibinfo {author} {\bibfnamefont
  {J}~\bibnamefont {Jiang}}, \bibinfo {author} {\bibfnamefont {K}~\bibnamefont
  {Shimada}}, \bibinfo {author} {\bibfnamefont {H}~\bibnamefont {Namatame}},
  \bibinfo {author} {\bibfnamefont {M}~\bibnamefont {Taniguchi}}, \ and\
  \bibinfo {author} {\bibfnamefont {Y}~\bibnamefont {Aiura}},\ }\bibfield
  {title} {\enquote {\bibinfo {title} {{Interplay among Coulomb Interaction,
  Spin-Orbit Interaction, and Multiple Electron-Boson Interactions in
  ${\mathrm{Sr}}_{2}{\mathrm{RuO}}_{4}$}},}\ }\href
  {https://link.aps.org/doi/10.1103/PhysRevLett.105.226406} {\bibfield
  {journal} {\bibinfo  {journal} {Phys. Rev. Lett.}\ }\textbf {\bibinfo
  {volume} {105}},\ \bibinfo {pages} {226406} (\bibinfo {year}
  {2010})}\BibitemShut {NoStop}%
\bibitem [{\citenamefont {Iwasawa}\ \emph {et~al.}(2012)\citenamefont
  {Iwasawa}, \citenamefont {Yoshida}, \citenamefont {Hase}, \citenamefont
  {Shimada}, \citenamefont {Namatame}, \citenamefont {Taniguchi},\ and\
  \citenamefont {Aiura}}]{Iwasawa2012}%
  \BibitemOpen
  \bibfield  {author} {\bibinfo {author} {\bibfnamefont {H}~\bibnamefont
  {Iwasawa}}, \bibinfo {author} {\bibfnamefont {Y}~\bibnamefont {Yoshida}},
  \bibinfo {author} {\bibfnamefont {I}~\bibnamefont {Hase}}, \bibinfo {author}
  {\bibfnamefont {K}~\bibnamefont {Shimada}}, \bibinfo {author} {\bibfnamefont
  {H}~\bibnamefont {Namatame}}, \bibinfo {author} {\bibfnamefont
  {M}~\bibnamefont {Taniguchi}}, \ and\ \bibinfo {author} {\bibfnamefont
  {Y}~\bibnamefont {Aiura}},\ }\bibfield  {title} {\enquote {\bibinfo {title}
  {{High-Energy Anomaly in the Band Dispersion of the Ruthenate
  Superconductor}},}\ }\href {\doibase 10.1103/PhysRevLett.109.066404}
  {\bibfield  {journal} {\bibinfo  {journal} {Phys. Rev. Lett.}\ }\textbf
  {\bibinfo {volume} {109}},\ \bibinfo {pages} {066404} (\bibinfo {year}
  {2012})}\BibitemShut {NoStop}%
\bibitem [{\citenamefont {Stricker}\ \emph {et~al.}(2014)\citenamefont
  {Stricker}, \citenamefont {Mravlje}, \citenamefont {Berthod}, \citenamefont
  {Fittipaldi}, \citenamefont {Vecchione}, \citenamefont {Georges},\ and\
  \citenamefont {van~der Marel}}]{Stricker2014}%
  \BibitemOpen
  \bibfield  {author} {\bibinfo {author} {\bibfnamefont {D}~\bibnamefont
  {Stricker}}, \bibinfo {author} {\bibfnamefont {J}~\bibnamefont {Mravlje}},
  \bibinfo {author} {\bibfnamefont {C}~\bibnamefont {Berthod}}, \bibinfo
  {author} {\bibfnamefont {R}~\bibnamefont {Fittipaldi}}, \bibinfo {author}
  {\bibfnamefont {A}~\bibnamefont {Vecchione}}, \bibinfo {author}
  {\bibfnamefont {A}~\bibnamefont {Georges}}, \ and\ \bibinfo {author}
  {\bibfnamefont {D}~\bibnamefont {van~der Marel}},\ }\bibfield  {title}
  {\enquote {\bibinfo {title} {{Optical Response of
  ${\mathrm{Sr}}_{2}{\mathrm{RuO}}_{4}$ Reveals Universal Fermi-Liquid Scaling
  and Quasiparticles Beyond Landau Theory}},}\ }\href
  {https://link.aps.org/doi/10.1103/PhysRevLett.113.087404} {\bibfield
  {journal} {\bibinfo  {journal} {Phys. Rev. Lett.}\ }\textbf {\bibinfo
  {volume} {113}},\ \bibinfo {pages} {087404} (\bibinfo {year}
  {2014})}\BibitemShut {NoStop}%
\bibitem [{\citenamefont {Tamai}\ \emph {et~al.}(2019)\citenamefont {Tamai},
  \citenamefont {Zingl}, \citenamefont {Rozbicki}, \citenamefont {Cappelli},
  \citenamefont {Ricc\`o}, \citenamefont {de~la Torre}, \citenamefont
  {McKeown~Walker}, \citenamefont {Bruno}, \citenamefont {King}, \citenamefont
  {Meevasana}, \citenamefont {Shi}, \citenamefont
  {Radovi\ifmmode~\acute{c}\else \'{c}\fi{}}, \citenamefont {Plumb},
  \citenamefont {Gibbs}, \citenamefont {Mackenzie}, \citenamefont {Berthod},
  \citenamefont {Strand}, \citenamefont {Kim}, \citenamefont {Georges},\ and\
  \citenamefont {Baumberger}}]{Tamai2019}%
  \BibitemOpen
  \bibfield  {author} {\bibinfo {author} {\bibfnamefont {A.}~\bibnamefont
  {Tamai}}, \bibinfo {author} {\bibfnamefont {M.}~\bibnamefont {Zingl}},
  \bibinfo {author} {\bibfnamefont {E.}~\bibnamefont {Rozbicki}}, \bibinfo
  {author} {\bibfnamefont {E.}~\bibnamefont {Cappelli}}, \bibinfo {author}
  {\bibfnamefont {S.}~\bibnamefont {Ricc\`o}}, \bibinfo {author} {\bibfnamefont
  {A.}~\bibnamefont {de~la Torre}}, \bibinfo {author} {\bibfnamefont
  {S.}~\bibnamefont {McKeown~Walker}}, \bibinfo {author} {\bibfnamefont
  {F.~Y.}\ \bibnamefont {Bruno}}, \bibinfo {author} {\bibfnamefont {P.~D.~C.}\
  \bibnamefont {King}}, \bibinfo {author} {\bibfnamefont {W.}~\bibnamefont
  {Meevasana}}, \bibinfo {author} {\bibfnamefont {M.}~\bibnamefont {Shi}},
  \bibinfo {author} {\bibfnamefont {M.}~\bibnamefont
  {Radovi\ifmmode~\acute{c}\else \'{c}\fi{}}}, \bibinfo {author} {\bibfnamefont
  {N.~C.}\ \bibnamefont {Plumb}}, \bibinfo {author} {\bibfnamefont {A.~S.}\
  \bibnamefont {Gibbs}}, \bibinfo {author} {\bibfnamefont {A.~P.}\ \bibnamefont
  {Mackenzie}}, \bibinfo {author} {\bibfnamefont {C.}~\bibnamefont {Berthod}},
  \bibinfo {author} {\bibfnamefont {H.~U.~R.}\ \bibnamefont {Strand}}, \bibinfo
  {author} {\bibfnamefont {M.}~\bibnamefont {Kim}}, \bibinfo {author}
  {\bibfnamefont {A.}~\bibnamefont {Georges}}, \ and\ \bibinfo {author}
  {\bibfnamefont {F.}~\bibnamefont {Baumberger}},\ }\bibfield  {title}
  {\enquote {\bibinfo {title} {{High-Resolution Photoemission on Sr$_2$RuO$_4$
  Reveals Correlation-Enhanced Effective Spin-Orbit Coupling and Dominantly
  Local Self-Energies}},}\ }\href {\doibase 10.1103/PhysRevX.9.021048}
  {\bibfield  {journal} {\bibinfo  {journal} {Phys. Rev. X}\ }\textbf {\bibinfo
  {volume} {9}},\ \bibinfo {pages} {021048} (\bibinfo {year}
  {2019})}\BibitemShut {NoStop}%
\bibitem [{\citenamefont {Maeno}\ \emph {et~al.}(1997)\citenamefont {Maeno},
  \citenamefont {Yoshida}, \citenamefont {Hashimoto}, \citenamefont
  {Nishizaki}, \citenamefont {Ikeda}, \citenamefont {Nohara}, \citenamefont
  {Fujita}, \citenamefont {Mackenzie}, \citenamefont {Hussey}, \citenamefont
  {Bednorz},\ and\ \citenamefont {Lichtenberg}}]{Maeno1997}%
  \BibitemOpen
  \bibfield  {author} {\bibinfo {author} {\bibfnamefont {Yoshiteru}\
  \bibnamefont {Maeno}}, \bibinfo {author} {\bibfnamefont {Koji}\ \bibnamefont
  {Yoshida}}, \bibinfo {author} {\bibfnamefont {Hiroaki}\ \bibnamefont
  {Hashimoto}}, \bibinfo {author} {\bibfnamefont {Shuji}\ \bibnamefont
  {Nishizaki}}, \bibinfo {author} {\bibfnamefont {Shin-ichi}\ \bibnamefont
  {Ikeda}}, \bibinfo {author} {\bibfnamefont {Minoru}\ \bibnamefont {Nohara}},
  \bibinfo {author} {\bibfnamefont {Toshizo}\ \bibnamefont {Fujita}}, \bibinfo
  {author} {\bibfnamefont {Andrew~P.}\ \bibnamefont {Mackenzie}}, \bibinfo
  {author} {\bibfnamefont {Nigel~E.}\ \bibnamefont {Hussey}}, \bibinfo {author}
  {\bibfnamefont {J.~Georg}\ \bibnamefont {Bednorz}}, \ and\ \bibinfo {author}
  {\bibfnamefont {Frank}\ \bibnamefont {Lichtenberg}},\ }\bibfield  {title}
  {\enquote {\bibinfo {title} {{Two-Dimensional Fermi Liquid Behavior of the
  Superconductor Sr$_2$RuO$_4$}},}\ }\href {\doibase 10.1143/JPSJ.66.1405}
  {\bibfield  {journal} {\bibinfo  {journal} {J. Phys. Soc. Japan}\ }\textbf
  {\bibinfo {volume} {66}},\ \bibinfo {pages} {1405--1408} (\bibinfo {year}
  {1997})}\BibitemShut {NoStop}%
\bibitem [{\citenamefont {Hussey}\ \emph {et~al.}(1998)\citenamefont {Hussey},
  \citenamefont {Mackenzie}, \citenamefont {Cooper}, \citenamefont {Maeno},
  \citenamefont {Nishizaki},\ and\ \citenamefont {Fujita}}]{Hussey1998}%
  \BibitemOpen
  \bibfield  {author} {\bibinfo {author} {\bibfnamefont {N~E}\ \bibnamefont
  {Hussey}}, \bibinfo {author} {\bibfnamefont {A~P}\ \bibnamefont {Mackenzie}},
  \bibinfo {author} {\bibfnamefont {J~R}\ \bibnamefont {Cooper}}, \bibinfo
  {author} {\bibfnamefont {Y}~\bibnamefont {Maeno}}, \bibinfo {author}
  {\bibfnamefont {S}~\bibnamefont {Nishizaki}}, \ and\ \bibinfo {author}
  {\bibfnamefont {T}~\bibnamefont {Fujita}},\ }\bibfield  {title} {\enquote
  {\bibinfo {title} {{Normal-state magnetoresistance of
  ${\mathrm{Sr}}_{2}{\mathrm{RuO}}_{4}$}},}\ }\href
  {https://link.aps.org/doi/10.1103/PhysRevB.57.5505} {\bibfield  {journal}
  {\bibinfo  {journal} {Phys. Rev. B}\ }\textbf {\bibinfo {volume} {57}},\
  \bibinfo {pages} {5505} (\bibinfo {year} {1998})}\BibitemShut {NoStop}%
\bibitem [{\citenamefont {Tyler}\ \emph {et~al.}(1998)\citenamefont {Tyler},
  \citenamefont {Mackenzie}, \citenamefont {NishiZaki},\ and\ \citenamefont
  {Maeno}}]{Tyler1998}%
  \BibitemOpen
  \bibfield  {author} {\bibinfo {author} {\bibfnamefont {A~W}\ \bibnamefont
  {Tyler}}, \bibinfo {author} {\bibfnamefont {A~P}\ \bibnamefont {Mackenzie}},
  \bibinfo {author} {\bibfnamefont {S}~\bibnamefont {NishiZaki}}, \ and\
  \bibinfo {author} {\bibfnamefont {Y}~\bibnamefont {Maeno}},\ }\bibfield
  {title} {\enquote {\bibinfo {title} {{High-temperature resistivity of
  ${\mathrm{Sr}}_{2}{\mathrm{RuO}}_{4}:$ Bad metallic transport in a good
  metal}},}\ }\href {\doibase 10.1103/PhysRevB.58.R10107} {\bibfield  {journal}
  {\bibinfo  {journal} {Phys. Rev. B}\ }\textbf {\bibinfo {volume} {58}},\
  \bibinfo {pages} {R10107(R)} (\bibinfo {year} {1998})}\BibitemShut {NoStop}%
\bibitem [{\citenamefont {Maeno}\ \emph {et~al.}(1994)\citenamefont {Maeno},
  \citenamefont {Hashimoto}, \citenamefont {Yoshida}, \citenamefont
  {Nishizaki}, \citenamefont {Fujita}, \citenamefont {Bednorz},\ and\
  \citenamefont {Lichtenberg}}]{Maeno1994}%
  \BibitemOpen
  \bibfield  {author} {\bibinfo {author} {\bibfnamefont {Y}~\bibnamefont
  {Maeno}}, \bibinfo {author} {\bibfnamefont {H}~\bibnamefont {Hashimoto}},
  \bibinfo {author} {\bibfnamefont {K}~\bibnamefont {Yoshida}}, \bibinfo
  {author} {\bibfnamefont {S}~\bibnamefont {Nishizaki}}, \bibinfo {author}
  {\bibfnamefont {T}~\bibnamefont {Fujita}}, \bibinfo {author} {\bibfnamefont
  {J~G}\ \bibnamefont {Bednorz}}, \ and\ \bibinfo {author} {\bibfnamefont
  {F}~\bibnamefont {Lichtenberg}},\ }\bibfield  {title} {\enquote {\bibinfo
  {title} {{Superconductivity in a layered perovskite without copper}},}\
  }\href {http://dx.doi.org/10.1038/372532a0} {\bibfield  {journal} {\bibinfo
  {journal} {Nature}\ }\textbf {\bibinfo {volume} {372}},\ \bibinfo {pages}
  {532--534} (\bibinfo {year} {1994})}\BibitemShut {NoStop}%
\bibitem [{\citenamefont {Rice}\ and\ \citenamefont
  {Sigrist}(1995)}]{Rice1995}%
  \BibitemOpen
  \bibfield  {author} {\bibinfo {author} {\bibfnamefont {T~M}\ \bibnamefont
  {Rice}}\ and\ \bibinfo {author} {\bibfnamefont {M}~\bibnamefont {Sigrist}},\
  }\bibfield  {title} {\enquote {\bibinfo {title} {{Sr$_2$RuO$_4$ : an
  electronic analogue of $^3$He?}}}\ }\href {\doibase
  10.1088/0953-8984/7/47/002} {\bibfield  {journal} {\bibinfo  {journal} {J.
  Phys. Condens. Matter Matter}\ }\textbf {\bibinfo {volume} {7}},\ \bibinfo
  {pages} {L643--L648} (\bibinfo {year} {1995})}\BibitemShut {NoStop}%
\bibitem [{\citenamefont {Mackenzie}\ \emph {et~al.}(1998)\citenamefont
  {Mackenzie}, \citenamefont {Haselwimmer}, \citenamefont {Tyler},
  \citenamefont {Lonzarich}, \citenamefont {Mori}, \citenamefont {Nishizaki},\
  and\ \citenamefont {Maeno}}]{Mackenzie1998}%
  \BibitemOpen
  \bibfield  {author} {\bibinfo {author} {\bibfnamefont {A~P}\ \bibnamefont
  {Mackenzie}}, \bibinfo {author} {\bibfnamefont {R~K~W}\ \bibnamefont
  {Haselwimmer}}, \bibinfo {author} {\bibfnamefont {A~W}\ \bibnamefont
  {Tyler}}, \bibinfo {author} {\bibfnamefont {G~G}\ \bibnamefont {Lonzarich}},
  \bibinfo {author} {\bibfnamefont {Y}~\bibnamefont {Mori}}, \bibinfo {author}
  {\bibfnamefont {S}~\bibnamefont {Nishizaki}}, \ and\ \bibinfo {author}
  {\bibfnamefont {Y}~\bibnamefont {Maeno}},\ }\bibfield  {title} {\enquote
  {\bibinfo {title} {{Extremely Strong Dependence of Superconductivity on
  Disorder in ${\mathrm{Sr}}_{2}{\mathrm{RuO}}_{4}$}},}\ }\href
  {https://link.aps.org/doi/10.1103/PhysRevLett.80.161} {\bibfield  {journal}
  {\bibinfo  {journal} {Phys. Rev. Lett.}\ }\textbf {\bibinfo {volume} {80}},\
  \bibinfo {pages} {161--164} (\bibinfo {year} {1998})}\BibitemShut {NoStop}%
\bibitem [{\citenamefont {Ishida}\ \emph {et~al.}(1998)\citenamefont {Ishida},
  \citenamefont {Mukuda}, \citenamefont {Kitaoka}, \citenamefont {Asayama},
  \citenamefont {Mao}, \citenamefont {Mori},\ and\ \citenamefont
  {Maeno}}]{Ishida1998}%
  \BibitemOpen
  \bibfield  {author} {\bibinfo {author} {\bibfnamefont {K}~\bibnamefont
  {Ishida}}, \bibinfo {author} {\bibfnamefont {H}~\bibnamefont {Mukuda}},
  \bibinfo {author} {\bibfnamefont {Y}~\bibnamefont {Kitaoka}}, \bibinfo
  {author} {\bibfnamefont {K}~\bibnamefont {Asayama}}, \bibinfo {author}
  {\bibfnamefont {Z~Q}\ \bibnamefont {Mao}}, \bibinfo {author} {\bibfnamefont
  {Y}~\bibnamefont {Mori}}, \ and\ \bibinfo {author} {\bibfnamefont
  {Y}~\bibnamefont {Maeno}},\ }\bibfield  {title} {\enquote {\bibinfo {title}
  {{Spin-triplet superconductivity in Sr$_2$RuO$_4$ identified by $^{17}$O
  Knight shift}},}\ }\href {\doibase 10.1038/25315} {\bibfield  {journal}
  {\bibinfo  {journal} {Nature}\ }\textbf {\bibinfo {volume} {396}},\ \bibinfo
  {pages} {658--660} (\bibinfo {year} {1998})}\BibitemShut {NoStop}%
\bibitem [{\citenamefont {Mackenzie}\ and\ \citenamefont
  {Maeno}(2003)}]{Mackenzie2003}%
  \BibitemOpen
  \bibfield  {author} {\bibinfo {author} {\bibfnamefont {Andrew~Peter}\
  \bibnamefont {Mackenzie}}\ and\ \bibinfo {author} {\bibfnamefont {Yoshiteru}\
  \bibnamefont {Maeno}},\ }\bibfield  {title} {\enquote {\bibinfo {title} {{The
  superconductivity of ${\mathrm{Sr}}_{2}{\mathrm{RuO}}_{4}$ and the physics of
  spin-triplet pairing}},}\ }\href {\doibase 10.1103/RevModPhys.75.657}
  {\bibfield  {journal} {\bibinfo  {journal} {Rev. Mod. Phys.}\ }\textbf
  {\bibinfo {volume} {75}},\ \bibinfo {pages} {657--712} (\bibinfo {year}
  {2003})}\BibitemShut {NoStop}%
\bibitem [{\citenamefont {Mackenzie}\ \emph {et~al.}(2017)\citenamefont
  {Mackenzie}, \citenamefont {Scaffidi}, \citenamefont {Hicks},\ and\
  \citenamefont {Maeno}}]{Mackenzie2017}%
  \BibitemOpen
  \bibfield  {author} {\bibinfo {author} {\bibfnamefont {Andrew~P}\
  \bibnamefont {Mackenzie}}, \bibinfo {author} {\bibfnamefont {Thomas}\
  \bibnamefont {Scaffidi}}, \bibinfo {author} {\bibfnamefont {Clifford~W}\
  \bibnamefont {Hicks}}, \ and\ \bibinfo {author} {\bibfnamefont {Yoshiteru}\
  \bibnamefont {Maeno}},\ }\bibfield  {title} {\enquote {\bibinfo {title}
  {{Even odder after twenty-three years: the superconducting order parameter
  puzzle of ${\mathrm{Sr}}_{2}{\mathrm{RuO}}_{4}$}},}\ }\href {\doibase
  10.1038/s41535-017-0045-4} {\bibfield  {journal} {\bibinfo  {journal} {npj
  Quantum Mater.}\ }\textbf {\bibinfo {volume} {2}},\ \bibinfo {pages} {40}
  (\bibinfo {year} {2017})}\BibitemShut {NoStop}%
\bibitem [{\citenamefont {Pustogow}\ \emph {et~al.}(2019)\citenamefont
  {Pustogow}, \citenamefont {Luo}, \citenamefont {Chronister}, \citenamefont
  {Su}, \citenamefont {Sokolov}, \citenamefont {Jerzembeck}, \citenamefont
  {Mackenzie}, \citenamefont {Hicks}, \citenamefont {Kikugawa}, \citenamefont
  {Raghu}, \citenamefont {Bauer},\ and\ \citenamefont {Brown}}]{Pustogow2019}%
  \BibitemOpen
  \bibfield  {author} {\bibinfo {author} {\bibfnamefont {A}~\bibnamefont
  {Pustogow}}, \bibinfo {author} {\bibfnamefont {Yongkang}\ \bibnamefont
  {Luo}}, \bibinfo {author} {\bibfnamefont {A}~\bibnamefont {Chronister}},
  \bibinfo {author} {\bibfnamefont {Y.-S.}\ \bibnamefont {Su}}, \bibinfo
  {author} {\bibfnamefont {D~A}\ \bibnamefont {Sokolov}}, \bibinfo {author}
  {\bibfnamefont {F}~\bibnamefont {Jerzembeck}}, \bibinfo {author}
  {\bibfnamefont {A~P}\ \bibnamefont {Mackenzie}}, \bibinfo {author}
  {\bibfnamefont {C~W}\ \bibnamefont {Hicks}}, \bibinfo {author} {\bibfnamefont
  {N}~\bibnamefont {Kikugawa}}, \bibinfo {author} {\bibfnamefont
  {S}~\bibnamefont {Raghu}}, \bibinfo {author} {\bibfnamefont {E~D}\
  \bibnamefont {Bauer}}, \ and\ \bibinfo {author} {\bibfnamefont {S~E}\
  \bibnamefont {Brown}},\ }\bibfield  {title} {\enquote {\bibinfo {title}
  {{Constraints on the superconducting order parameter in Sr$_2$RuO$_4$ from
  oxygen-17 nuclear magnetic resonance}},}\ }\href {\doibase
  10.1038/s41586-019-1596-2} {\bibfield  {journal} {\bibinfo  {journal}
  {Nature}\ }\textbf {\bibinfo {volume} {574}},\ \bibinfo {pages} {72--75}
  (\bibinfo {year} {2019})}\BibitemShut {NoStop}%
\bibitem [{\citenamefont {Hicks}\ \emph {et~al.}(2014)\citenamefont {Hicks},
  \citenamefont {Brodsky}, \citenamefont {Yelland}, \citenamefont {Gibbs},
  \citenamefont {Bruin}, \citenamefont {Barber}, \citenamefont {Edkins},
  \citenamefont {Nishimura}, \citenamefont {Yonezawa}, \citenamefont {Maeno},\
  and\ \citenamefont {Mackenzie}}]{Hicks2014}%
  \BibitemOpen
  \bibfield  {author} {\bibinfo {author} {\bibfnamefont {Clifford~W}\
  \bibnamefont {Hicks}}, \bibinfo {author} {\bibfnamefont {Daniel~O}\
  \bibnamefont {Brodsky}}, \bibinfo {author} {\bibfnamefont {Edward~A}\
  \bibnamefont {Yelland}}, \bibinfo {author} {\bibfnamefont {Alexandra~S}\
  \bibnamefont {Gibbs}}, \bibinfo {author} {\bibfnamefont {Jan A~N}\
  \bibnamefont {Bruin}}, \bibinfo {author} {\bibfnamefont {Mark~E}\
  \bibnamefont {Barber}}, \bibinfo {author} {\bibfnamefont {Stephen~D}\
  \bibnamefont {Edkins}}, \bibinfo {author} {\bibfnamefont {Keigo}\
  \bibnamefont {Nishimura}}, \bibinfo {author} {\bibfnamefont {Shingo}\
  \bibnamefont {Yonezawa}}, \bibinfo {author} {\bibfnamefont {Yoshiteru}\
  \bibnamefont {Maeno}}, \ and\ \bibinfo {author} {\bibfnamefont {Andrew~P}\
  \bibnamefont {Mackenzie}},\ }\bibfield  {title} {\enquote {\bibinfo {title}
  {{Strong Increase of $T_c$ of Sr$_2$RuO$_4$ Under Both Tensile and
  Compressive Strain}},}\ }\href {\doibase 10.1126/science.1248292} {\bibfield
  {journal} {\bibinfo  {journal} {Science}\ }\textbf {\bibinfo {volume}
  {344}},\ \bibinfo {pages} {283} (\bibinfo {year} {2014})}\BibitemShut
  {NoStop}%
\bibitem [{\citenamefont {Steppke}\ \emph {et~al.}(2017)\citenamefont
  {Steppke}, \citenamefont {Zhao}, \citenamefont {Barber}, \citenamefont
  {Scaffidi}, \citenamefont {Jerzembeck}, \citenamefont {Rosner}, \citenamefont
  {Gibbs}, \citenamefont {Maeno}, \citenamefont {Simon}, \citenamefont
  {Mackenzie},\ and\ \citenamefont {Hicks}}]{Steppke2017}%
  \BibitemOpen
  \bibfield  {author} {\bibinfo {author} {\bibfnamefont {Alexander}\
  \bibnamefont {Steppke}}, \bibinfo {author} {\bibfnamefont {Lishan}\
  \bibnamefont {Zhao}}, \bibinfo {author} {\bibfnamefont {Mark~E}\ \bibnamefont
  {Barber}}, \bibinfo {author} {\bibfnamefont {Thomas}\ \bibnamefont
  {Scaffidi}}, \bibinfo {author} {\bibfnamefont {Fabian}\ \bibnamefont
  {Jerzembeck}}, \bibinfo {author} {\bibfnamefont {Helge}\ \bibnamefont
  {Rosner}}, \bibinfo {author} {\bibfnamefont {Alexandra~S}\ \bibnamefont
  {Gibbs}}, \bibinfo {author} {\bibfnamefont {Yoshiteru}\ \bibnamefont
  {Maeno}}, \bibinfo {author} {\bibfnamefont {Steven~H}\ \bibnamefont {Simon}},
  \bibinfo {author} {\bibfnamefont {Andrew~P}\ \bibnamefont {Mackenzie}}, \
  and\ \bibinfo {author} {\bibfnamefont {Clifford~W}\ \bibnamefont {Hicks}},\
  }\bibfield  {title} {\enquote {\bibinfo {title} {{Strong peak in $T_c$ of
  Sr$_2$RuO$_4$ under uniaxial pressure}},}\ }\href
  {http://science.sciencemag.org/content/355/6321/eaaf9398.abstract} {\bibfield
   {journal} {\bibinfo  {journal} {Science}\ }\textbf {\bibinfo {volume}
  {355}},\ \bibinfo {pages} {eaaf9398} (\bibinfo {year} {2017})}\BibitemShut
  {NoStop}%
\bibitem [{\citenamefont {Barber}\ \emph {et~al.}(2018)\citenamefont {Barber},
  \citenamefont {Gibbs}, \citenamefont {Maeno}, \citenamefont {Mackenzie},\
  and\ \citenamefont {Hicks}}]{Barber2018}%
  \BibitemOpen
  \bibfield  {author} {\bibinfo {author} {\bibfnamefont {M.~E.}\ \bibnamefont
  {Barber}}, \bibinfo {author} {\bibfnamefont {A.~S.}\ \bibnamefont {Gibbs}},
  \bibinfo {author} {\bibfnamefont {Y}~\bibnamefont {Maeno}}, \bibinfo {author}
  {\bibfnamefont {A.~P.}\ \bibnamefont {Mackenzie}}, \ and\ \bibinfo {author}
  {\bibfnamefont {C.~W.}\ \bibnamefont {Hicks}},\ }\bibfield  {title} {\enquote
  {\bibinfo {title} {{Resistivity in the Vicinity of a van Hove Singularity:
  ${\mathrm{Sr}}_{2}{\mathrm{RuO}}_{4}$ under Uniaxial Pressure}},}\ }\href
  {\doibase 10.1103/PhysRevLett.120.076602} {\bibfield  {journal} {\bibinfo
  {journal} {Phys. Rev. Lett.}\ }\textbf {\bibinfo {volume} {120}},\ \bibinfo
  {pages} {076602} (\bibinfo {year} {2018})}\BibitemShut {NoStop}%
\bibitem [{\citenamefont {Luo}\ \emph {et~al.}(2019)\citenamefont {Luo},
  \citenamefont {Pustogow}, \citenamefont {Guzman}, \citenamefont {Dioguardi},
  \citenamefont {Thomas}, \citenamefont {Ronning}, \citenamefont {Kikugawa},
  \citenamefont {Sokolov}, \citenamefont {Jerzembeck}, \citenamefont
  {Mackenzie}, \citenamefont {Hicks}, \citenamefont {Bauer}, \citenamefont
  {Mazin},\ and\ \citenamefont {Brown}}]{Luo2019}%
  \BibitemOpen
  \bibfield  {author} {\bibinfo {author} {\bibfnamefont {Yongkang}\
  \bibnamefont {Luo}}, \bibinfo {author} {\bibfnamefont {A.}~\bibnamefont
  {Pustogow}}, \bibinfo {author} {\bibfnamefont {P.}~\bibnamefont {Guzman}},
  \bibinfo {author} {\bibfnamefont {A.~P.}\ \bibnamefont {Dioguardi}}, \bibinfo
  {author} {\bibfnamefont {S.~M.}\ \bibnamefont {Thomas}}, \bibinfo {author}
  {\bibfnamefont {F.}~\bibnamefont {Ronning}}, \bibinfo {author} {\bibfnamefont
  {N.}~\bibnamefont {Kikugawa}}, \bibinfo {author} {\bibfnamefont {D.~A.}\
  \bibnamefont {Sokolov}}, \bibinfo {author} {\bibfnamefont {F.}~\bibnamefont
  {Jerzembeck}}, \bibinfo {author} {\bibfnamefont {A.~P.}\ \bibnamefont
  {Mackenzie}}, \bibinfo {author} {\bibfnamefont {C.~W.}\ \bibnamefont
  {Hicks}}, \bibinfo {author} {\bibfnamefont {E.~D.}\ \bibnamefont {Bauer}},
  \bibinfo {author} {\bibfnamefont {I.~I.}\ \bibnamefont {Mazin}}, \ and\
  \bibinfo {author} {\bibfnamefont {S.~E.}\ \bibnamefont {Brown}},\ }\bibfield
  {title} {\enquote {\bibinfo {title} {{Normal State $^{17}\mathrm{O}$ NMR
  Studies of ${\mathrm{Sr}}_{2}{\mathrm{RuO}}_{4}$ under Uniaxial Stress}},}\
  }\href {\doibase 10.1103/physrevx.9.021044} {\bibfield  {journal} {\bibinfo
  {journal} {Phys. Rev. X}\ }\textbf {\bibinfo {volume} {9}},\ \bibinfo {pages}
  {021044} (\bibinfo {year} {2019})}\BibitemShut {NoStop}%
\bibitem [{\citenamefont {Barber}\ \emph {et~al.}(2019)\citenamefont {Barber},
  \citenamefont {Lechermann}, \citenamefont {Streltsov}, \citenamefont
  {Skornyakov}, \citenamefont {Ghosh}, \citenamefont {Ramshaw}, \citenamefont
  {Kikugawa}, \citenamefont {Sokolov}, \citenamefont {Mackenzie}, \citenamefont
  {Hicks},\ and\ \citenamefont {Mazin}}]{Barber2019}%
  \BibitemOpen
  \bibfield  {author} {\bibinfo {author} {\bibfnamefont {Mark~E}\ \bibnamefont
  {Barber}}, \bibinfo {author} {\bibfnamefont {Frank}\ \bibnamefont
  {Lechermann}}, \bibinfo {author} {\bibfnamefont {Sergey~V}\ \bibnamefont
  {Streltsov}}, \bibinfo {author} {\bibfnamefont {Sergey~L}\ \bibnamefont
  {Skornyakov}}, \bibinfo {author} {\bibfnamefont {Sayak}\ \bibnamefont
  {Ghosh}}, \bibinfo {author} {\bibfnamefont {B~J}\ \bibnamefont {Ramshaw}},
  \bibinfo {author} {\bibfnamefont {Naoki}\ \bibnamefont {Kikugawa}}, \bibinfo
  {author} {\bibfnamefont {Dmitry~A}\ \bibnamefont {Sokolov}}, \bibinfo
  {author} {\bibfnamefont {Andrew~P}\ \bibnamefont {Mackenzie}}, \bibinfo
  {author} {\bibfnamefont {Clifford~W}\ \bibnamefont {Hicks}}, \ and\ \bibinfo
  {author} {\bibfnamefont {I~I}\ \bibnamefont {Mazin}},\ }\bibfield  {title}
  {\enquote {\bibinfo {title} {{Role of correlations in determining the Van
  Hove strain in ${\mathrm{Sr}}_{2}{\mathrm{RuO}}_{4}$}},}\ }\href {\doibase
  10.1103/PhysRevB.100.245139} {\bibfield  {journal} {\bibinfo  {journal}
  {Phys. Rev. B}\ }\textbf {\bibinfo {volume} {100}},\ \bibinfo {pages}
  {245139} (\bibinfo {year} {2019})}\BibitemShut {NoStop}%
\bibitem [{\citenamefont {Han}\ \emph {et~al.}(2016)\citenamefont {Han},
  \citenamefont {Dang},\ and\ \citenamefont {Millis}}]{Han2016}%
  \BibitemOpen
  \bibfield  {author} {\bibinfo {author} {\bibfnamefont {Qiang}\ \bibnamefont
  {Han}}, \bibinfo {author} {\bibfnamefont {Hung~T}\ \bibnamefont {Dang}}, \
  and\ \bibinfo {author} {\bibfnamefont {A~J}\ \bibnamefont {Millis}},\
  }\bibfield  {title} {\enquote {\bibinfo {title} {{Ferromagnetism and
  correlation strength in cubic barium ruthenate in comparison to strontium and
  calcium ruthenate: A dynamical mean-field study}},}\ }\href {\doibase
  10.1103/PhysRevB.93.155103} {\bibfield  {journal} {\bibinfo  {journal} {Phys.
  Rev. B}\ }\textbf {\bibinfo {volume} {93}},\ \bibinfo {pages} {155103}
  (\bibinfo {year} {2016})}\BibitemShut {NoStop}%
\bibitem [{\citenamefont {Kikugawa}\ \emph
  {et~al.}(2004{\natexlab{a}})\citenamefont {Kikugawa}, \citenamefont
  {Mackenzie}, \citenamefont {Bergemann}, \citenamefont {Borzi}, \citenamefont
  {Grigera},\ and\ \citenamefont {Maeno}}]{Kikugawa2004PRBR}%
  \BibitemOpen
  \bibfield  {author} {\bibinfo {author} {\bibfnamefont {N}~\bibnamefont
  {Kikugawa}}, \bibinfo {author} {\bibfnamefont {A~P}\ \bibnamefont
  {Mackenzie}}, \bibinfo {author} {\bibfnamefont {C}~\bibnamefont {Bergemann}},
  \bibinfo {author} {\bibfnamefont {R~A}\ \bibnamefont {Borzi}}, \bibinfo
  {author} {\bibfnamefont {S~A}\ \bibnamefont {Grigera}}, \ and\ \bibinfo
  {author} {\bibfnamefont {Y}~\bibnamefont {Maeno}},\ }\bibfield  {title}
  {\enquote {\bibinfo {title} {{Rigid-band shift of the Fermi level in the
  strongly correlated metal:
  ${\mathrm{Sr}}_{2\ensuremath{-}y}{\mathrm{La}}_{y}{\mathrm{RuO}}_{4}$}},}\
  }\href {\doibase 10.1103/PhysRevB.70.060508} {\bibfield  {journal} {\bibinfo
  {journal} {Phys. Rev. B}\ }\textbf {\bibinfo {volume} {70}},\ \bibinfo
  {pages} {060508(R)} (\bibinfo {year} {2004}{\natexlab{a}})}\BibitemShut
  {NoStop}%
\bibitem [{\citenamefont {Kikugawa}\ \emph
  {et~al.}(2004{\natexlab{b}})\citenamefont {Kikugawa}, \citenamefont
  {Bergemann}, \citenamefont {Mackenzie},\ and\ \citenamefont
  {Maeno}}]{Kikugawa2004}%
  \BibitemOpen
  \bibfield  {author} {\bibinfo {author} {\bibfnamefont {Naoki}\ \bibnamefont
  {Kikugawa}}, \bibinfo {author} {\bibfnamefont {Christoph}\ \bibnamefont
  {Bergemann}}, \bibinfo {author} {\bibfnamefont {Andrew~Peter}\ \bibnamefont
  {Mackenzie}}, \ and\ \bibinfo {author} {\bibfnamefont {Yoshiteru}\
  \bibnamefont {Maeno}},\ }\bibfield  {title} {\enquote {\bibinfo {title}
  {{Band-selective modification of the magnetic fluctuations in
  ${\mathrm{Sr}}_{2}{\mathrm{RuO}}_{4}$: A study of substitution effects}},}\
  }\href {\doibase 10.1103/PhysRevB.70.134520} {\bibfield  {journal} {\bibinfo
  {journal} {Phys. Rev. B}\ }\textbf {\bibinfo {volume} {70}},\ \bibinfo
  {pages} {134520} (\bibinfo {year} {2004}{\natexlab{b}})}\BibitemShut
  {NoStop}%
\bibitem [{\citenamefont {Burganov}\ \emph {et~al.}(2016)\citenamefont
  {Burganov}, \citenamefont {Adamo}, \citenamefont {Mulder}, \citenamefont
  {Uchida}, \citenamefont {King}, \citenamefont {Harter}, \citenamefont {Shai},
  \citenamefont {Gibbs}, \citenamefont {Mackenzie}, \citenamefont {Uecker},
  \citenamefont {Bruetzam}, \citenamefont {Beasley}, \citenamefont {Fennie},
  \citenamefont {Schlom},\ and\ \citenamefont {Shen}}]{Burganov2016}%
  \BibitemOpen
  \bibfield  {author} {\bibinfo {author} {\bibfnamefont {B}~\bibnamefont
  {Burganov}}, \bibinfo {author} {\bibfnamefont {C}~\bibnamefont {Adamo}},
  \bibinfo {author} {\bibfnamefont {A}~\bibnamefont {Mulder}}, \bibinfo
  {author} {\bibfnamefont {M}~\bibnamefont {Uchida}}, \bibinfo {author}
  {\bibfnamefont {P.~D.~C.}\ \bibnamefont {King}}, \bibinfo {author}
  {\bibfnamefont {J.~W.}\ \bibnamefont {Harter}}, \bibinfo {author}
  {\bibfnamefont {D.~E.}\ \bibnamefont {Shai}}, \bibinfo {author}
  {\bibfnamefont {A.~S.}\ \bibnamefont {Gibbs}}, \bibinfo {author}
  {\bibfnamefont {A.~P.}\ \bibnamefont {Mackenzie}}, \bibinfo {author}
  {\bibfnamefont {R}~\bibnamefont {Uecker}}, \bibinfo {author} {\bibfnamefont
  {M}~\bibnamefont {Bruetzam}}, \bibinfo {author} {\bibfnamefont {M.~R.}\
  \bibnamefont {Beasley}}, \bibinfo {author} {\bibfnamefont {C.~J.}\
  \bibnamefont {Fennie}}, \bibinfo {author} {\bibfnamefont {D.~G.}\
  \bibnamefont {Schlom}}, \ and\ \bibinfo {author} {\bibfnamefont {K.~M.}\
  \bibnamefont {Shen}},\ }\bibfield  {title} {\enquote {\bibinfo {title}
  {{Strain Control of Fermiology and Many-Body Interactions in Two-Dimensional
  Ruthenates}},}\ }\href
  {https://link.aps.org/doi/10.1103/PhysRevLett.116.197003} {\bibfield
  {journal} {\bibinfo  {journal} {Phys. Rev. Lett.}\ }\textbf {\bibinfo
  {volume} {116}},\ \bibinfo {pages} {197003} (\bibinfo {year}
  {2016})}\BibitemShut {NoStop}%
\bibitem [{\citenamefont {Herman}\ \emph {et~al.}(2019)\citenamefont {Herman},
  \citenamefont {Buhmann}, \citenamefont {Fischer},\ and\ \citenamefont
  {Sigrist}}]{Herman2019}%
  \BibitemOpen
  \bibfield  {author} {\bibinfo {author} {\bibfnamefont {Franti\v{s}ek}\
  \bibnamefont {Herman}}, \bibinfo {author} {\bibfnamefont {Jonathan}\
  \bibnamefont {Buhmann}}, \bibinfo {author} {\bibfnamefont {Mark~H}\
  \bibnamefont {Fischer}}, \ and\ \bibinfo {author} {\bibfnamefont {Manfred}\
  \bibnamefont {Sigrist}},\ }\bibfield  {title} {\enquote {\bibinfo {title}
  {{Deviation from Fermi-liquid transport behavior in the vicinity of a Van
  Hove singularity}},}\ }\href {\doibase 10.1103/PhysRevB.99.184107} {\bibfield
   {journal} {\bibinfo  {journal} {Phys. Rev. B}\ }\textbf {\bibinfo {volume}
  {99}},\ \bibinfo {pages} {184107} (\bibinfo {year} {2019})}\BibitemShut
  {NoStop}%
\bibitem [{\citenamefont {Liebsch}\ and\ \citenamefont
  {Lichtenstein}(2000)}]{Liebsch2000}%
  \BibitemOpen
  \bibfield  {author} {\bibinfo {author} {\bibfnamefont {A}~\bibnamefont
  {Liebsch}}\ and\ \bibinfo {author} {\bibfnamefont {A}~\bibnamefont
  {Lichtenstein}},\ }\bibfield  {title} {\enquote {\bibinfo {title}
  {{Photoemission Quasiparticle Spectra of
  ${\mathrm{Sr}}_{2}{\mathrm{RuO}}_{4}$}},}\ }\href
  {https://link.aps.org/doi/10.1103/PhysRevLett.84.1591} {\bibfield  {journal}
  {\bibinfo  {journal} {Phys. Rev. Lett.}\ }\textbf {\bibinfo {volume} {84}},\
  \bibinfo {pages} {1591--1594} (\bibinfo {year} {2000})}\BibitemShut {NoStop}%
\bibitem [{\citenamefont {Pchelkina}\ \emph {et~al.}(2007)\citenamefont
  {Pchelkina}, \citenamefont {Nekrasov}, \citenamefont {Pruschke},
  \citenamefont {Sekiyama}, \citenamefont {Suga}, \citenamefont {Anisimov},\
  and\ \citenamefont {Vollhardt}}]{Pchelkina2007}%
  \BibitemOpen
  \bibfield  {author} {\bibinfo {author} {\bibfnamefont {Z~V}\ \bibnamefont
  {Pchelkina}}, \bibinfo {author} {\bibfnamefont {I~A}\ \bibnamefont
  {Nekrasov}}, \bibinfo {author} {\bibfnamefont {Th.}\ \bibnamefont
  {Pruschke}}, \bibinfo {author} {\bibfnamefont {A}~\bibnamefont {Sekiyama}},
  \bibinfo {author} {\bibfnamefont {S}~\bibnamefont {Suga}}, \bibinfo {author}
  {\bibfnamefont {V~I}\ \bibnamefont {Anisimov}}, \ and\ \bibinfo {author}
  {\bibfnamefont {D}~\bibnamefont {Vollhardt}},\ }\bibfield  {title} {\enquote
  {\bibinfo {title} {{Evidence for strong electronic correlations in the
  spectra of ${\mathrm{Sr}}_{2}\mathrm{Ru}{\mathrm{O}}_{4}$}},}\ }\href
  {https://link.aps.org/doi/10.1103/PhysRevB.75.035122} {\bibfield  {journal}
  {\bibinfo  {journal} {Phys. Rev. B}\ }\textbf {\bibinfo {volume} {75}},\
  \bibinfo {pages} {35122} (\bibinfo {year} {2007})}\BibitemShut {NoStop}%
\bibitem [{\citenamefont {Mravlje}\ \emph {et~al.}(2011)\citenamefont
  {Mravlje}, \citenamefont {Aichhorn}, \citenamefont {Miyake}, \citenamefont
  {Haule}, \citenamefont {Kotliar},\ and\ \citenamefont
  {Georges}}]{Mravlje2011}%
  \BibitemOpen
  \bibfield  {author} {\bibinfo {author} {\bibfnamefont {Jernej}\ \bibnamefont
  {Mravlje}}, \bibinfo {author} {\bibfnamefont {Markus}\ \bibnamefont
  {Aichhorn}}, \bibinfo {author} {\bibfnamefont {Takashi}\ \bibnamefont
  {Miyake}}, \bibinfo {author} {\bibfnamefont {Kristjan}\ \bibnamefont
  {Haule}}, \bibinfo {author} {\bibfnamefont {Gabriel}\ \bibnamefont
  {Kotliar}}, \ and\ \bibinfo {author} {\bibfnamefont {Antoine}\ \bibnamefont
  {Georges}},\ }\bibfield  {title} {\enquote {\bibinfo {title}
  {{Coherence-Incoherence Crossover and the Mass-Renormalization Puzzles in
  ${\mathrm{Sr}}_{2}{\mathrm{RuO}}_{4}$}},}\ }\href
  {https://link.aps.org/doi/10.1103/PhysRevLett.106.096401} {\bibfield
  {journal} {\bibinfo  {journal} {Phys. Rev. Lett.}\ }\textbf {\bibinfo
  {volume} {106}},\ \bibinfo {pages} {096401} (\bibinfo {year}
  {2011})}\BibitemShut {NoStop}%
\bibitem [{\citenamefont {Deng}\ \emph {et~al.}(2016)\citenamefont {Deng},
  \citenamefont {Haule},\ and\ \citenamefont {Kotliar}}]{Deng2016}%
  \BibitemOpen
  \bibfield  {author} {\bibinfo {author} {\bibfnamefont {Xiaoyu}\ \bibnamefont
  {Deng}}, \bibinfo {author} {\bibfnamefont {Kristjan}\ \bibnamefont {Haule}},
  \ and\ \bibinfo {author} {\bibfnamefont {Gabriel}\ \bibnamefont {Kotliar}},\
  }\bibfield  {title} {\enquote {\bibinfo {title} {{Transport Properties of
  Metallic Ruthenates: A $\mathrm{DFT}+\mathrm{DMFT}$ Investigation}},}\ }\href
  {https://link.aps.org/doi/10.1103/PhysRevLett.116.256401} {\bibfield
  {journal} {\bibinfo  {journal} {Phys. Rev. Lett.}\ }\textbf {\bibinfo
  {volume} {116}},\ \bibinfo {pages} {256401} (\bibinfo {year}
  {2016})}\BibitemShut {NoStop}%
\bibitem [{\citenamefont {Mravlje}\ and\ \citenamefont
  {Georges}(2016)}]{Mravlje2016}%
  \BibitemOpen
  \bibfield  {author} {\bibinfo {author} {\bibfnamefont {Jernej}\ \bibnamefont
  {Mravlje}}\ and\ \bibinfo {author} {\bibfnamefont {Antoine}\ \bibnamefont
  {Georges}},\ }\bibfield  {title} {\enquote {\bibinfo {title} {{Thermopower
  and Entropy: Lessons from ${\mathrm{Sr}}_{2}{\mathrm{RuO}}_{4}$}},}\ }\href
  {https://link.aps.org/doi/10.1103/PhysRevLett.117.036401} {\bibfield
  {journal} {\bibinfo  {journal} {Phys. Rev. Lett.}\ }\textbf {\bibinfo
  {volume} {117}},\ \bibinfo {pages} {036401} (\bibinfo {year}
  {2016})}\BibitemShut {NoStop}%
\bibitem [{\citenamefont {Zhang}\ \emph {et~al.}(2016)\citenamefont {Zhang},
  \citenamefont {Gorelov}, \citenamefont {Sarvestani},\ and\ \citenamefont
  {Pavarini}}]{Zhang2016}%
  \BibitemOpen
  \bibfield  {author} {\bibinfo {author} {\bibfnamefont {Guoren}\ \bibnamefont
  {Zhang}}, \bibinfo {author} {\bibfnamefont {Evgeny}\ \bibnamefont {Gorelov}},
  \bibinfo {author} {\bibfnamefont {Esmaeel}\ \bibnamefont {Sarvestani}}, \
  and\ \bibinfo {author} {\bibfnamefont {Eva}\ \bibnamefont {Pavarini}},\
  }\bibfield  {title} {\enquote {\bibinfo {title} {{Fermi Surface of
  Sr$_2$RuO$_4$: Spin-Orbit and Anisotropic Coulomb Interaction Effects}},}\
  }\href {\doibase 10.1103/PhysRevLett.116.106402} {\bibfield  {journal}
  {\bibinfo  {journal} {Phys. Rev. Lett.}\ }\textbf {\bibinfo {volume} {116}},\
  \bibinfo {pages} {106402} (\bibinfo {year} {2016})}\BibitemShut {NoStop}%
\bibitem [{\citenamefont {Sarvestani}\ \emph {et~al.}(2018)\citenamefont
  {Sarvestani}, \citenamefont {Zhang}, \citenamefont {Gorelov},\ and\
  \citenamefont {Pavarini}}]{Sarvestani2018}%
  \BibitemOpen
  \bibfield  {author} {\bibinfo {author} {\bibfnamefont {Esmaeel}\ \bibnamefont
  {Sarvestani}}, \bibinfo {author} {\bibfnamefont {Gouren}\ \bibnamefont
  {Zhang}}, \bibinfo {author} {\bibfnamefont {Evgeny}\ \bibnamefont {Gorelov}},
  \ and\ \bibinfo {author} {\bibfnamefont {Eva}\ \bibnamefont {Pavarini}},\
  }\bibfield  {title} {\enquote {\bibinfo {title} {{Effective masses,
  lifetimes, and optical conductivity in ${\mathrm{Sr}}_{2}{\mathrm{RuO}}_{4}$
  and ${\mathrm{Sr}}_{3}{\mathrm{Ru}}_{2}{\mathrm{O}}_{7}$: Interplay of
  spin-orbit, crystal-field, and Coulomb tetragonal tensor interactions}},}\
  }\href {\doibase 10.1103/PhysRevB.97.085141} {\bibfield  {journal} {\bibinfo
  {journal} {Phys. Rev. B}\ }\textbf {\bibinfo {volume} {97}},\ \bibinfo
  {pages} {085141} (\bibinfo {year} {2018})}\BibitemShut {NoStop}%
\bibitem [{\citenamefont {Kim}\ \emph {et~al.}(2018)\citenamefont {Kim},
  \citenamefont {Mravlje}, \citenamefont {Ferrero}, \citenamefont {Parcollet},\
  and\ \citenamefont {Georges}}]{MKim2018}%
  \BibitemOpen
  \bibfield  {author} {\bibinfo {author} {\bibfnamefont {Minjae}\ \bibnamefont
  {Kim}}, \bibinfo {author} {\bibfnamefont {Jernej}\ \bibnamefont {Mravlje}},
  \bibinfo {author} {\bibfnamefont {Michel}\ \bibnamefont {Ferrero}}, \bibinfo
  {author} {\bibfnamefont {Olivier}\ \bibnamefont {Parcollet}}, \ and\ \bibinfo
  {author} {\bibfnamefont {Antoine}\ \bibnamefont {Georges}},\ }\bibfield
  {title} {\enquote {\bibinfo {title} {{Spin-Orbit Coupling and Electronic
  Correlations in ${\mathrm{Sr}}_{2}{\mathrm{RuO}}_{4}$}},}\ }\href {\doibase
  10.1103/PhysRevLett.120.126401} {\bibfield  {journal} {\bibinfo  {journal}
  {Phys. Rev. Lett.}\ }\textbf {\bibinfo {volume} {120}},\ \bibinfo {pages}
  {126401} (\bibinfo {year} {2018})}\BibitemShut {NoStop}%
\bibitem [{\citenamefont {Facio}\ \emph {et~al.}(2018)\citenamefont {Facio},
  \citenamefont {Mravlje}, \citenamefont {Pourovskii}, \citenamefont
  {Cornaglia},\ and\ \citenamefont {Vildosola}}]{Facio2018}%
  \BibitemOpen
  \bibfield  {author} {\bibinfo {author} {\bibfnamefont {Jorge~I}\ \bibnamefont
  {Facio}}, \bibinfo {author} {\bibfnamefont {Jernej}\ \bibnamefont {Mravlje}},
  \bibinfo {author} {\bibfnamefont {Leonid}\ \bibnamefont {Pourovskii}},
  \bibinfo {author} {\bibfnamefont {Pablo~S}\ \bibnamefont {Cornaglia}}, \ and\
  \bibinfo {author} {\bibfnamefont {V}~\bibnamefont {Vildosola}},\ }\bibfield
  {title} {\enquote {\bibinfo {title} {{Spin-orbit and anisotropic strain
  effects on the electronic correlations in
  ${\mathrm{Sr}}_{2}{\mathrm{RuO}}_{4}$}},}\ }\href {\doibase
  10.1103/PhysRevB.98.085121} {\bibfield  {journal} {\bibinfo  {journal} {Phys.
  Rev. B}\ }\textbf {\bibinfo {volume} {98}},\ \bibinfo {pages} {085121}
  (\bibinfo {year} {2018})}\BibitemShut {NoStop}%
\bibitem [{\citenamefont {Deng}\ \emph {et~al.}(2019)\citenamefont {Deng},
  \citenamefont {Stadler}, \citenamefont {Haule}, \citenamefont {Weichselbaum},
  \citenamefont {von Delft},\ and\ \citenamefont {Kotliar}}]{Deng2019}%
  \BibitemOpen
  \bibfield  {author} {\bibinfo {author} {\bibfnamefont {Xiaoyu}\ \bibnamefont
  {Deng}}, \bibinfo {author} {\bibfnamefont {Katharina~M}\ \bibnamefont
  {Stadler}}, \bibinfo {author} {\bibfnamefont {Kristjan}\ \bibnamefont
  {Haule}}, \bibinfo {author} {\bibfnamefont {Andreas}\ \bibnamefont
  {Weichselbaum}}, \bibinfo {author} {\bibfnamefont {Jan}\ \bibnamefont {von
  Delft}}, \ and\ \bibinfo {author} {\bibfnamefont {Gabriel}\ \bibnamefont
  {Kotliar}},\ }\bibfield  {title} {\enquote {\bibinfo {title} {{Signatures of
  Mottness and Hundness in archetypal correlated metals}},}\ }\href {\doibase
  10.1038/s41467-019-10257-2} {\bibfield  {journal} {\bibinfo  {journal} {Nat.
  Commun.}\ }\textbf {\bibinfo {volume} {10}},\ \bibinfo {pages} {2721}
  (\bibinfo {year} {2019})}\BibitemShut {NoStop}%
\bibitem [{\citenamefont {Zingl}\ \emph {et~al.}(2019)\citenamefont {Zingl},
  \citenamefont {Mravlje}, \citenamefont {Aichhorn}, \citenamefont
  {Parcollet},\ and\ \citenamefont {Georges}}]{Zingl2019}%
  \BibitemOpen
  \bibfield  {author} {\bibinfo {author} {\bibfnamefont {Manuel}\ \bibnamefont
  {Zingl}}, \bibinfo {author} {\bibfnamefont {Jernej}\ \bibnamefont {Mravlje}},
  \bibinfo {author} {\bibfnamefont {Markus}\ \bibnamefont {Aichhorn}}, \bibinfo
  {author} {\bibfnamefont {Olivier}\ \bibnamefont {Parcollet}}, \ and\ \bibinfo
  {author} {\bibfnamefont {Antoine}\ \bibnamefont {Georges}},\ }\bibfield
  {title} {\enquote {\bibinfo {title} {{Hall coefficient signals orbital
  differentiation in the Hund's metal Sr$_2$RuO$_4$}},}\ }\href {\doibase
  10.1038/s41535-019-0175-y} {\bibfield  {journal} {\bibinfo  {journal} {npj
  Quantum Mater.}\ }\textbf {\bibinfo {volume} {4}},\ \bibinfo {pages} {35}
  (\bibinfo {year} {2019})}\BibitemShut {NoStop}%
\bibitem [{\citenamefont {Strand}\ \emph {et~al.}(2019)\citenamefont {Strand},
  \citenamefont {Zingl}, \citenamefont {Wentzell}, \citenamefont {Parcollet},\
  and\ \citenamefont {Georges}}]{Strand2019}%
  \BibitemOpen
  \bibfield  {author} {\bibinfo {author} {\bibfnamefont {Hugo U~R}\
  \bibnamefont {Strand}}, \bibinfo {author} {\bibfnamefont {Manuel}\
  \bibnamefont {Zingl}}, \bibinfo {author} {\bibfnamefont {Nils}\ \bibnamefont
  {Wentzell}}, \bibinfo {author} {\bibfnamefont {Olivier}\ \bibnamefont
  {Parcollet}}, \ and\ \bibinfo {author} {\bibfnamefont {Antoine}\ \bibnamefont
  {Georges}},\ }\bibfield  {title} {\enquote {\bibinfo {title} {{Magnetic
  response of ${\mathrm{Sr}}_{2}{\mathrm{RuO}}_{4}$: Quasi-local spin
  fluctuations due to Hund's coupling}},}\ }\href {\doibase
  10.1103/PhysRevB.100.125120} {\bibfield  {journal} {\bibinfo  {journal}
  {Phys. Rev. B}\ }\textbf {\bibinfo {volume} {100}},\ \bibinfo {pages}
  {125120} (\bibinfo {year} {2019})}\BibitemShut {NoStop}%
\bibitem [{\citenamefont {Gingras}\ \emph {et~al.}(2019)\citenamefont
  {Gingras}, \citenamefont {Nourafkan}, \citenamefont {Tremblay},\ and\
  \citenamefont {C{\^{o}}t{\'{e}}}}]{Gingras2019}%
  \BibitemOpen
  \bibfield  {author} {\bibinfo {author} {\bibfnamefont {O}~\bibnamefont
  {Gingras}}, \bibinfo {author} {\bibfnamefont {R}~\bibnamefont {Nourafkan}},
  \bibinfo {author} {\bibfnamefont {A.-M.~S}\ \bibnamefont {Tremblay}}, \ and\
  \bibinfo {author} {\bibfnamefont {M}~\bibnamefont {C{\^{o}}t{\'{e}}}},\
  }\bibfield  {title} {\enquote {\bibinfo {title} {{Superconducting Symmetries
  of ${\mathrm{Sr}}_{2}{\mathrm{RuO}}_{4}$ from First-Principles Electronic
  Structure}},}\ }\href {\doibase 10.1103/PhysRevLett.123.217005} {\bibfield
  {journal} {\bibinfo  {journal} {Phys. Rev. Lett.}\ }\textbf {\bibinfo
  {volume} {123}},\ \bibinfo {pages} {217005} (\bibinfo {year}
  {2019})}\BibitemShut {NoStop}%
\bibitem [{\citenamefont {Kugler}\ \emph {et~al.}(2020)\citenamefont {Kugler},
  \citenamefont {Zingl}, \citenamefont {Strand}, \citenamefont {Lee},
  \citenamefont {von Delft},\ and\ \citenamefont {Georges}}]{Kugler2020}%
  \BibitemOpen
  \bibfield  {author} {\bibinfo {author} {\bibfnamefont {Fabian~B}\
  \bibnamefont {Kugler}}, \bibinfo {author} {\bibfnamefont {Manuel}\
  \bibnamefont {Zingl}}, \bibinfo {author} {\bibfnamefont {Hugo U~R}\
  \bibnamefont {Strand}}, \bibinfo {author} {\bibfnamefont {Seung-Sup~B}\
  \bibnamefont {Lee}}, \bibinfo {author} {\bibfnamefont {Jan}\ \bibnamefont
  {von Delft}}, \ and\ \bibinfo {author} {\bibfnamefont {Antoine}\ \bibnamefont
  {Georges}},\ }\bibfield  {title} {\enquote {\bibinfo {title} {{Strongly
  Correlated Materials from a Numerical Renormalization Group Perspective: How
  the Fermi-Liquid State of ${\mathrm{Sr}}_{2}{\mathrm{RuO}}_{4}$ Emerges}},}\
  }\href {\doibase 10.1103/PhysRevLett.124.016401} {\bibfield  {journal}
  {\bibinfo  {journal} {Phys. Rev. Lett.}\ }\textbf {\bibinfo {volume} {124}},\
  \bibinfo {pages} {16401} (\bibinfo {year} {2020})}\BibitemShut {NoStop}%
\bibitem [{\citenamefont {Linden}\ \emph {et~al.}(2020)\citenamefont {Linden},
  \citenamefont {Zingl}, \citenamefont {Hubig}, \citenamefont {Parcollet},\
  and\ \citenamefont {Schollw{\"{o}}ck}}]{Linden2020}%
  \BibitemOpen
  \bibfield  {author} {\bibinfo {author} {\bibfnamefont {Nils-Oliver}\
  \bibnamefont {Linden}}, \bibinfo {author} {\bibfnamefont {Manuel}\
  \bibnamefont {Zingl}}, \bibinfo {author} {\bibfnamefont {Claudius}\
  \bibnamefont {Hubig}}, \bibinfo {author} {\bibfnamefont {Olivier}\
  \bibnamefont {Parcollet}}, \ and\ \bibinfo {author} {\bibfnamefont {Ulrich}\
  \bibnamefont {Schollw{\"{o}}ck}},\ }\bibfield  {title} {\enquote {\bibinfo
  {title} {{Imaginary-time matrix product state impurity solver in a real
  material calculation: Spin-orbit coupling in
  $\mathrm{Sr}{}_{2}\mathrm{RuO}{}_{4}$}},}\ }\href {\doibase
  10.1103/PhysRevB.101.041101} {\bibfield  {journal} {\bibinfo  {journal}
  {Phys. Rev. B}\ }\textbf {\bibinfo {volume} {101}},\ \bibinfo {pages} {41101}
  (\bibinfo {year} {2020})}\BibitemShut {NoStop}%
\bibitem [{\citenamefont {Karp}\ \emph {et~al.}(2020)\citenamefont {Karp},
  \citenamefont {Bramberger}, \citenamefont {Grundner}, \citenamefont
  {Schollw{\"{o}}ck}, \citenamefont {Millis},\ and\ \citenamefont
  {Zingl}}]{Karp2020}%
  \BibitemOpen
  \bibfield  {author} {\bibinfo {author} {\bibfnamefont {Jonathan}\
  \bibnamefont {Karp}}, \bibinfo {author} {\bibfnamefont {Max}\ \bibnamefont
  {Bramberger}}, \bibinfo {author} {\bibfnamefont {Martin}\ \bibnamefont
  {Grundner}}, \bibinfo {author} {\bibfnamefont {Ulrich}\ \bibnamefont
  {Schollw{\"{o}}ck}}, \bibinfo {author} {\bibfnamefont {Andrew~J.}\
  \bibnamefont {Millis}}, \ and\ \bibinfo {author} {\bibfnamefont {Manuel}\
  \bibnamefont {Zingl}},\ }\href {http://arxiv.org/abs/2004.12515} {\enquote
  {\bibinfo {title} {{Sr$_2$MoO$_4$ and Sr$_2$RuO$_4$: Disentangling the Roles
  of Hund's and van Hove Physics}},}\ } (\bibinfo {year} {2020}),\ \Eprint
  {http://arxiv.org/abs/2004.12515} {arXiv:2004.12515} \BibitemShut {NoStop}%
\bibitem [{\citenamefont {Ryee}\ \emph {et~al.}(2016)\citenamefont {Ryee},
  \citenamefont {Jang}, \citenamefont {Kino}, \citenamefont {Kotani},\ and\
  \citenamefont {Han}}]{Ryee2016}%
  \BibitemOpen
  \bibfield  {author} {\bibinfo {author} {\bibfnamefont {Siheon}\ \bibnamefont
  {Ryee}}, \bibinfo {author} {\bibfnamefont {Seung~Woo}\ \bibnamefont {Jang}},
  \bibinfo {author} {\bibfnamefont {Hiori}\ \bibnamefont {Kino}}, \bibinfo
  {author} {\bibfnamefont {Takao}\ \bibnamefont {Kotani}}, \ and\ \bibinfo
  {author} {\bibfnamefont {Myung~Joon}\ \bibnamefont {Han}},\ }\bibfield
  {title} {\enquote {\bibinfo {title} {{Quasiparticle self-consistent GW
  calculation of Sr$_2$RuO$_4$ and SrRuO$_3$}},}\ }\href {\doibase
  10.1103/PhysRevB.93.075125} {\bibfield  {journal} {\bibinfo  {journal} {Phys.
  Rev. B}\ }\textbf {\bibinfo {volume} {93}},\ \bibinfo {pages} {075125}
  (\bibinfo {year} {2016})}\BibitemShut {NoStop}%
\bibitem [{\citenamefont {Cobo}\ \emph {et~al.}(2016)\citenamefont {Cobo},
  \citenamefont {Ahn}, \citenamefont {Eremin},\ and\ \citenamefont
  {Akbari}}]{Cobo2016}%
  \BibitemOpen
  \bibfield  {author} {\bibinfo {author} {\bibfnamefont {Sergio}\ \bibnamefont
  {Cobo}}, \bibinfo {author} {\bibfnamefont {Felix}\ \bibnamefont {Ahn}},
  \bibinfo {author} {\bibfnamefont {Ilya}\ \bibnamefont {Eremin}}, \ and\
  \bibinfo {author} {\bibfnamefont {Alireza}\ \bibnamefont {Akbari}},\
  }\bibfield  {title} {\enquote {\bibinfo {title} {{Anisotropic spin
  fluctuations in ${\mathrm{Sr}}_{2}{\mathrm{RuO}}_{4}$: Role of spin-orbit
  coupling and induced strain}},}\ }\href
  {https://link.aps.org/doi/10.1103/PhysRevB.94.224507} {\bibfield  {journal}
  {\bibinfo  {journal} {Phys. Rev. B}\ }\textbf {\bibinfo {volume} {94}},\
  \bibinfo {pages} {224507} (\bibinfo {year} {2016})}\BibitemShut {NoStop}%
\bibitem [{\citenamefont {Veenstra}\ \emph {et~al.}(2014)\citenamefont
  {Veenstra}, \citenamefont {Zhu}, \citenamefont {Raichle}, \citenamefont
  {Ludbrook}, \citenamefont {Nicolaou}, \citenamefont {Slomski}, \citenamefont
  {Landolt}, \citenamefont {Kittaka}, \citenamefont {Maeno}, \citenamefont
  {Dil}, \citenamefont {Elfimov}, \citenamefont {Haverkort},\ and\
  \citenamefont {Damascelli}}]{Veenstra2014}%
  \BibitemOpen
  \bibfield  {author} {\bibinfo {author} {\bibfnamefont {C.~N.}\ \bibnamefont
  {Veenstra}}, \bibinfo {author} {\bibfnamefont {Z.-H.}\ \bibnamefont {Zhu}},
  \bibinfo {author} {\bibfnamefont {M}~\bibnamefont {Raichle}}, \bibinfo
  {author} {\bibfnamefont {B.~M.}\ \bibnamefont {Ludbrook}}, \bibinfo {author}
  {\bibfnamefont {A}~\bibnamefont {Nicolaou}}, \bibinfo {author} {\bibfnamefont
  {B}~\bibnamefont {Slomski}}, \bibinfo {author} {\bibfnamefont
  {G}~\bibnamefont {Landolt}}, \bibinfo {author} {\bibfnamefont
  {S}~\bibnamefont {Kittaka}}, \bibinfo {author} {\bibfnamefont
  {Y}~\bibnamefont {Maeno}}, \bibinfo {author} {\bibfnamefont {J.~H.}\
  \bibnamefont {Dil}}, \bibinfo {author} {\bibfnamefont {I.~S.}\ \bibnamefont
  {Elfimov}}, \bibinfo {author} {\bibfnamefont {M.~W.}\ \bibnamefont
  {Haverkort}}, \ and\ \bibinfo {author} {\bibfnamefont {A}~\bibnamefont
  {Damascelli}},\ }\bibfield  {title} {\enquote {\bibinfo {title}
  {{Spin-Orbital Entanglement and the Breakdown of Singlets and Triplets in
  ${\mathrm{Sr}}_{2}{\mathrm{RuO}}_{4}$ Revealed by Spin- and Angle-Resolved
  Photoemission Spectroscopy}},}\ }\href
  {https://link.aps.org/doi/10.1103/PhysRevLett.112.127002} {\bibfield
  {journal} {\bibinfo  {journal} {Phys. Rev. Lett.}\ }\textbf {\bibinfo
  {volume} {112}},\ \bibinfo {pages} {127002} (\bibinfo {year}
  {2014})}\BibitemShut {NoStop}%
\bibitem [{\citenamefont {Kim}\ \emph {et~al.}(2008)\citenamefont {Kim},
  \citenamefont {Jin}, \citenamefont {Moon}, \citenamefont {Kim}, \citenamefont
  {Park}, \citenamefont {Leem}, \citenamefont {Yu}, \citenamefont {Noh},
  \citenamefont {Kim}, \citenamefont {Oh}, \citenamefont {Park}, \citenamefont
  {Durairaj}, \citenamefont {Cao},\ and\ \citenamefont
  {Rotenberg}}]{BJKim2008}%
  \BibitemOpen
  \bibfield  {author} {\bibinfo {author} {\bibfnamefont {B~J}\ \bibnamefont
  {Kim}}, \bibinfo {author} {\bibfnamefont {Hosub}\ \bibnamefont {Jin}},
  \bibinfo {author} {\bibfnamefont {S~J}\ \bibnamefont {Moon}}, \bibinfo
  {author} {\bibfnamefont {J.-Y.}\ \bibnamefont {Kim}}, \bibinfo {author}
  {\bibfnamefont {B.-G.}\ \bibnamefont {Park}}, \bibinfo {author}
  {\bibfnamefont {C~S}\ \bibnamefont {Leem}}, \bibinfo {author} {\bibfnamefont
  {Jaejun}\ \bibnamefont {Yu}}, \bibinfo {author} {\bibfnamefont {T~W}\
  \bibnamefont {Noh}}, \bibinfo {author} {\bibfnamefont {C}~\bibnamefont
  {Kim}}, \bibinfo {author} {\bibfnamefont {S.-J.}\ \bibnamefont {Oh}},
  \bibinfo {author} {\bibfnamefont {J.-H.}\ \bibnamefont {Park}}, \bibinfo
  {author} {\bibfnamefont {V}~\bibnamefont {Durairaj}}, \bibinfo {author}
  {\bibfnamefont {G}~\bibnamefont {Cao}}, \ and\ \bibinfo {author}
  {\bibfnamefont {E}~\bibnamefont {Rotenberg}},\ }\bibfield  {title} {\enquote
  {\bibinfo {title} {{Novel ${J}_{\mathrm{eff}}=1/2$ Mott State Induced by
  Relativistic Spin-Orbit Coupling in
  ${\mathrm{Sr}}_{2}{\mathrm{IrO}}_{4}$}},}\ }\href {\doibase
  10.1103/PhysRevLett.101.076402} {\bibfield  {journal} {\bibinfo  {journal}
  {Phys. Rev. Lett.}\ }\textbf {\bibinfo {volume} {101}},\ \bibinfo {pages}
  {76402} (\bibinfo {year} {2008})}\BibitemShut {NoStop}%
\bibitem [{\citenamefont {Pavarini}\ and\ \citenamefont
  {Mazin}(2006)}]{Pavarini2006}%
  \BibitemOpen
  \bibfield  {author} {\bibinfo {author} {\bibfnamefont {E}~\bibnamefont
  {Pavarini}}\ and\ \bibinfo {author} {\bibfnamefont {I.~I.}\ \bibnamefont
  {Mazin}},\ }\bibfield  {title} {\enquote {\bibinfo {title} {{First-principles
  study of spin-orbit effects and NMR in Sr2 Ru O4}},}\ }\href {\doibase
  10.1103/PhysRevB.74.035115} {\bibfield  {journal} {\bibinfo  {journal} {Phys.
  Rev. B}\ }\textbf {\bibinfo {volume} {74}},\ \bibinfo {pages} {035115}
  (\bibinfo {year} {2006})}\BibitemShut {NoStop}%
\bibitem [{\citenamefont {Haverkort}\ \emph {et~al.}(2008)\citenamefont
  {Haverkort}, \citenamefont {Elfimov}, \citenamefont {Tjeng}, \citenamefont
  {Sawatzky},\ and\ \citenamefont {Damascelli}}]{Haverkort2008}%
  \BibitemOpen
  \bibfield  {author} {\bibinfo {author} {\bibfnamefont {M~W}\ \bibnamefont
  {Haverkort}}, \bibinfo {author} {\bibfnamefont {I~S}\ \bibnamefont
  {Elfimov}}, \bibinfo {author} {\bibfnamefont {L~H}\ \bibnamefont {Tjeng}},
  \bibinfo {author} {\bibfnamefont {G~A}\ \bibnamefont {Sawatzky}}, \ and\
  \bibinfo {author} {\bibfnamefont {A}~\bibnamefont {Damascelli}},\ }\bibfield
  {title} {\enquote {\bibinfo {title} {{Strong spin-orbit coupling effects on
  the fermi surface of Sr$_2$RuO$_4$ and Sr$_2$RhO$_4$}},}\ }\href {\doibase
  10.1103/PhysRevLett.101.026406} {\bibfield  {journal} {\bibinfo  {journal}
  {Phys. Rev. Lett.}\ }\textbf {\bibinfo {volume} {101}},\ \bibinfo {pages}
  {026406} (\bibinfo {year} {2008})}\BibitemShut {NoStop}%
\bibitem [{\citenamefont {Gull}\ \emph {et~al.}(2011)\citenamefont {Gull},
  \citenamefont {Millis}, \citenamefont {Lichtenstein}, \citenamefont
  {Rubtsov}, \citenamefont {Troyer},\ and\ \citenamefont {Werner}}]{Gull2011}%
  \BibitemOpen
  \bibfield  {author} {\bibinfo {author} {\bibfnamefont {Emanuel}\ \bibnamefont
  {Gull}}, \bibinfo {author} {\bibfnamefont {Andrew~J}\ \bibnamefont {Millis}},
  \bibinfo {author} {\bibfnamefont {Alexander~I}\ \bibnamefont {Lichtenstein}},
  \bibinfo {author} {\bibfnamefont {Alexey~N}\ \bibnamefont {Rubtsov}},
  \bibinfo {author} {\bibfnamefont {Matthias}\ \bibnamefont {Troyer}}, \ and\
  \bibinfo {author} {\bibfnamefont {Philipp}\ \bibnamefont {Werner}},\
  }\bibfield  {title} {\enquote {\bibinfo {title} {{Continuous-time Monte Carlo
  methods for quantum impurity models}},}\ }\href {\doibase
  10.1103/RevModPhys.83.349} {\bibfield  {journal} {\bibinfo  {journal} {Rev.
  Mod. Phys.}\ }\textbf {\bibinfo {volume} {83}},\ \bibinfo {pages} {349--404}
  (\bibinfo {year} {2011})}\BibitemShut {NoStop}%
\bibitem [{\citenamefont {Georges}\ \emph {et~al.}(1996)\citenamefont
  {Georges}, \citenamefont {Kotliar}, \citenamefont {Krauth},\ and\
  \citenamefont {Rozenberg}}]{Georges1996}%
  \BibitemOpen
  \bibfield  {author} {\bibinfo {author} {\bibfnamefont {Antoine}\ \bibnamefont
  {Georges}}, \bibinfo {author} {\bibfnamefont {G}~\bibnamefont {Kotliar}},
  \bibinfo {author} {\bibfnamefont {Werner}\ \bibnamefont {Krauth}}, \ and\
  \bibinfo {author} {\bibfnamefont {M~J}\ \bibnamefont {Rozenberg}},\
  }\bibfield  {title} {\enquote {\bibinfo {title} {{Dynamical mean-field theory
  of strongly correlated fermion systems and the limit of infinite
  dimensions}},}\ }\href {\doibase http://dx.doi.org/10.1103/RevModPhys.68.13}
  {\bibfield  {journal} {\bibinfo  {journal} {Rev. Mod. Phys.}\ }\textbf
  {\bibinfo {volume} {68}},\ \bibinfo {pages} {13--125} (\bibinfo {year}
  {1996})}\BibitemShut {NoStop}%
\bibitem [{\citenamefont {Kotliar}\ \emph {et~al.}(2006)\citenamefont
  {Kotliar}, \citenamefont {Savrasov}, \citenamefont {Haule}, \citenamefont
  {Oudovenko}, \citenamefont {Parcollet},\ and\ \citenamefont
  {Marianetti}}]{Kotliar2006}%
  \BibitemOpen
  \bibfield  {author} {\bibinfo {author} {\bibfnamefont {G.}~\bibnamefont
  {Kotliar}}, \bibinfo {author} {\bibfnamefont {S.}~\bibnamefont {Savrasov}},
  \bibinfo {author} {\bibfnamefont {K.}~\bibnamefont {Haule}}, \bibinfo
  {author} {\bibfnamefont {V.}~\bibnamefont {Oudovenko}}, \bibinfo {author}
  {\bibfnamefont {O.}~\bibnamefont {Parcollet}}, \ and\ \bibinfo {author}
  {\bibfnamefont {C.}~\bibnamefont {Marianetti}},\ }\bibfield  {title}
  {\enquote {\bibinfo {title} {{Electronic structure calculations with
  dynamical mean-field theory}},}\ }\href {\doibase 10.1103/RevModPhys.78.865}
  {\bibfield  {journal} {\bibinfo  {journal} {Rev. Mod. Phys.}\ }\textbf
  {\bibinfo {volume} {78}},\ \bibinfo {pages} {865--951} (\bibinfo {year}
  {2006})}\BibitemShut {NoStop}%
\bibitem [{sup()}]{supp}%
  \BibitemOpen
  \href@noop {} {}\bibinfo {note} {See Supplementary Information for more
  details.}\BibitemShut {Stop}%
\bibitem [{\citenamefont {Mostofi}\ \emph {et~al.}(2014)\citenamefont
  {Mostofi}, \citenamefont {Yates}, \citenamefont {Pizzi}, \citenamefont {Lee},
  \citenamefont {Souza}, \citenamefont {Vanderbilt},\ and\ \citenamefont
  {Marzari}}]{Mostofi2014}%
  \BibitemOpen
  \bibfield  {author} {\bibinfo {author} {\bibfnamefont {Arash~A}\ \bibnamefont
  {Mostofi}}, \bibinfo {author} {\bibfnamefont {Jonathan~R}\ \bibnamefont
  {Yates}}, \bibinfo {author} {\bibfnamefont {Giovanni}\ \bibnamefont {Pizzi}},
  \bibinfo {author} {\bibfnamefont {Young-Su}\ \bibnamefont {Lee}}, \bibinfo
  {author} {\bibfnamefont {Ivo}\ \bibnamefont {Souza}}, \bibinfo {author}
  {\bibfnamefont {David}\ \bibnamefont {Vanderbilt}}, \ and\ \bibinfo {author}
  {\bibfnamefont {Nicola}\ \bibnamefont {Marzari}},\ }\bibfield  {title}
  {\enquote {\bibinfo {title} {{An updated version of wannier90: A tool for
  obtaining maximally-localised Wannier functions}},}\ }\href {\doibase
  https://doi.org/10.1016/j.cpc.2014.05.003} {\bibfield  {journal} {\bibinfo
  {journal} {Comput. Phys. Commun.}\ }\textbf {\bibinfo {volume} {185}},\
  \bibinfo {pages} {2309--2310} (\bibinfo {year} {2014})}\BibitemShut {NoStop}%
\bibitem [{\citenamefont {Kresse}\ and\ \citenamefont
  {Joubert}(1999)}]{Kresse1999}%
  \BibitemOpen
  \bibfield  {author} {\bibinfo {author} {\bibfnamefont {G}~\bibnamefont
  {Kresse}}\ and\ \bibinfo {author} {\bibfnamefont {D}~\bibnamefont
  {Joubert}},\ }\bibfield  {title} {\enquote {\bibinfo {title} {{From ultrasoft
  pseudopotentials to the projector augmented-wave method}},}\ }\href {\doibase
  10.1103/PhysRevB.59.1758} {\bibfield  {journal} {\bibinfo  {journal} {Phys.
  Rev. B}\ }\textbf {\bibinfo {volume} {59}},\ \bibinfo {pages} {1758--1775}
  (\bibinfo {year} {1999})}\BibitemShut {NoStop}%
\bibitem [{\citenamefont {Kanamori}(1963)}]{Kanamori1963}%
  \BibitemOpen
  \bibfield  {author} {\bibinfo {author} {\bibfnamefont {Junjiro}\ \bibnamefont
  {Kanamori}},\ }\bibfield  {title} {\enquote {\bibinfo {title} {{Electron
  Correlation and Ferromagnetism of Transition Metals}},}\ }\href {\doibase
  10.1143/PTP.30.275} {\bibfield  {journal} {\bibinfo  {journal} {Prog. Theor.
  Phys.}\ }\textbf {\bibinfo {volume} {30}},\ \bibinfo {pages} {275--289}
  (\bibinfo {year} {1963})}\BibitemShut {NoStop}%
\bibitem [{\citenamefont {Caffarel}\ and\ \citenamefont
  {Krauth}(1994)}]{Caffarel1994}%
  \BibitemOpen
  \bibfield  {author} {\bibinfo {author} {\bibfnamefont {Michel}\ \bibnamefont
  {Caffarel}}\ and\ \bibinfo {author} {\bibfnamefont {Werner}\ \bibnamefont
  {Krauth}},\ }\bibfield  {title} {\enquote {\bibinfo {title} {{Exact
  diagonalization approach to correlated fermions in infinite dimensions: Mott
  transition and superconductivity}},}\ }\href
  {https://link.aps.org/doi/10.1103/PhysRevLett.72.1545} {\bibfield  {journal}
  {\bibinfo  {journal} {Phys. Rev. Lett.}\ }\textbf {\bibinfo {volume} {72}},\
  \bibinfo {pages} {1545--1548} (\bibinfo {year} {1994})}\BibitemShut {NoStop}%
\bibitem [{\citenamefont {Koch}\ \emph {et~al.}(2008)\citenamefont {Koch},
  \citenamefont {Sangiovanni},\ and\ \citenamefont {Gunnarsson}}]{Koch2008}%
  \BibitemOpen
  \bibfield  {author} {\bibinfo {author} {\bibfnamefont {Erik}\ \bibnamefont
  {Koch}}, \bibinfo {author} {\bibfnamefont {Giorgio}\ \bibnamefont
  {Sangiovanni}}, \ and\ \bibinfo {author} {\bibfnamefont {Olle}\ \bibnamefont
  {Gunnarsson}},\ }\bibfield  {title} {\enquote {\bibinfo {title} {{Sum rules
  and bath parametrization for quantum cluster theories}},}\ }\href
  {https://link.aps.org/doi/10.1103/PhysRevB.78.115102} {\bibfield  {journal}
  {\bibinfo  {journal} {Phys. Rev. B}\ }\textbf {\bibinfo {volume} {78}},\
  \bibinfo {pages} {115102} (\bibinfo {year} {2008})}\BibitemShut {NoStop}%
\bibitem [{\citenamefont {S{\'{e}}n{\'{e}}chal}(2010)}]{Senechal2010}%
  \BibitemOpen
  \bibfield  {author} {\bibinfo {author} {\bibfnamefont {David}\ \bibnamefont
  {S{\'{e}}n{\'{e}}chal}},\ }\bibfield  {title} {\enquote {\bibinfo {title}
  {{Bath optimization in the cellular dynamical mean-field theory}},}\ }\href
  {https://link.aps.org/doi/10.1103/PhysRevB.81.235125} {\bibfield  {journal}
  {\bibinfo  {journal} {Phys. Rev. B}\ }\textbf {\bibinfo {volume} {81}},\
  \bibinfo {pages} {235125} (\bibinfo {year} {2010})}\BibitemShut {NoStop}%
\bibitem [{\citenamefont {Liebsch}\ and\ \citenamefont
  {Ishida}(2012)}]{Liebsch2012}%
  \BibitemOpen
  \bibfield  {author} {\bibinfo {author} {\bibfnamefont {Ansgar}\ \bibnamefont
  {Liebsch}}\ and\ \bibinfo {author} {\bibfnamefont {Hiroshi}\ \bibnamefont
  {Ishida}},\ }\bibfield  {title} {\enquote {\bibinfo {title} {{Temperature and
  bath size in exact diagonalization dynamical mean field theory}},}\ }\href
  {http://stacks.iop.org/0953-8984/24/i=5/a=053201} {\bibfield  {journal}
  {\bibinfo  {journal} {J. Phys. Condens. Matter}\ }\textbf {\bibinfo {volume}
  {24}},\ \bibinfo {pages} {53201} (\bibinfo {year} {2012})}\BibitemShut
  {NoStop}%
\bibitem [{\citenamefont {Go}\ and\ \citenamefont {Millis}(2015)}]{Go2015}%
  \BibitemOpen
  \bibfield  {author} {\bibinfo {author} {\bibfnamefont {Ara}\ \bibnamefont
  {Go}}\ and\ \bibinfo {author} {\bibfnamefont {Andrew~J}\ \bibnamefont
  {Millis}},\ }\bibfield  {title} {\enquote {\bibinfo {title} {{Spatial
  Correlations and the Insulating Phase of the High-$T_c$ Cuprates: Insights
  from a Configuration-Interaction-Based Solver for Dynamical Mean Field
  Theory}},}\ }\href {http://link.aps.org/doi/10.1103/PhysRevLett.114.016402}
  {\bibfield  {journal} {\bibinfo  {journal} {Phys. Rev. Lett.}\ }\textbf
  {\bibinfo {volume} {114}},\ \bibinfo {pages} {016402} (\bibinfo {year}
  {2015})}\BibitemShut {NoStop}%
\bibitem [{\citenamefont {Fertitta}\ and\ \citenamefont
  {Booth}(2018)}]{Fertitta2018}%
  \BibitemOpen
  \bibfield  {author} {\bibinfo {author} {\bibfnamefont {Edoardo}\ \bibnamefont
  {Fertitta}}\ and\ \bibinfo {author} {\bibfnamefont {George~H}\ \bibnamefont
  {Booth}},\ }\bibfield  {title} {\enquote {\bibinfo {title} {{Rigorous wave
  function embedding with dynamical fluctuations}},}\ }\href {\doibase
  10.1103/PhysRevB.98.235132} {\bibfield  {journal} {\bibinfo  {journal} {Phys.
  Rev. B}\ }\textbf {\bibinfo {volume} {98}},\ \bibinfo {pages} {235132}
  (\bibinfo {year} {2018})}\BibitemShut {NoStop}%
\bibitem [{\citenamefont {Go}\ and\ \citenamefont {Millis}(2017)}]{Go2017}%
  \BibitemOpen
  \bibfield  {author} {\bibinfo {author} {\bibfnamefont {Ara}\ \bibnamefont
  {Go}}\ and\ \bibinfo {author} {\bibfnamefont {Andrew~J}\ \bibnamefont
  {Millis}},\ }\bibfield  {title} {\enquote {\bibinfo {title} {{Adaptively
  truncated Hilbert space based impurity solver for dynamical mean-field
  theory}},}\ }\href {https://link.aps.org/doi/10.1103/PhysRevB.96.085139}
  {\bibfield  {journal} {\bibinfo  {journal} {Phys. Rev. B}\ }\textbf {\bibinfo
  {volume} {96}},\ \bibinfo {pages} {085139} (\bibinfo {year}
  {2017})}\BibitemShut {NoStop}%
\bibitem [{\citenamefont {Stadler}\ \emph {et~al.}(2015)\citenamefont
  {Stadler}, \citenamefont {Yin}, \citenamefont {von Delft}, \citenamefont
  {Kotliar},\ and\ \citenamefont {Weichselbaum}}]{Stadler2015}%
  \BibitemOpen
  \bibfield  {author} {\bibinfo {author} {\bibfnamefont {K.~M.}\ \bibnamefont
  {Stadler}}, \bibinfo {author} {\bibfnamefont {Z.~P.}\ \bibnamefont {Yin}},
  \bibinfo {author} {\bibfnamefont {J}~\bibnamefont {von Delft}}, \bibinfo
  {author} {\bibfnamefont {G}~\bibnamefont {Kotliar}}, \ and\ \bibinfo {author}
  {\bibfnamefont {A}~\bibnamefont {Weichselbaum}},\ }\bibfield  {title}
  {\enquote {\bibinfo {title} {{Dynamical Mean-Field Theory Plus Numerical
  Renormalization-Group Study of Spin-Orbital Separation in a Three-Band Hund
  Metal}},}\ }\href {\doibase 10.1103/PhysRevLett.115.136401} {\bibfield
  {journal} {\bibinfo  {journal} {Phy. Rev. Lett.}\ }\textbf {\bibinfo {volume}
  {115}},\ \bibinfo {pages} {136401} (\bibinfo {year} {2015})}\BibitemShut
  {NoStop}%
\bibitem [{\citenamefont {Kowalski}\ \emph {et~al.}(2019)\citenamefont
  {Kowalski}, \citenamefont {Hausoel}, \citenamefont {Wallerberger},
  \citenamefont {Gunacker},\ and\ \citenamefont {Sangiovanni}}]{Kowalski2019}%
  \BibitemOpen
  \bibfield  {author} {\bibinfo {author} {\bibfnamefont {Alexander}\
  \bibnamefont {Kowalski}}, \bibinfo {author} {\bibfnamefont {Andreas}\
  \bibnamefont {Hausoel}}, \bibinfo {author} {\bibfnamefont {Markus}\
  \bibnamefont {Wallerberger}}, \bibinfo {author} {\bibfnamefont {Patrik}\
  \bibnamefont {Gunacker}}, \ and\ \bibinfo {author} {\bibfnamefont {Giorgio}\
  \bibnamefont {Sangiovanni}},\ }\bibfield  {title} {\enquote {\bibinfo {title}
  {{State and superstate sampling in hybridization-expansion continuous-time
  quantum Monte Carlo}},}\ }\href {\doibase 10.1103/PhysRevB.99.155112}
  {\bibfield  {journal} {\bibinfo  {journal} {Phys. Rev. B}\ }\textbf {\bibinfo
  {volume} {99}},\ \bibinfo {pages} {155112} (\bibinfo {year}
  {2019})}\BibitemShut {NoStop}%
\bibitem [{\citenamefont {Oguchi}(1995)}]{Oguchi1995}%
  \BibitemOpen
  \bibfield  {author} {\bibinfo {author} {\bibfnamefont {Tamio}\ \bibnamefont
  {Oguchi}},\ }\bibfield  {title} {\enquote {\bibinfo {title} {{Electronic band
  structure of the superconductor Sr$_2$RuO$_4$}},}\ }\href
  {https://link.aps.org/doi/10.1103/PhysRevB.51.1385} {\bibfield  {journal}
  {\bibinfo  {journal} {Phy. Rev. B}\ }\textbf {\bibinfo {volume} {51}},\
  \bibinfo {pages} {1385--1388} (\bibinfo {year} {1995})}\BibitemShut {NoStop}%
\bibitem [{\citenamefont {Singh}(1995)}]{Singh1995}%
  \BibitemOpen
  \bibfield  {author} {\bibinfo {author} {\bibfnamefont {David~J}\ \bibnamefont
  {Singh}},\ }\bibfield  {title} {\enquote {\bibinfo {title} {{Relationship of
  Sr$_2$RuO$_4$ to the superconducting layered cuprates}},}\ }\href
  {https://link.aps.org/doi/10.1103/PhysRevB.52.1358} {\bibfield  {journal}
  {\bibinfo  {journal} {Phys. Rev. B}\ }\textbf {\bibinfo {volume} {52}},\
  \bibinfo {pages} {1358--1361} (\bibinfo {year} {1995})}\BibitemShut {NoStop}%
\bibitem [{Note1()}]{Note1}%
  \BibitemOpen
  \bibinfo {note} {We obtain $(m^*_{(\protect \frac {3}{2},\protect \frac
  {1}{2})} , m^*_{(\protect \frac {1}{2},\protect \frac {1}{2})} ,
  m^*_{(\protect \frac {3}{2},\protect \frac {3}{2})})\simeq $ (5.60, 5.09,
  3.94) at $T=0$ K and (4.56, 4.27, 3.61) at $T=11$ K. We use a fourth order
  polynomial with lowest six Matsubara frequencies and fit imaginary part of
  self-energy. See Supplementary Information for more details.}\BibitemShut
  {Stop}%
\bibitem [{\citenamefont {Kondo}\ \emph {et~al.}(2016)\citenamefont {Kondo},
  \citenamefont {Ochi}, \citenamefont {Nakayama}, \citenamefont {Taniguchi},
  \citenamefont {Akebi}, \citenamefont {Kuroda}, \citenamefont {Arita},
  \citenamefont {Sakai}, \citenamefont {Namatame}, \citenamefont {Taniguchi},
  \citenamefont {Maeno}, \citenamefont {Arita},\ and\ \citenamefont
  {Shin}}]{Kondo2016}%
  \BibitemOpen
  \bibfield  {author} {\bibinfo {author} {\bibfnamefont {Takeshi}\ \bibnamefont
  {Kondo}}, \bibinfo {author} {\bibfnamefont {M}~\bibnamefont {Ochi}}, \bibinfo
  {author} {\bibfnamefont {M}~\bibnamefont {Nakayama}}, \bibinfo {author}
  {\bibfnamefont {H}~\bibnamefont {Taniguchi}}, \bibinfo {author}
  {\bibfnamefont {S}~\bibnamefont {Akebi}}, \bibinfo {author} {\bibfnamefont
  {K}~\bibnamefont {Kuroda}}, \bibinfo {author} {\bibfnamefont {M}~\bibnamefont
  {Arita}}, \bibinfo {author} {\bibfnamefont {S}~\bibnamefont {Sakai}},
  \bibinfo {author} {\bibfnamefont {H}~\bibnamefont {Namatame}}, \bibinfo
  {author} {\bibfnamefont {M}~\bibnamefont {Taniguchi}}, \bibinfo {author}
  {\bibfnamefont {Y}~\bibnamefont {Maeno}}, \bibinfo {author} {\bibfnamefont
  {R}~\bibnamefont {Arita}}, \ and\ \bibinfo {author} {\bibfnamefont
  {S}~\bibnamefont {Shin}},\ }\bibfield  {title} {\enquote {\bibinfo {title}
  {{Orbital-Dependent Band Narrowing Revealed in an Extremely Correlated Hund's
  Metal Emerging on the Topmost Layer of Sr$_2$RuO$_4$}},}\ }\href {\doibase
  10.1103/PhysRevLett.117.247001} {\bibfield  {journal} {\bibinfo  {journal}
  {Phys. Rev. Lett.}\ }\textbf {\bibinfo {volume} {117}},\ \bibinfo {pages}
  {247001} (\bibinfo {year} {2016})}\BibitemShut {NoStop}%
\bibitem [{\citenamefont {Liu}\ \emph {et~al.}(2008)\citenamefont {Liu},
  \citenamefont {Antonov}, \citenamefont {Jepsen},\ and\ \citenamefont
  {Andersen.}}]{Liu2008}%
  \BibitemOpen
  \bibfield  {author} {\bibinfo {author} {\bibfnamefont {Guo-Qiang}\
  \bibnamefont {Liu}}, \bibinfo {author} {\bibfnamefont {V~N}\ \bibnamefont
  {Antonov}}, \bibinfo {author} {\bibfnamefont {O}~\bibnamefont {Jepsen}}, \
  and\ \bibinfo {author} {\bibfnamefont {O~K}\ \bibnamefont {Andersen.}},\
  }\bibfield  {title} {\enquote {\bibinfo {title} {{Coulomb-Enhanced Spin-Orbit
  Splitting: The Missing Piece in the ${\mathrm{Sr}}_{2}{\mathrm{RhO}}_{4}$
  Puzzle}},}\ }\href {https://link.aps.org/doi/10.1103/PhysRevLett.101.026408}
  {\bibfield  {journal} {\bibinfo  {journal} {Phys. Rev. Lett.}\ }\textbf
  {\bibinfo {volume} {101}},\ \bibinfo {pages} {26408} (\bibinfo {year}
  {2008})}\BibitemShut {NoStop}%
\bibitem [{\citenamefont {B{\"{u}}nemann}\ \emph {et~al.}(2016)\citenamefont
  {B{\"{u}}nemann}, \citenamefont {Linneweber}, \citenamefont {L{\"{o}}w},
  \citenamefont {Anders},\ and\ \citenamefont {Gebhard}}]{Bunemann2016}%
  \BibitemOpen
  \bibfield  {author} {\bibinfo {author} {\bibfnamefont {J{\"{o}}rg}\
  \bibnamefont {B{\"{u}}nemann}}, \bibinfo {author} {\bibfnamefont {Thorben}\
  \bibnamefont {Linneweber}}, \bibinfo {author} {\bibfnamefont {Ute}\
  \bibnamefont {L{\"{o}}w}}, \bibinfo {author} {\bibfnamefont {Frithjof~B}\
  \bibnamefont {Anders}}, \ and\ \bibinfo {author} {\bibfnamefont {Florian}\
  \bibnamefont {Gebhard}},\ }\bibfield  {title} {\enquote {\bibinfo {title}
  {{Interplay of Coulomb interaction and spin-orbit coupling}},}\ }\href
  {https://link.aps.org/doi/10.1103/PhysRevB.94.035116} {\bibfield  {journal}
  {\bibinfo  {journal} {Phy. Rev. B}\ }\textbf {\bibinfo {volume} {94}},\
  \bibinfo {pages} {035116} (\bibinfo {year} {2016})}\BibitemShut {NoStop}%
\bibitem [{\citenamefont {Hoshino}\ and\ \citenamefont
  {Werner}(2015)}]{Hoshino2015}%
  \BibitemOpen
  \bibfield  {author} {\bibinfo {author} {\bibfnamefont {Shintaro}\
  \bibnamefont {Hoshino}}\ and\ \bibinfo {author} {\bibfnamefont {Philipp}\
  \bibnamefont {Werner}},\ }\bibfield  {title} {\enquote {\bibinfo {title}
  {{Superconductivity from Emerging Magnetic Moments}},}\ }\href
  {https://link.aps.org/doi/10.1103/PhysRevLett.115.247001} {\bibfield
  {journal} {\bibinfo  {journal} {Phys. Rev. Lett.}\ }\textbf {\bibinfo
  {volume} {115}},\ \bibinfo {pages} {247001} (\bibinfo {year}
  {2015})}\BibitemShut {NoStop}%
\bibitem [{\citenamefont {Zgid}\ and\ \citenamefont {Chan}(2011)}]{Zgid2011}%
  \BibitemOpen
  \bibfield  {author} {\bibinfo {author} {\bibfnamefont {Dominika}\
  \bibnamefont {Zgid}}\ and\ \bibinfo {author} {\bibfnamefont {Garnet Kin-Lic}\
  \bibnamefont {Chan}},\ }\bibfield  {title} {\enquote {\bibinfo {title}
  {{Dynamical mean-field theory from a quantum chemical perspective}},}\ }\href
  {\doibase 10.1063/1.3556707} {\bibfield  {journal} {\bibinfo  {journal} {J.
  Chem. Phys.}\ }\textbf {\bibinfo {volume} {134}},\ \bibinfo {pages} {94115}
  (\bibinfo {year} {2011})}\BibitemShut {NoStop}%
\bibitem [{\citenamefont {Lu}\ \emph {et~al.}(2014)\citenamefont {Lu},
  \citenamefont {H{\"{o}}ppner}, \citenamefont {Gunnarsson},\ and\
  \citenamefont {Haverkort}}]{Lu2014}%
  \BibitemOpen
  \bibfield  {author} {\bibinfo {author} {\bibfnamefont {Y}~\bibnamefont {Lu}},
  \bibinfo {author} {\bibfnamefont {M}~\bibnamefont {H{\"{o}}ppner}}, \bibinfo
  {author} {\bibfnamefont {O}~\bibnamefont {Gunnarsson}}, \ and\ \bibinfo
  {author} {\bibfnamefont {M~W}\ \bibnamefont {Haverkort}},\ }\bibfield
  {title} {\enquote {\bibinfo {title} {{Efficient real-frequency solver for
  dynamical mean-field theory}},}\ }\href {\doibase 10.1103/PhysRevB.90.085102}
  {\bibfield  {journal} {\bibinfo  {journal} {Phys. Rev. B}\ }\textbf {\bibinfo
  {volume} {90}},\ \bibinfo {pages} {085102} (\bibinfo {year}
  {2014})}\BibitemShut {NoStop}%
\end{thebibliography}%


%merlin.mbs apsrev4-1.bst 2010-07-25 4.21a (PWD, AO, DPC) hacked
%Control: key (0)
%Control: author (72) initials jnrlst
%Control: editor formatted (1) identically to author
%Control: production of article title (-1) disabled
%Control: page (0) single
%Control: year (1) truncated
%Control: production of eprint (0) enabled
\begin{thebibliography}{12}%
\makeatletter
\providecommand \@ifxundefined [1]{%
 \@ifx{#1\undefined}
}%
\providecommand \@ifnum [1]{%
 \ifnum #1\expandafter \@firstoftwo
 \else \expandafter \@secondoftwo
 \fi
}%
\providecommand \@ifx [1]{%
 \ifx #1\expandafter \@firstoftwo
 \else \expandafter \@secondoftwo
 \fi
}%
\providecommand \natexlab [1]{#1}%
\providecommand \enquote  [1]{``#1''}%
\providecommand \bibnamefont  [1]{#1}%
\providecommand \bibfnamefont [1]{#1}%
\providecommand \citenamefont [1]{#1}%
\providecommand \href@noop [0]{\@secondoftwo}%
\providecommand \href [0]{\begingroup \@sanitize@url \@href}%
\providecommand \@href[1]{\@@startlink{#1}\@@href}%
\providecommand \@@href[1]{\endgroup#1\@@endlink}%
\providecommand \@sanitize@url [0]{\catcode `\\12\catcode `\$12\catcode
  `\&12\catcode `\#12\catcode `\^12\catcode `\_12\catcode `\%12\relax}%
\providecommand \@@startlink[1]{}%
\providecommand \@@endlink[0]{}%
\providecommand \url  [0]{\begingroup\@sanitize@url \@url }%
\providecommand \@url [1]{\endgroup\@href {#1}{\urlprefix }}%
\providecommand \urlprefix  [0]{URL }%
\providecommand \Eprint [0]{\href }%
\providecommand \doibase [0]{http://dx.doi.org/}%
\providecommand \selectlanguage [0]{\@gobble}%
\providecommand \bibinfo  [0]{\@secondoftwo}%
\providecommand \bibfield  [0]{\@secondoftwo}%
\providecommand \translation [1]{[#1]}%
\providecommand \BibitemOpen [0]{}%
\providecommand \bibitemStop [0]{}%
\providecommand \bibitemNoStop [0]{.\EOS\space}%
\providecommand \EOS [0]{\spacefactor3000\relax}%
\providecommand \BibitemShut  [1]{\csname bibitem#1\endcsname}%
\let\auto@bib@innerbib\@empty
%</preamble>
\bibitem [{\citenamefont {Mravlje}\ \emph {et~al.}(2011)\citenamefont
  {Mravlje}, \citenamefont {Aichhorn}, \citenamefont {Miyake}, \citenamefont
  {Haule}, \citenamefont {Kotliar},\ and\ \citenamefont
  {Georges}}]{Mravlje2011}%
  \BibitemOpen
  \bibfield  {author} {\bibinfo {author} {\bibfnamefont {J.}~\bibnamefont
  {Mravlje}}, \bibinfo {author} {\bibfnamefont {M.}~\bibnamefont {Aichhorn}},
  \bibinfo {author} {\bibfnamefont {T.}~\bibnamefont {Miyake}}, \bibinfo
  {author} {\bibfnamefont {K.}~\bibnamefont {Haule}}, \bibinfo {author}
  {\bibfnamefont {G.}~\bibnamefont {Kotliar}}, \ and\ \bibinfo {author}
  {\bibfnamefont {A.}~\bibnamefont {Georges}},\ }\href
  {https://link.aps.org/doi/10.1103/PhysRevLett.106.096401} {\bibfield
  {journal} {\bibinfo  {journal} {Phys. Rev. Lett.}\ }\textbf {\bibinfo
  {volume} {106}},\ \bibinfo {pages} {096401} (\bibinfo {year}
  {2011})}\BibitemShut {NoStop}%
\bibitem [{\citenamefont {Kim}\ \emph {et~al.}(2018)\citenamefont {Kim},
  \citenamefont {Mravlje}, \citenamefont {Ferrero}, \citenamefont {Parcollet},\
  and\ \citenamefont {Georges}}]{MKim2018}%
  \BibitemOpen
  \bibfield  {author} {\bibinfo {author} {\bibfnamefont {M.}~\bibnamefont
  {Kim}}, \bibinfo {author} {\bibfnamefont {J.}~\bibnamefont {Mravlje}},
  \bibinfo {author} {\bibfnamefont {M.}~\bibnamefont {Ferrero}}, \bibinfo
  {author} {\bibfnamefont {O.}~\bibnamefont {Parcollet}}, \ and\ \bibinfo
  {author} {\bibfnamefont {A.}~\bibnamefont {Georges}},\ }\href {\doibase
  10.1103/PhysRevLett.120.126401} {\bibfield  {journal} {\bibinfo  {journal}
  {Phys. Rev. Lett.}\ }\textbf {\bibinfo {volume} {120}},\ \bibinfo {pages}
  {126401} (\bibinfo {year} {2018})}\BibitemShut {NoStop}%
\bibitem [{\citenamefont {Tamai}\ \emph {et~al.}(2019)\citenamefont {Tamai},
  \citenamefont {Zingl}, \citenamefont {Rozbicki}, \citenamefont {Cappelli},
  \citenamefont {Ricc\`o}, \citenamefont {de~la Torre}, \citenamefont
  {McKeown~Walker}, \citenamefont {Bruno}, \citenamefont {King}, \citenamefont
  {Meevasana}, \citenamefont {Shi}, \citenamefont
  {Radovi\ifmmode~\acute{c}\else \'{c}\fi{}}, \citenamefont {Plumb},
  \citenamefont {Gibbs}, \citenamefont {Mackenzie}, \citenamefont {Berthod},
  \citenamefont {Strand}, \citenamefont {Kim}, \citenamefont {Georges},\ and\
  \citenamefont {Baumberger}}]{Tamai2019}%
  \BibitemOpen
  \bibfield  {author} {\bibinfo {author} {\bibfnamefont {A.}~\bibnamefont
  {Tamai}}, \bibinfo {author} {\bibfnamefont {M.}~\bibnamefont {Zingl}},
  \bibinfo {author} {\bibfnamefont {E.}~\bibnamefont {Rozbicki}}, \bibinfo
  {author} {\bibfnamefont {E.}~\bibnamefont {Cappelli}}, \bibinfo {author}
  {\bibfnamefont {S.}~\bibnamefont {Ricc\`o}}, \bibinfo {author} {\bibfnamefont
  {A.}~\bibnamefont {de~la Torre}}, \bibinfo {author} {\bibfnamefont
  {S.}~\bibnamefont {McKeown~Walker}}, \bibinfo {author} {\bibfnamefont
  {F.~Y.}\ \bibnamefont {Bruno}}, \bibinfo {author} {\bibfnamefont {P.~D.~C.}\
  \bibnamefont {King}}, \bibinfo {author} {\bibfnamefont {W.}~\bibnamefont
  {Meevasana}}, \bibinfo {author} {\bibfnamefont {M.}~\bibnamefont {Shi}},
  \bibinfo {author} {\bibfnamefont {M.}~\bibnamefont
  {Radovi\ifmmode~\acute{c}\else \'{c}\fi{}}}, \bibinfo {author} {\bibfnamefont
  {N.~C.}\ \bibnamefont {Plumb}}, \bibinfo {author} {\bibfnamefont {A.~S.}\
  \bibnamefont {Gibbs}}, \bibinfo {author} {\bibfnamefont {A.~P.}\ \bibnamefont
  {Mackenzie}}, \bibinfo {author} {\bibfnamefont {C.}~\bibnamefont {Berthod}},
  \bibinfo {author} {\bibfnamefont {H.~U.~R.}\ \bibnamefont {Strand}}, \bibinfo
  {author} {\bibfnamefont {M.}~\bibnamefont {Kim}}, \bibinfo {author}
  {\bibfnamefont {A.}~\bibnamefont {Georges}}, \ and\ \bibinfo {author}
  {\bibfnamefont {F.}~\bibnamefont {Baumberger}},\ }\href {\doibase
  10.1103/PhysRevX.9.021048} {\bibfield  {journal} {\bibinfo  {journal} {Phys.
  Rev. X}\ }\textbf {\bibinfo {volume} {9}},\ \bibinfo {pages} {021048}
  (\bibinfo {year} {2019})}\BibitemShut {NoStop}%
\bibitem [{\citenamefont {Kugler}\ \emph {et~al.}(2020)\citenamefont {Kugler},
  \citenamefont {Zingl}, \citenamefont {Strand}, \citenamefont {Lee},
  \citenamefont {von Delft},\ and\ \citenamefont {Georges}}]{Kugler2020}%
  \BibitemOpen
  \bibfield  {author} {\bibinfo {author} {\bibfnamefont {F.~B.}\ \bibnamefont
  {Kugler}}, \bibinfo {author} {\bibfnamefont {M.}~\bibnamefont {Zingl}},
  \bibinfo {author} {\bibfnamefont {H.~U.~R.}\ \bibnamefont {Strand}}, \bibinfo
  {author} {\bibfnamefont {S.-S.~B.}\ \bibnamefont {Lee}}, \bibinfo {author}
  {\bibfnamefont {J.}~\bibnamefont {von Delft}}, \ and\ \bibinfo {author}
  {\bibfnamefont {A.}~\bibnamefont {Georges}},\ }\href {\doibase
  10.1103/PhysRevLett.124.016401} {\bibfield  {journal} {\bibinfo  {journal}
  {Phys. Rev. Lett.}\ }\textbf {\bibinfo {volume} {124}},\ \bibinfo {pages}
  {16401} (\bibinfo {year} {2020})}\BibitemShut {NoStop}%
\bibitem [{\citenamefont {Linden}\ \emph {et~al.}(2020)\citenamefont {Linden},
  \citenamefont {Zingl}, \citenamefont {Hubig}, \citenamefont {Parcollet},\
  and\ \citenamefont {Schollw{\"{o}}ck}}]{Linden2020}%
  \BibitemOpen
  \bibfield  {author} {\bibinfo {author} {\bibfnamefont {N.-O.}\ \bibnamefont
  {Linden}}, \bibinfo {author} {\bibfnamefont {M.}~\bibnamefont {Zingl}},
  \bibinfo {author} {\bibfnamefont {C.}~\bibnamefont {Hubig}}, \bibinfo
  {author} {\bibfnamefont {O.}~\bibnamefont {Parcollet}}, \ and\ \bibinfo
  {author} {\bibfnamefont {U.}~\bibnamefont {Schollw{\"{o}}ck}},\ }\href
  {\doibase 10.1103/PhysRevB.101.041101} {\bibfield  {journal} {\bibinfo
  {journal} {Phys. Rev. B}\ }\textbf {\bibinfo {volume} {101}},\ \bibinfo
  {pages} {41101} (\bibinfo {year} {2020})}\BibitemShut {NoStop}%
\bibitem [{\citenamefont {Karp}\ \emph {et~al.}(2020)\citenamefont {Karp},
  \citenamefont {Bramberger}, \citenamefont {Grundner}, \citenamefont
  {Schollw{\"{o}}ck}, \citenamefont {Millis},\ and\ \citenamefont
  {Zingl}}]{Karp2020}%
  \BibitemOpen
  \bibfield  {author} {\bibinfo {author} {\bibfnamefont {J.}~\bibnamefont
  {Karp}}, \bibinfo {author} {\bibfnamefont {M.}~\bibnamefont {Bramberger}},
  \bibinfo {author} {\bibfnamefont {M.}~\bibnamefont {Grundner}}, \bibinfo
  {author} {\bibfnamefont {U.}~\bibnamefont {Schollw{\"{o}}ck}}, \bibinfo
  {author} {\bibfnamefont {A.~J.}\ \bibnamefont {Millis}}, \ and\ \bibinfo
  {author} {\bibfnamefont {M.}~\bibnamefont {Zingl}},\ }\href
  {http://arxiv.org/abs/2004.12515} {\enquote {\bibinfo {title} {{Sr$_2$MoO$_4$
  and Sr$_2$RuO$_4$: Disentangling the Roles of Hund's and van Hove
  Physics}},}\ } (\bibinfo {year} {2020}),\ \Eprint
  {http://arxiv.org/abs/2004.12515} {arXiv:2004.12515} \BibitemShut {NoStop}%
\bibitem [{\citenamefont {Bergemann}\ \emph {et~al.}(2003)\citenamefont
  {Bergemann}, \citenamefont {Mackenzie}, \citenamefont {Julian}, \citenamefont
  {Forsythe},\ and\ \citenamefont {Ohmichi}}]{Bergemann2003}%
  \BibitemOpen
  \bibfield  {author} {\bibinfo {author} {\bibfnamefont {C.}~\bibnamefont
  {Bergemann}}, \bibinfo {author} {\bibfnamefont {A.~P.}\ \bibnamefont
  {Mackenzie}}, \bibinfo {author} {\bibfnamefont {S.~R.}\ \bibnamefont
  {Julian}}, \bibinfo {author} {\bibfnamefont {D.}~\bibnamefont {Forsythe}}, \
  and\ \bibinfo {author} {\bibfnamefont {E.}~\bibnamefont {Ohmichi}},\ }\href
  {\doibase 10.1080/00018730310001621737} {\bibfield  {journal} {\bibinfo
  {journal} {Adv. Phys.}\ }\textbf {\bibinfo {volume} {52}},\ \bibinfo {pages}
  {639} (\bibinfo {year} {2003})}\BibitemShut {NoStop}%
\bibitem [{\citenamefont {Liebsch}\ and\ \citenamefont
  {Ishida}(2012)}]{Liebsch2012}%
  \BibitemOpen
  \bibfield  {author} {\bibinfo {author} {\bibfnamefont {A.}~\bibnamefont
  {Liebsch}}\ and\ \bibinfo {author} {\bibfnamefont {H.}~\bibnamefont
  {Ishida}},\ }\href {http://stacks.iop.org/0953-8984/24/i=5/a=053201}
  {\bibfield  {journal} {\bibinfo  {journal} {J. Phys. Condens. Matter}\
  }\textbf {\bibinfo {volume} {24}},\ \bibinfo {pages} {53201} (\bibinfo {year}
  {2012})}\BibitemShut {NoStop}%
\bibitem [{\citenamefont {Koch}\ \emph {et~al.}(2008)\citenamefont {Koch},
  \citenamefont {Sangiovanni},\ and\ \citenamefont {Gunnarsson}}]{Koch2008}%
  \BibitemOpen
  \bibfield  {author} {\bibinfo {author} {\bibfnamefont {E.}~\bibnamefont
  {Koch}}, \bibinfo {author} {\bibfnamefont {G.}~\bibnamefont {Sangiovanni}}, \
  and\ \bibinfo {author} {\bibfnamefont {O.}~\bibnamefont {Gunnarsson}},\
  }\href {https://link.aps.org/doi/10.1103/PhysRevB.78.115102} {\bibfield
  {journal} {\bibinfo  {journal} {Phys. Rev. B}\ }\textbf {\bibinfo {volume}
  {78}},\ \bibinfo {pages} {115102} (\bibinfo {year} {2008})}\BibitemShut
  {NoStop}%
\bibitem [{\citenamefont {S{\'{e}}n{\'{e}}chal}(2010)}]{Senechal2010}%
  \BibitemOpen
  \bibfield  {author} {\bibinfo {author} {\bibfnamefont {D.}~\bibnamefont
  {S{\'{e}}n{\'{e}}chal}},\ }\href
  {https://link.aps.org/doi/10.1103/PhysRevB.81.235125} {\bibfield  {journal}
  {\bibinfo  {journal} {Phys. Rev. B}\ }\textbf {\bibinfo {volume} {81}},\
  \bibinfo {pages} {235125} (\bibinfo {year} {2010})}\BibitemShut {NoStop}%
\bibitem [{\citenamefont {Go}\ and\ \citenamefont {Millis}(2015)}]{Go2015}%
  \BibitemOpen
  \bibfield  {author} {\bibinfo {author} {\bibfnamefont {A.}~\bibnamefont
  {Go}}\ and\ \bibinfo {author} {\bibfnamefont {A.~J.}\ \bibnamefont
  {Millis}},\ }\href {http://link.aps.org/doi/10.1103/PhysRevLett.114.016402}
  {\bibfield  {journal} {\bibinfo  {journal} {Phys. Rev. Lett.}\ }\textbf
  {\bibinfo {volume} {114}},\ \bibinfo {pages} {016402} (\bibinfo {year}
  {2015})}\BibitemShut {NoStop}%
\bibitem [{\citenamefont {Fertitta}\ and\ \citenamefont
  {Booth}(2018)}]{Fertitta2018}%
  \BibitemOpen
  \bibfield  {author} {\bibinfo {author} {\bibfnamefont {E.}~\bibnamefont
  {Fertitta}}\ and\ \bibinfo {author} {\bibfnamefont {G.~H.}\ \bibnamefont
  {Booth}},\ }\href {\doibase 10.1103/PhysRevB.98.235132} {\bibfield  {journal}
  {\bibinfo  {journal} {Phys. Rev. B}\ }\textbf {\bibinfo {volume} {98}},\
  \bibinfo {pages} {235132} (\bibinfo {year} {2018})}\BibitemShut {NoStop}%
\end{thebibliography}%

\end{document}

\title{
	Supplemental Material for \\
	``Fermi liquid behavior and van Hove singularity in Sr$_2$RuO$_4$''
}

\author{Hyeong~Jun \surname{Lee}}
%\email[]{jun5921@snu.ac.kr}

\affiliation{Center for Theoretical Physics of Complex Systems, Institute for Basic Science (IBS), Daejeon 34126, Republic of Korea}
\affiliation{Center for Correlated Electron Systems, Institute for Basic Science (IBS), Seoul 08826, Republic of Korea}
\affiliation{Department of Physics and Astronomy, Seoul National University, Seoul 08826, Republic of Korea}

\author{Choong~H. \surname{Kim}}
\email[]{chkim82@snu.ac.kr}

\affiliation{Center for Correlated Electron Systems, Institute for Basic Science (IBS), Seoul 08826, Republic of Korea}
\affiliation{Department of Physics and Astronomy, Seoul National University, Seoul 08826, Republic of Korea}

\author{Ara \surname{Go}}
\email[]{arago@ibs.re.kr}

\affiliation{Center for Theoretical Physics of Complex Systems, Institute for Basic Science (IBS), Daejeon 34126, Republic of Korea}

%\date{\today}
\maketitle

\beginsupplement

\subsection*{Tight-binding Hamiltonian}

\begin{figure}[hp]
	\includegraphics[width=0.5\columnwidth]{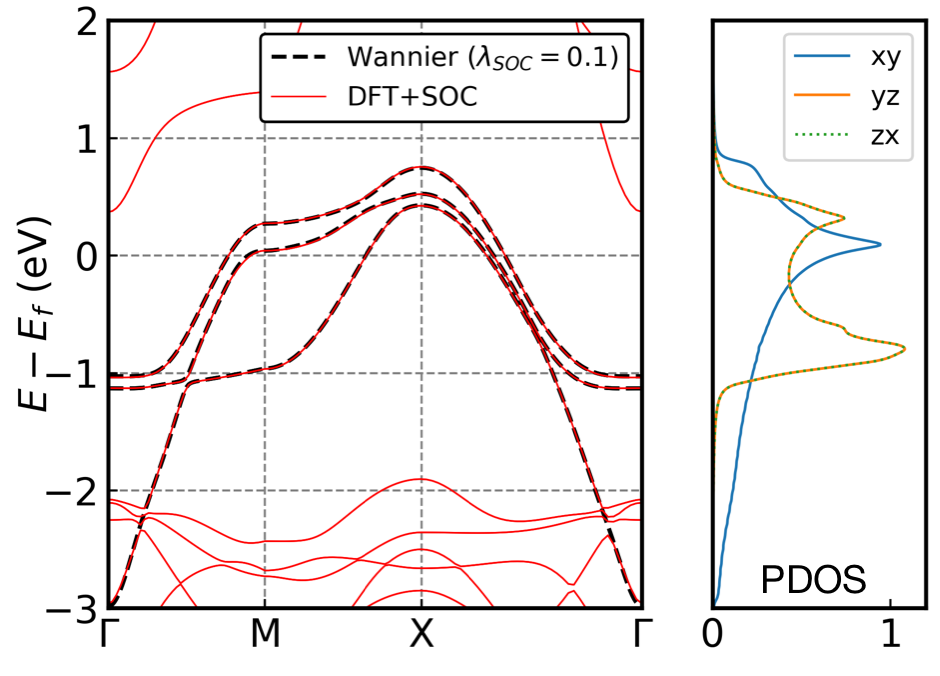}
	\caption{
	(a) DFT and MLWF bands of \SRO along the high-symmetry lines.
	(b) Density of states of MLWF bands on \ttg.
	}
\label{fig:dft}
\end{figure}

We obtain the tight-binding (TB) Hamiltonian $\mathcal{H}_{\mathrm{TB}} =  \sum_{i,j} \sum_{\mu,\beta} t_{\mu\nu}^{ij} \cre{i\mu} \can{j\nu} $ on the maximally localized Wannier functions (MLWF)
from density functional theory (DFT) calculation with the PBEsol functional. The indices $\mu$ and $\nu$ denote the $t_{2g}$ orbitals.
Some parameters between nearest neighbors and next-nearest neighbors are obtained as 
$	t^{\pm1,0,0}_{xy,xy} = -0.388	$, 
$	t^{\pm1,0,0}_{yz,yz} = -0.325	$,
$	t^{0,\pm1,0}_{yz,yz} = -0.048	$,
$	t^{\pm1,\pm1,0}_{xy,xy} = -0.130	$,
$	t^{\pm1,\pm1,0}_{yz,yz} = -0.017	$, and
$	t^{\pm1,\pm1,0}_{xy,yz} =  0.010	$.
We introduce spin-orbit coupling (SOC) with $\lsoc \sum_i \bm{l}_i \cdot \bm{s}_i$, where $i$ is the index for electrons on a site, in the TB Hamiltonian.
From fitting the band structure of the Wannier to the DFT bands having SOC, the strength of the SOC, $\lsoc$, is set to 0.1 eV (as shown in Fig.~\ref{fig:dft}).
We perform dynamical mean-field theory (DMFT) calculation based on the top of this TB Hamiltonian with $\jeff$-basis states.

We also perform local-density approximation (LDA) calculations to obtain the MLWF and TB Hamiltonian. 
With slightly reduced bandwidths compared to that from PBEsol, this gives only a minor correction on the magnitudes of the self-energies for fixed $U$ and $J_H$ of our interest in the DMFT calculation.
We confirm that this correction does not significantly affect our main result; it is expected to lead to a small shift in the critical Hund's coupling strength associated with the resulting correlation effects.

\subsection*{Orbital-dependent correlations}
\begin{figure}[!tbh]
	\includegraphics[width=0.9\columnwidth]{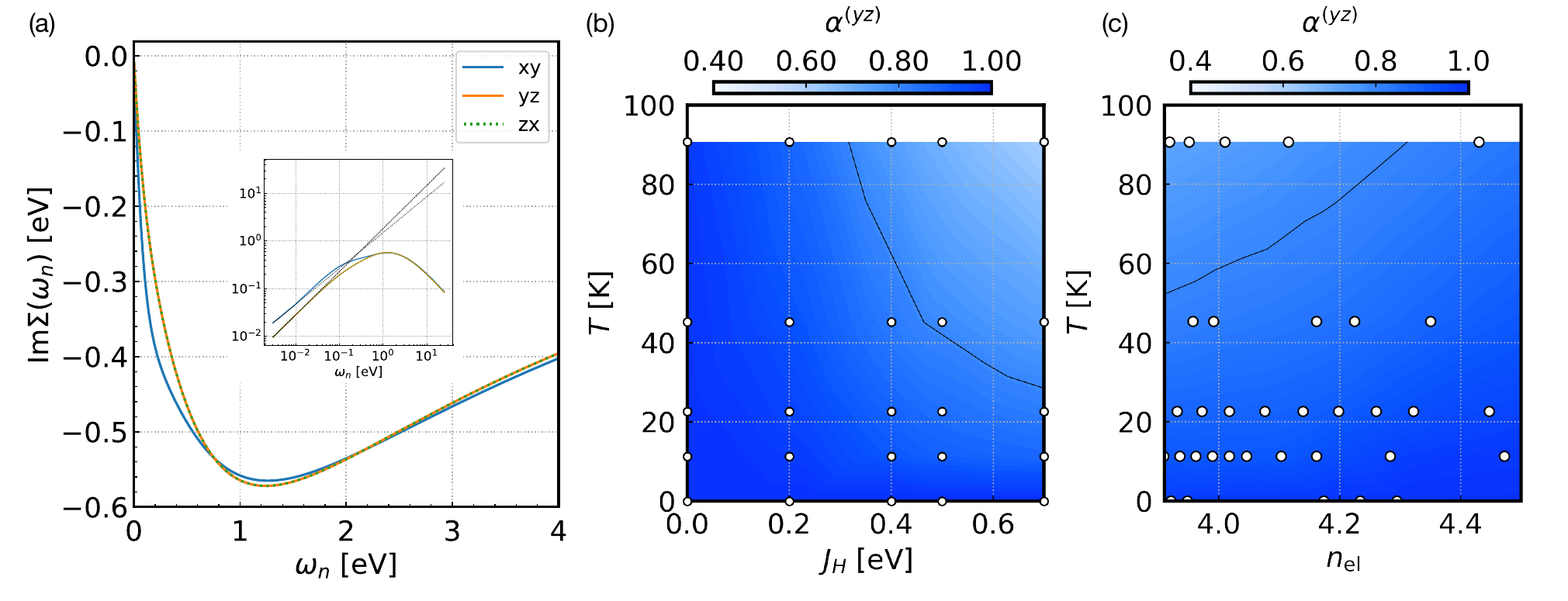}
	\caption{
	(a) Imaginary part of the Matsubara self-energy projected on the \ttg-basis in \SRO at $T\simeq10$ K.
	The inset is a log-log plot with fitting lines showing different power exponents: $-0.75$ for \dxy and $0.91$ for \dyz/\dzx.
	(b,c) Power exponent $\alpha$ of the \dyz component in the $T$-$J_H$ plane (b) and in the $T$-$n$ plane (c).
	}
\label{fig:self}
\end{figure}
%\begin{itemize}
%	\item Comparsion of xy's and yz's
%	\item basis-dependence
%	\item Temperature-dependence %	\item One example showing the power devitations above
%	\item Effective masses
%\end{itemize}
From the DFT band structure and density of states, the effective model of \SRO can be described by two degenerate \dyz and \dzx bands along with the other \dxy band.
This difference between the bands enables \SRO to gain orbital selectivity, which is one of the remarkable aspects of the correlation effects driven by Hund's coupling.
Figure~\ref{fig:self}(a) shows the orbital-dependent correlations in the self-energy on \ttg.
The imaginary part of the self-energy of the \dxy band is more enhanced than that of \dyz / \dzx near $\omega=0$.
This difference exists noticeably in low-energy excitations and maintains up to 0.5 eV.

The Matsubara-frequency self-energy near $\omega=0$ is associated with renormalized mass and a renormalization factor when its imaginary part follows linear (or Fermi liquid) behavior in $\omega_n$.
From the $T$ = 0 data, we compute and compare the renormalized masses $m^*$ on \ttg and $\jeff$ defined as  $m^*=\left( 1-\partial_\omega \mathrm{Im}\Sigma(i\omega) |_{\omega \rightarrow 0} \right)^{-1}$.
We obtain $(m^*_{\threeh} , m^*_{\halfh} , m^*_{\threet})\simeq$ (5.60, 5.09, 3.94) at $T=0$ K and (4.56, 4.27, 3.61) at $T=11$ K.
We use a fourth order polynomial with the lowest six Matsubara frequencies and fit the imaginary part of self-energy;
comparisons to $m^*$ from an experiment and previous DMFT studies with the \ttg and $\jeff$ basis are given in Table~\ref{tab:mass}.

\begin{table}[hbt]
\caption{Renormalized masses on the \ttg and $\jeff$ basis at various temperatures compared to previous Sr$_2$RuO$_4$ studies.}
\label{tab:mass}
\begin{tabular}{ l|l l l|l  }
\hline
								& $m_{xy}$ & $m_{yz/zx}$ & & Etc. \\
								& ($m_{\threeh}$ & $m_{\halfh}$ & $m_{\threet}$) & \\ \hline
Mravlje \textit{et al.} 2011PRL	\cite{Mravlje2011}		& 4.5 & 3.3 & & imag. freq. $T\gtrsim50$ K \\ \hline
Kim \textit{et al.} 2018PRL \cite{MKim2018}			& 4.0 & 3.1 & & real freq. $T=230$ K ($\lambda_{\mathrm{SOC}}=0.1$ eV) \\ \hline
Tamai \textit{et al.} 2019PRX \cite{Tamai2019}			& 5.55(30) & 3.33(30) & & real freq. $T=29$ K \\ \hline
Kugler \textit{et al.} 2020PRL \cite{Kugler2020}		& $\sim$5.9 & $\sim$3.6 & & real freq. (saturating below 10 K or 40 K) \\ \hline
Linden \textit{et al.} 2020PRB \cite{Linden2020}		& 5.26 & 3.45 & & imag freq. (probably at $T=58$ K)  \\ 
								& 5.00 & 3.70 & & imag freq. (probably at $T=58$ K, $\lambda_{\mathrm{SOC}}=0.1$ eV) \\ \hline
Karp \textit{et al.} 2020arXiv \cite{Karp2020}			& $\sim$5.3 & $\sim$3.4 & & imag. freq. at 0 K, MPS ($\beta^{\mathrm{fit}}=200$ eV${^{-1}}$) \\ 
								& $\sim$5.0 & $\sim$3.5 & & imag. freq. at 29 K, CTHYB \\ \hline
Ours								& 4.86 & 3.50 & & imag. freq. at 0 K    ($t_{2g}$, $\beta^{\mathrm{fit}}=256$ eV${^{-1}}$) \\
($\lambda_{\mathrm{SOC}}=0$ eV)					& 6.51 & 3.82 & & imag. freq. at 0 K    ($t_{2g}$, $\beta^{\mathrm{fit}}\rightarrow \infty$) \\
								& 5.38 & 3.63 & & imag. freq. at 11 K  ($t_{2g}$) \\ 
								& 4.76 & 3.48 & & imag. freq. at 23 K  ($t_{2g}$) \\ \hline
Ours								& 6.69 & 4.04 & & imag. freq. at 0 K    ($t_{2g}$) \\ 
($\lambda_{\mathrm{SOC}}=0.1$ eV)				& 5.03 & 3.67 & & imag. freq. at 11 K  ($t_{2g}$)   \\
								& 5.60 & 5.09 & 3.94 & imag. freq. at 0 K    ($j_{\mathrm{eff}}$)  \\ 
								& 4.56 & 4.27 & 3.61 & imag. freq. at 11 K  ($j_{\mathrm{eff}}$) \\ \hline
Experiment (dHvA) \cite{Bergemann2003}				& 5.5 & 3.5 & 3.0 &  ($m_{\gamma}$ , $m_{\beta}$ , $m_{\alpha}$) \\
\hline
\end{tabular}
\end{table}

%\begin{figure}[!h]
%	\includegraphics[width=0.7\columnwidth]{supp_mass_Z_T}
%	\caption{
%	}
%\label{fig:massT}
%\end{figure}

\clearpage

\subsection*{Self-energy with electron-doping and power exponent $\alpha$}
\begin{figure}[!h]
	\includegraphics[width=0.35\columnwidth]{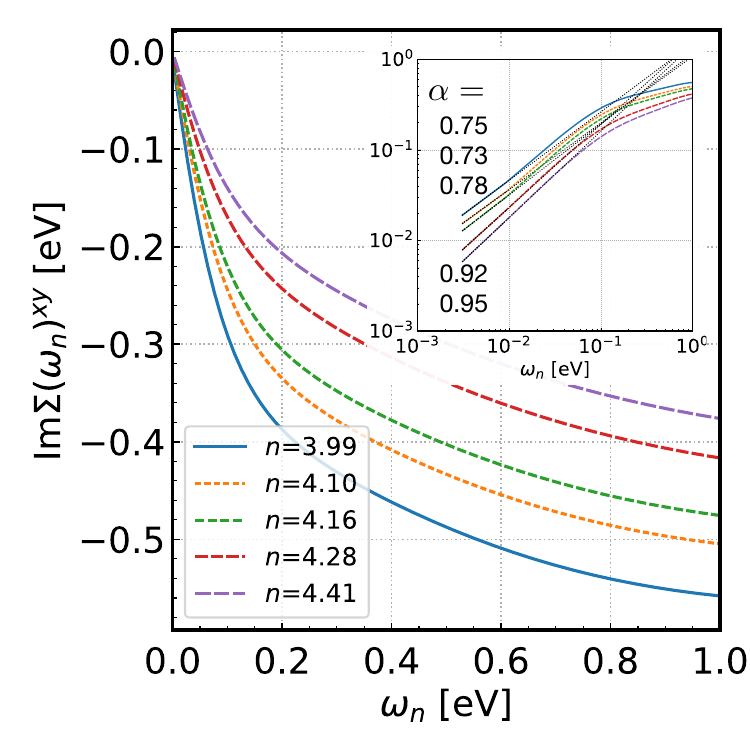}
	\caption{
		Imaginary part of the self-energies in the Matsubara frequencies for various electron occupations with $(U, J_H, \lambda_{\mathrm{SOC}})=(2.5, 0.4, 0.1)$ eV at $T=11$ K.
		The inset is a log-log plot of the same data, their fitting lines of $(\omega_n)^{\alpha}$, and their corresponding $\alpha$ values.
	}
\label{fig:self_n}
\end{figure}

\subsection*{Comparison to the case in the absence of SOC}
\begin{figure}[!h]
	\includegraphics[width=0.9\columnwidth]{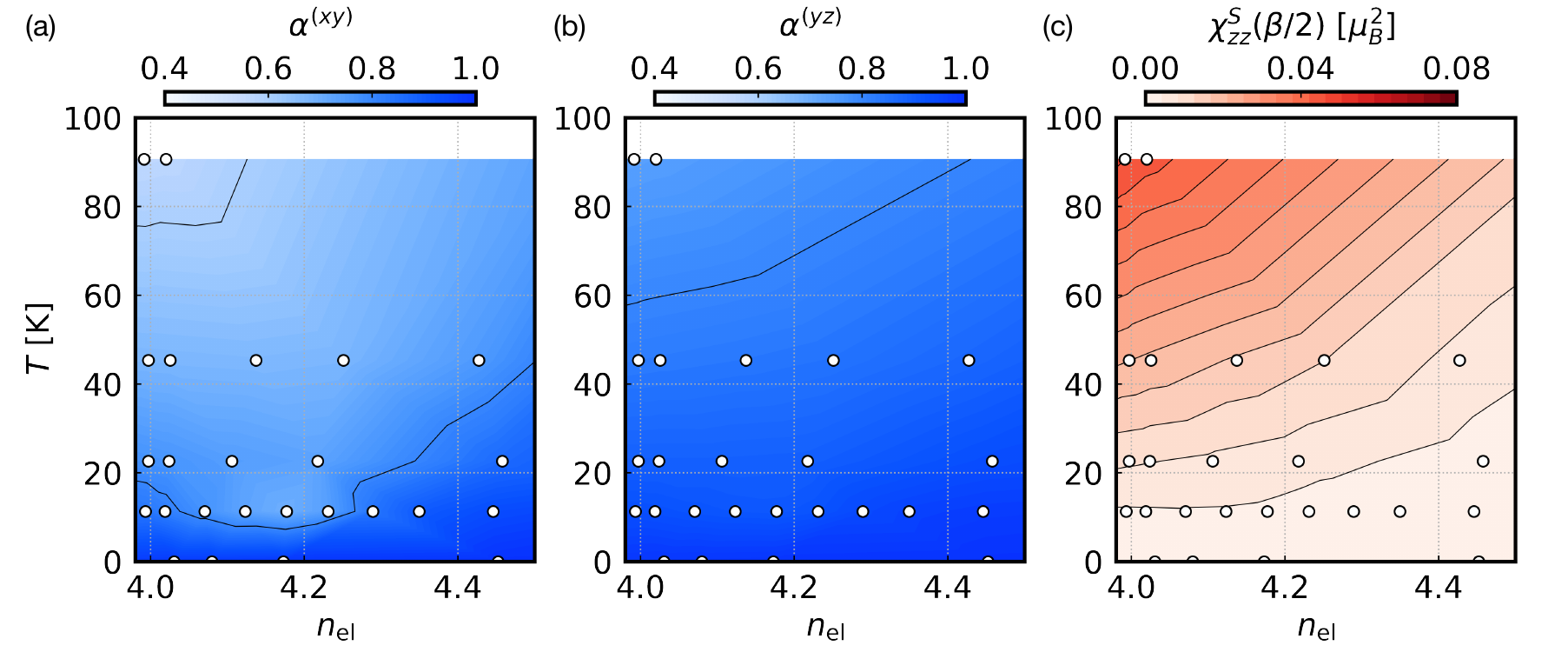}
	\caption{
	(a,b) Power exponent $\alpha$ of the imaginary part of the Matsubara self-energy
	of the \dxy component (a) and the \dyz component (b).
	(c) Long-time correlator $\chi^{S}_{zz}(\tau=\beta/2)$ in the $T$-$n$ plane.
	}
\label{fig:zerosoc}
\end{figure}

\subsection*{Fitting bath parameters in the ED solver}
One important aspect of the exact diagonalization (ED) solver is how to fit the bath parameters in an impurity model.
All the data presented in the main text come from minimizing a non-uniformly weighted cost function, such that
\begin{align}
        \chi^{\mathrm{cost}} (\{\epsilon_l,V_{\mu l}\})
        = \sum_n \sum_{\mu\nu} W(i\omega_n) \left|[\mathcal{G}^{-1}_{0,\mathrm{target}}(i\omega_n)]_{\mu\nu} - [\mathcal{G}^{-1}_0(i\omega_n;\{\epsilon_l,V_{\mu l}\})]_{\mu\nu} \right|^2,
\end{align}
where $\epsilon_l$ is the $l$-th bath orbital energy, $V_{\mu l}$ is the hybridization strength between the $l$-th bath and $\mu$-th correlated orbitals,
       $\mathcal{G}^{-1}_0$ is a non-interacting Green function,
      and $W(i\omega_n) = \frac{1}{\mathcal{N}} (\omega_n)^{-\gamma} $ where $\mathcal{N}$ is a normalization constant.
We set $\gamma=1$ to weight low-frequency, which is believed to be more reliable in strongly correlated metals~\cite{Liebsch2012, Linden2020}.
We use three (six, considering spin) correlated and nine (eighteen) bath orbitals to satisfy the minimum requirement to describe multi-orbital physics~\cite{Koch2008,Senechal2010,Liebsch2012,Go2015,Fertitta2018}.
Futher consideration of the number of bath orbitals, namely, more than three per correlated orbital, is expceted not to improve the spectra near the Fermi level but rather the resolution of high-frequency spectra~\cite{Go2015,Fertitta2018}.
%It would be also an interesting future work to analyze various cases of more complex multi-orbital correlated systems such that VHS or SOC play an important role.

%%%%%%%%%%%%%%%%%%%%%%%%%%%%%%%%%%%%%%%%%%%%%%%%%%%%%%%%%%%
\bibliography{supp_SRO214}